\pdfoutput=1

\documentclass[11pt]{article}

\usepackage[final]{acl}

\usepackage{times}
\usepackage{svg}
\usepackage{latexsym}
\usepackage[T1]{fontenc}
\usepackage[utf8]{inputenc}
\usepackage{microtype}
\usepackage{inconsolata}
\usepackage{graphicx}
\usepackage{amsmath}
\usepackage{cleveref}
\usepackage{booktabs}
\usepackage{multirow}
\usepackage{array}
\usepackage{longtable}
\usepackage{geometry}
\usepackage{colortbl}
\usepackage{url}
\usepackage{amssymb}
\usepackage{xspace}
\usepackage{booktabs}
\usepackage{tcolorbox}
\usepackage{xcolor}
\usepackage{enumitem}

\definecolor{promptbg}{RGB}{240, 248, 255}  
\definecolor{promptborder}{RGB}{30, 144, 255}  

\title{Evaluating the Effectiveness and Scalability of LLM-Based Data Augmentation for Retrieval}

\author{\parbox{0.9\linewidth}{\centering{Pranjal A. Chitale \quad Bishal Santra \quad Yashoteja Prabhu \quad Amit Sharma \\
  {\rm Microsoft Research India \\ } {\tt \small pranjalchitale@gmail.com, \{bsantra, yprabhu, amshar\}@microsoft.com}}}}

\begin{document}
\maketitle

\begin{abstract}
Compact dual-encoder models are widely used for retrieval owing to their efficiency and scalability. However, such models often underperform compared to their Large Language Model (LLM)-based retrieval counterparts, likely due to their limited world knowledge. While LLM-based data augmentation has been proposed as a strategy to bridge this performance gap, there is insufficient understanding of its effectiveness and scalability to real-world retrieval problems. Existing research does not systematically explore key factors such as the optimal augmentation scale, the necessity of using large augmentation models, and whether diverse augmentations improve generalization, particularly in out-of-distribution (OOD) settings.
This work presents a comprehensive study of the effectiveness of LLM augmentation for retrieval, comprising over 100 distinct experimental settings of retrieval models, augmentation models and augmentation strategies. We find that, while augmentation enhances retrieval performance, its benefits diminish beyond a certain augmentation scale, even with diverse augmentation strategies. Surprisingly, we observe that augmentation with smaller LLMs can achieve performance competitive with larger augmentation models. Moreover, we examine how augmentation effectiveness varies with retrieval model pre-training, revealing that augmentation provides the most benefit to models which are not well pre-trained. Our insights pave the way for more judicious and efficient augmentation strategies, thus enabling informed decisions and maximising retrieval performance while being more cost-effective.\footnote{Code and augmented datasets accompanying this work are publicly available at \url{https://aka.ms/DAGR}.}
\end{abstract}

\section{Introduction}
\label{sec:introduction}
Recent advances in LLMs have significantly influenced information retrieval (IR), with LLM-based retrievers achieving state-of-the-art performance across benchmarks \citep{wang-etal-2024-improving-text,behnamghader2024llmvec}. However, the high computational cost of LLMs renders them impractical for real-world deployment. Consequently, retrieval systems often employ compact dual-encoder models, which, while efficient, typically underperform compared to LLM-based retrievers due to limited world knowledge. LLM-based augmentation of synthetic training pairs to enhance smaller retrievers has emerged as a promising strategy to address this performance gap. 

\begin{figure*}[!ht]
    \centering
    \includegraphics[width=\linewidth]{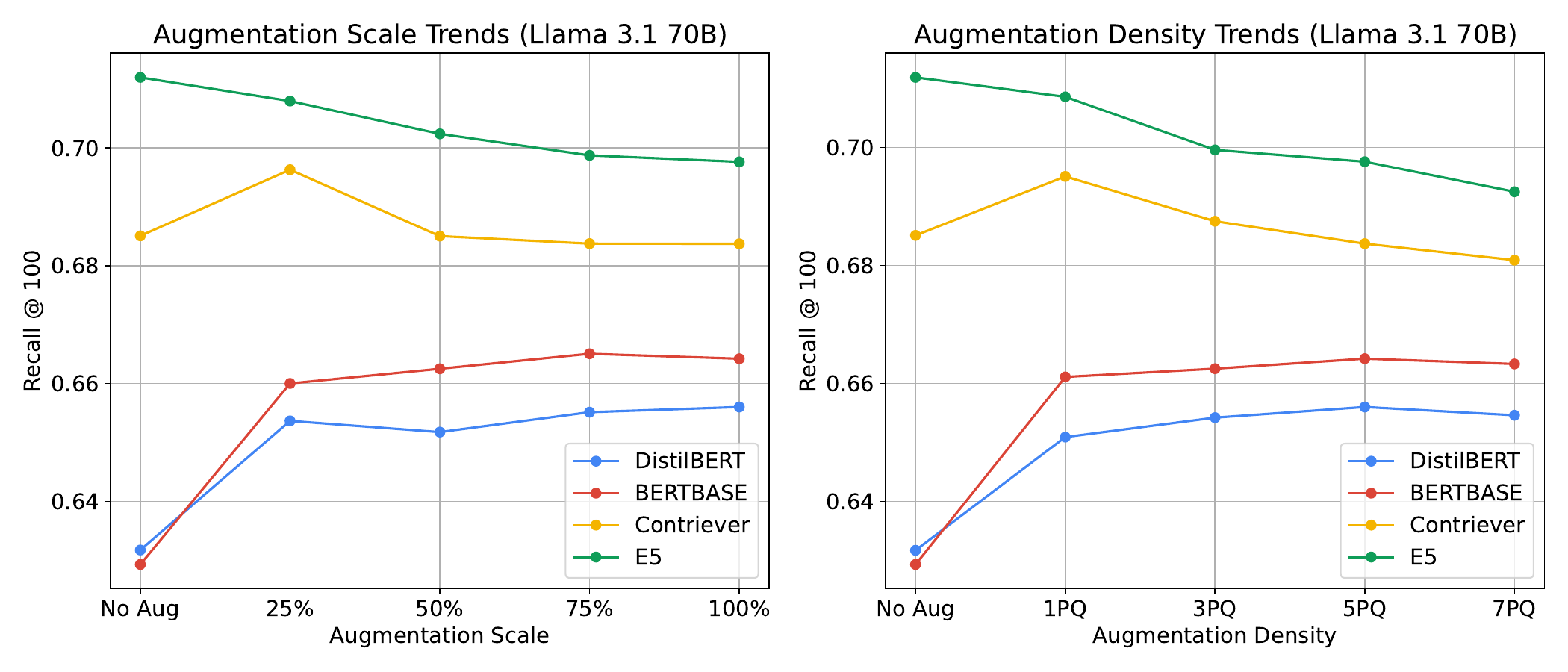}
    \caption{Recall @ 100 trends on the BEIR benchmark, demonstrating the efficacy of augmentation when using Llama 3.1 70B as the augmentation model - Augmentation scale (No. of documents - left) with a fixed density of 5, and Augmentation Density (Augmentations per document - right) with a fixed data scale of 8.8M passages,  across several downstream retrievers.
    Across both axes: augmentation scale and augmentation density, we see a saturation effect. 2) the benefits of augmentation is much lower in case of strongly pre-trained models such as E5 or Contriever compared to standard models like BERTBASE or DistilBERT. PQ: Pseudo-queries per document.}
    \label{fig:general_trends}
\end{figure*}

Despite growing interest in LLM-based augmentation \citep{inpars,inparsv2,dai2023promptagator}, several fundamental questions about their effectiveness and scalability to large-scale real-world retrieval corpora remain unanswered: How does gains in retrieval performance vary with the amount of augmentation? Are large and expensive augmentation models essential for achieving the best retrieval performance?
How much augmentation is necessary? Does retrieval performance scale proportionally with increased augmented data, or does it eventually plateau? Most prior works select random small-scale subsets of the corpus for augmentation, but a systematic study on scaling up augmentation and its effect on retrieval performance is still missing.

Another open question is whether large-scale LLMs are essential for effective augmentation. The existing literature is inconclusive; many studies demonstrate that augmentation with large models leads to significant retrieval performance gains \citep{inpars,dai2023promptagator,wang-etal-2024-improving-text}, while recent work suggests that smaller models can generate data competitively\citep{chen2024littlegiantssynthesizinghighquality}. However, findings of \citet{chen2024littlegiantssynthesizinghighquality} are based on large 7B-scale LLM-based retrievers, making it difficult to generalize results to smaller, more deployable retrieval models. Thus, it remains unclear how augmentation dynamics evolve with scale for smaller retrieval models, particularly dual-encoders in the 100M–300M parameter range commonly used in production systems.

Given that modern web-scale corpora often contain millions of documents, large-scale augmentation is computationally expensive, raising fundamental efficiency questions: Are smaller augmentation models nearly as effective as larger ones? Is it necessary to augment the entire document corpus, or can retrieval models achieve comparable gains with either a random or strategically chosen subset of the corpus?


Through a comprehensive experimental study involving four downstream retrieval models, ten augmentation scales and densities, and two augmentation models resulting in nearly \textbf{100 configurations}, we  conclusively demonstrate nuanced trends which answer the above questions about the scalability and effectiveness of augmentation.

Our findings challenge common assumptions about augmentation. First, while augmentation improves retrieval models, its benefits do not scale indefinitely; beyond a certain point, additional augmentation leads to diminishing returns (refer \Cref{fig:general_trends}). This suggests that augmenting all documents is not necessary and that a smaller subset can be just as effective. Second, we find that off-the-shelf 7B / 8B-scale LLMs are sufficient for Doc2Query-style \citep{nogueira2019documentexpansionqueryprediction} augmentation, with minimal performance differences compared to larger 70B-scale models. Lastly, we establish that augmentation benefits depend heavily on the retriever’s level of pretraining. As demonstrated in \Cref{fig:general_trends}, retrievers with strong pretraining show diminishing gains from augmentation, indicating that augmentation may be most impactful for models with limited pretraining. 

Thus, our study investigates the scalability of augmentation while balancing efficiency, providing actionable insights for more judicious and efficient augmentation strategies. This enables smaller retrieval models to achieve strong performance while maintaining scalability and deployment feasibility.

\section{Related Works}
\label{sec:related_works}

\subsection{Dual-Encoder and LLM-Based Retrieval Models}
Embedding models, particularly Dual Encoders based on BERT \citep{devlin-etal-2019-bert} and T5 \citep{2020t5}, have become standard for retrieval tasks \citep{karpukhin-etal-2020-dense,reimers-gurevych-2019-sentence,ni-etal-2022-large}. Contrastive pre-training on large-scale unsupervised corpora, as demonstrated by models such as Contriever \citep{izacard2022unsupervised}, SimCSE \citep{gao-etal-2021-simcse}, and E5 \citep{wang2024textembeddingsweaklysupervisedcontrastive}, has enhanced unsupervised retrieval without requiring extensive labeled data. While these dual encoder models are efficient and practical for real-world applications, these generally underperform compared to LLMs.

LLMs such as GPT \citep{brown2020,openai2024gpt4technicalreport}, Llama \citep{llama1,llama2,llama3}, and Mistral \citep{jiang2023mistral7b,jiang2024mixtralexperts} exhibit strong generalization capabilities but are computationally expensive for large-scale retrieval. Recent approaches like E5-Mistral \citep{wang-etal-2024-improving-text} and LLM2Vec \citep{behnamghader2024llmvec}, have employed contrastive fine-tuning to enhance LLMs for retrieval tasks, achieving state-of-the-art performance albeit at high deployment costs. This study investigates whether scaling labeled data through LLM-based data augmentation can bridge the performance gap between standard dual encoders and LLM-based retrieval models while maintaining efficiency.

\subsection{Data Augmentation for Improving Retrieval Models}
Synthetic data augmentation has been extensively explored to enhance retrieval models. Early approaches like Doc2Query \citep{nogueira2019documentexpansionqueryprediction} utilized transformers to generate queries from documents, while GPL \citep{wang-etal-2022-gpl} improved this by mining additional documents and incorporating a cross-encoder filter. Recent methods, such as Promptgator \citep{dai2023promptagator}, InParsV1 \citep{inpars}, and InParsV2 \citep{inparsv2}, leverage LLMs to generate pseudo-queries directly. 

These techniques, however, often expose the target domain by using target documents \citep{inpars,inparsv2} or employing few-shot prompting \citep{dai2023promptagator} with examples from the target dataset. Their augmentation scales - 100K for \citet{inpars,inparsv2} and 1M per corpus for \citet{dai2023promptagator} are fixed and they do not systematically assess the impact of scaling up augmentation. GECKO \citep{gecko} improves upon GPL \citep{wang-etal-2022-gpl} by replacing both the query generator and the cross-encoder filter with an LLM. 
However, both the retrieval model as well as the augmentation model are proprietary,  limiting any insights about the scalability of augmentation from the aforementioned work.

 Alternative strategies, such as Query2Doc \citep{wang-etal-2023-query2doc}, generate documents from queries using few-shot prompting. Other approaches bypass passage-based augmentation and instead generate anchor-positive-negative triplets via LLM prompting. For example, \citet{wang-etal-2024-improving-text} uses GPT-4 for triplet generation, while \citet{chen2024littlegiantssynthesizinghighquality} synthesizes seed data with GPT-4 and fine-tunes a Llama 3.1 8B model \citep{llama3} as an efficient data generator.  \citet{chen2024littlegiantssynthesizinghighquality} does present some experiments to analyze the effect of large-scale augmentation and observe diminishing returns from increased augmentation. However, since this study primarily employs the Mistral 7B \citep{jiang2023mistral7b} model as the retriever model, it remains unclear whether saturation is observed because the model is well-pretrained or does it reflect a broader trend across different retrieval architectures.

Prior work has \textbf{not systematically compared }different augmentation models at scale. Our study explicitly assesses LLMs of varying scales for data augmentation, assessing their impact on improving smaller retrieval models. Through controlled scaling experiments, we analyze the extent to which large-scale augmentation benefits small dual-encoder based retrieval models and whether gains diminish beyond a certain threshold. This systematic exploration of scaling effects and its impact on retrieval performance is a key contribution of our work, distinguishing it from prior studies that employ fixed-scale augmentation without assessing the upper limits of its effectiveness.

\section{Methodology}
\label{sec:methodology}
\begin{table*}[ht]
\scriptsize
  \centering
  \begin{tabular}{ll}
    \toprule
    \textbf{Category} & \textbf{Task Name} \\
    \midrule
    \textbf{Information Seeking} & 
    CQADupstackAndroidRetrieval \citep{hoogeveen2015}, FiQA2018 \citep{fiqa2018}, HotpotQA \citep{yang-etal-2018-hotpotqa}, \\
    & QuoraRetrieval, NFCorpus \citep{boteva2016}, NQ \citep{kwiatkowski-etal-2019-natural} \\
    \midrule
    \textbf{Fact Verification} & 
    ClimateFEVER \citep{leippold2020climatefever}, FEVER \citep{thorne-etal-2018-fever}, SciFact \citep{wadden-etal-2020-fact} \\
    \midrule
    \textbf{Argument Retrieval} & 
    ArguAna \citep{wachsmuth-etal-2018-retrieval} \\
    \midrule
    \textbf{Citation Prediction} & 
    SCIDOCS \citep{cohan-etal-2020-specter} \\
    \midrule
    \textbf{Entity Retrieval} & 
    DBPedia \citep{dbpedia} \\
    \midrule
    \textbf{In-Domain} & 
    MS MARCO \citep{bajaj2018msmarcohumangenerated} \\
    \bottomrule
  \end{tabular}
  \caption{BEIR Task Classification for Evaluation}
  \label{tab:task_classification}
\end{table*}

\subsection{Choosing the Augmentation Strategy}
As outlined in \Cref{sec:related_works}, prior work on data augmentation for retrieval has explored techniques such as Query2Doc, Doc2Query, and synthetic query-document pair generation. Generating direct synthetic query-document pairs often lacks diversity, leading to redundant training signals, while Query2Doc relies on the availability of queries, which may not always be feasible at scale. Given these constraints, we adopt the most widely used Doc2Query paradigm, leveraging LLMs to generate pseudo-queries from unlabeled passages, that are typically available in abundance. 

The generated pseudo-queries are then paired with their respective passages to create augmented training data. 
For our experimentation, the objective was to evaluate the efficacy and scaling dynamics of leveraging an LLM as an augmentation model for generating positives and facilitating knowledge transfer. As a result, we did not use the LLM to generate negative examples or introduce an additional LLM call as a filtering module. This setup serves as a viable testbed for assessing the LLM’s effectiveness as an augmenter.

\subsection{Pseudo-Query Generation and Augmentation Type}
The LLM is prompted to generate up to ten pseudo-queries per passage, although augmentation density is treated as an ablation axis in our experiments. For augmentation scaling experiments, where we analyze the effect of varying the proportion of augmented passages, we fix the number of pseudo-queries at five per passage to ensure consistency.
By default, pseudo-queries are generated with an "Information Seeking" intent, optimized for question-answering. However, to ensure robustness and improvements across diverse retrieval tasks, we conduct an additional ablation study by diversifying the augmentation to include various intents like \textit{Information Seeking, Fact Verification, Argument Retrieval, Citation Prediction, and Entity Retrieval.}
\Cref{app:prompt} presents the prompts used to generate diverse pseudo-queries, while \Cref{app:qualitative_example} includes an example pseudo-query for each task.


\subsection{Rigorous OOD Evaluation setup}
While prior research has demonstrated the efficacy of synthetic augmentation for retrieval, evaluation setups often lack rigorous out-of-distribution (OOD) assessments. Most studies evaluate performance improvements on target-domain datasets, often using augmented data derived from the same distribution \citep{dai2023promptagator,inpars,inparsv2}, which risks overestimating effectiveness as the gains are expected. In contrast, our work focuses exclusively on augmenting the MS MARCO corpus \citep{bajaj2018msmarcohumangenerated} while ensuring the target evaluation datasets remain unseen. This approach allows us to measure both in-domain and genuine OOD performance, providing a more stringent test of augmentation efficacy. Unlike previous studies that enhance target-domain performance by exposing models to relevant data, our evaluation framework isolates the impact of augmentation on retrieval generalization.

\textbf{Analysis of Augmentation Effects}
Large-scale data augmentation has been widely adopted to enhance retrieval models, yet its scalability, efficiency,remain understudied. While prior work has explored augmentation with large LLMs \citep{dai2023promptagator,inpars,wang-etal-2024-improving-text}, none of these studies have systematically examined how the effectiveness of augmentation scales, and whether smaller augmentation models can achieve similar gains at lower computational cost.

To understand the impact of augmentation on retrieval models, we conduct experiments along the following dimensions:

\noindent\textbf{Augmentation Scale vs. Performance Gain:} We investigate the impact of varying the proportion of augmented passages while maintaining a fixed augmentation density of 5 pseudo-queries. \\ 
\noindent \textbf{Augmentation Density vs. Performance Gain:} We analyze the effect of increasing the number of pseudo-queries generated per passage while augmenting all the passages. \\ 
\noindent\textbf{Choice of Augmentation Model:} We compare a smaller (8B) and a larger (70B) model from the same family to study the effect of scale, and also include another smaller 7B model from a different family for comparison.\\ 
\noindent \textbf{Model Pre-training vs. Augmentation Benefits:} 
We study the augmentation requirements of retrieval models to determine whether there is a relationship between pretraining strength and the extent of augmentation needed for performance gains. \\ 
\noindent \textbf{Distillation for Augmentation Efficiency:} We explore whether knowledge distillation \citep{kim-rush-2016-sequence} from larger models can help smaller models leverage augmentation more effectively, potentially reducing the computational overhead of large-scale augmentation.
\section{Experimental Setup}
\label{sec:experiments}
In this section, we outline the experimental setup used to evaluate the impact of data augmentation on dense retrieval models. 

\subsection{Data Generation}
The prompt formulation for standard as well as task diversity based pseudo-query generation is detailed in \Cref{app:prompt}.
We utilize the \textbf{Llama 3.1 8B Instruct, Llama 3.1 70B Instruct}, and a fine-tuned version of the Llama 3.1 8B Instruct model for pseudo-query generation. Decoding is performed with parameters: $top_p = 0.9$, temperature = 0.7, and a 512-token limit. All references to Llama models refer to their Instruct variants. We also include an ablation experiment with the \textbf{Mistral 7B Instruct v0.3 model} \citep{jiang2023mistral7b} for standard pseudo-query generation.

\subsection{Scales of the Augmented Data}
To study the impact of augmentation scale, we progressively augment 25\%, 50\%, 75\%, and 100\% of the MS MARCO corpus (8.8M passages in total), generating 5 pseudo-queries per passage. Combined with the original 532K examples, this results in 0.5M (No Aug), 11.5M, 22.5M, 33.5M, and 44.5M training pairs respectively. For augmentation density, we augment the entire set of 8.8M passages but vary the number of pseudo-queries per passage (0, 1, 3, 5, 7), which including the original dataset yields 0.5M, 9.3M, 26.9M, 44.5M, and 62.1M training pairs respectively.

\subsection{Retrieval Model Training}
Our experiments focus on the MS MARCO Passage corpus\footnote{\url{https://microsoft.github.io/msmarco/Datasets.html}}\citep{bajaj2018msmarcohumangenerated} as the base dataset, with various augmentation strategies applied to its passages, with the rest of the tasks evaluated upon forming a zero-shot evaluation set. We leverage the provided (532K) anchor-positive pairs for training with in-batch negatives alongside the augmented data (if applicable). 
We evaluate the impact of augmentation by training a diverse set of downstream retriever models using the SentenceTransformers repository\footnote{\url{https://github.com/UKPLab/sentence-transformers}} \citep{reimers-gurevych-2019-sentence}.

The set of downstream retriever models include :-
\begin{itemize}[noitemsep, topsep=0pt]
\item \textbf{BERTBASE} – \citep{devlin-etal-2019-bert}, a widely-used baseline encoder.
\item \textbf{DistilBERT} – \citep{distilbert} a lightweight and faster alternative to BERTBASE.
\item \textbf{Contriever} – \citep{izacard2022unsupervised} a model that undergoes unsupervised contrastive pre-training and is optimized for generalization.
\item \textbf{E5-BASE-Unsupervised} – \citep{wang2024textembeddingsweaklysupervisedcontrastive} an unsupervised embedding model with more extensive contrastive pre-training than Contriever.
\end{itemize}
The training hyperparameters for all the retrieval models are listed in \Cref{tab:hyperparams}.

\subsection{Evaluation Benchmark}
Evaluation is performed using the BEIR dataset \citep{thakur2021beir} which offers several out-of-domain (OOD) tasks listed in \Cref{tab:task_classification} as the primary benchmark. Additionally, we further assess generalization using nine zero-shot tasks from MTEB \citep{muennighoff-etal-2023-mteb} namely COSQA \citep{huang-etal-2021-cosqa}, HellaSwag \citep{zellers-etal-2019-hellaswag}, MedicalQARetrieval \citep{BenAbacha-BMC-2019}, NarrativeQARetrieval \citep{kocisky-etal-2018-narrativeqa}, PIQA \citep{PIQA}, PublicHealthQA \citep{enevoldsen2025mmteb}, SIQA \citep{sap-etal-2019-social}, StackOverflowQA \citep{li2025coir}, SyntheticText2SQL\citep{gretel-synthetic-text-to-sql-2024} all of which fall under the Information Seeking category. 

\subsection{Metrics}
The motivation behind the use of LLM-based augmentation is to transfer knowledge from LLMs to enhance the retriever’s ability to capture diverse intents. Recall$@100$ effectively evaluates this diversity by ensuring that all the intents are represented in the retrieved set. While this diversity can occasionally lower NDCG due to uniform relevance weighting across pseudo-queries, re-ranking or weighted augmentation strategies can mitigate this, which we leave to future work, as retrievers can always be paired with re-rankers to optimize for both NDCG alongside strong recall. For completeness, we also report NDCG$@10$ and MRR$@10$ to analyze ranking performance.

\begin{figure*}[t]
    \centering
    \includegraphics[width=0.85\linewidth]{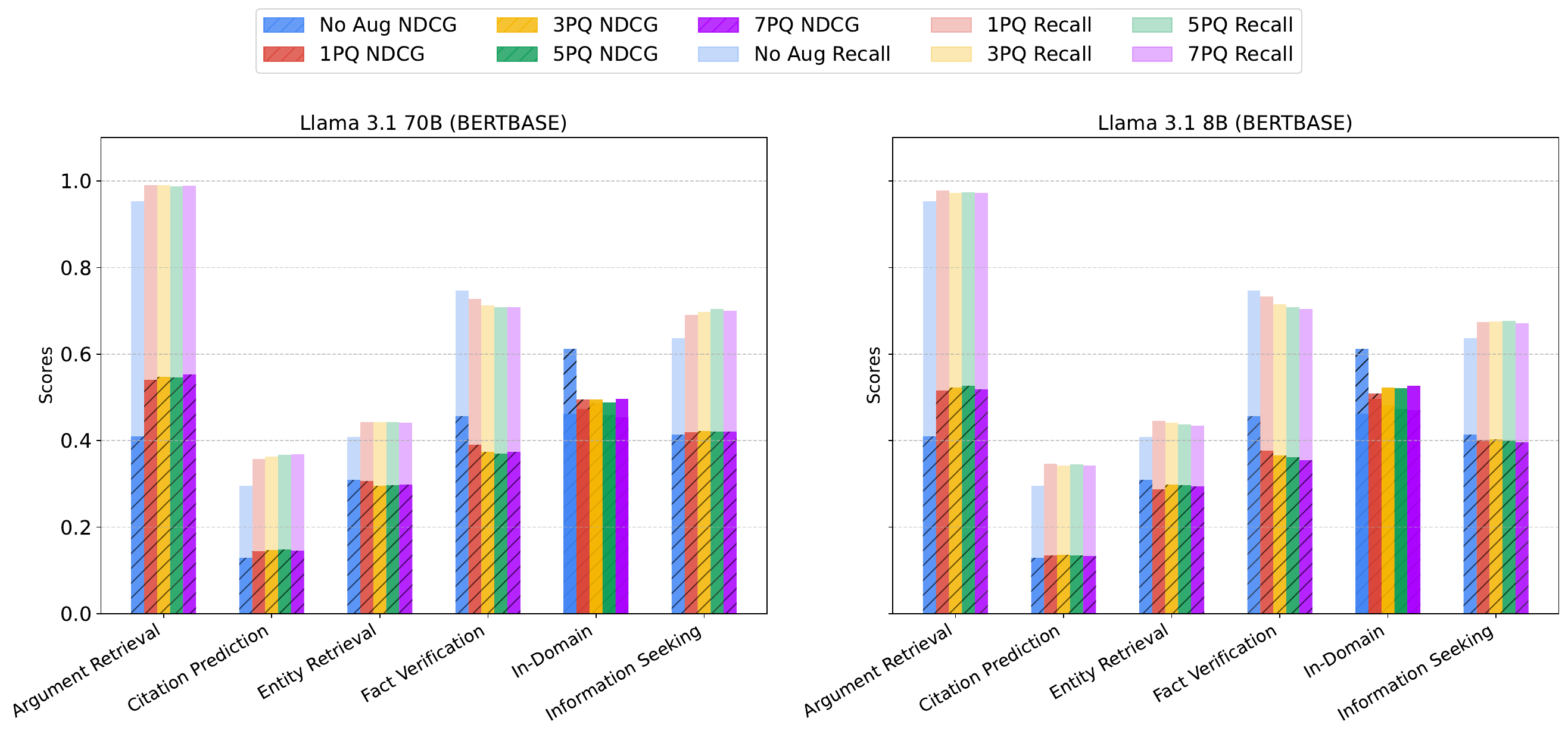}
    \caption{Category-wise Recall @ 100 and NDCG @10 scores on the BEIR benchmark, comparing BERTBASE baselines leveraging different augmentation densities using Llama variants 3.1 70B (left) and 3.1 8B (right).}
    \label{fig:density_bert}
\end{figure*}

\begin{figure*}[t]
    \centering
    \includegraphics[width=0.85\linewidth]{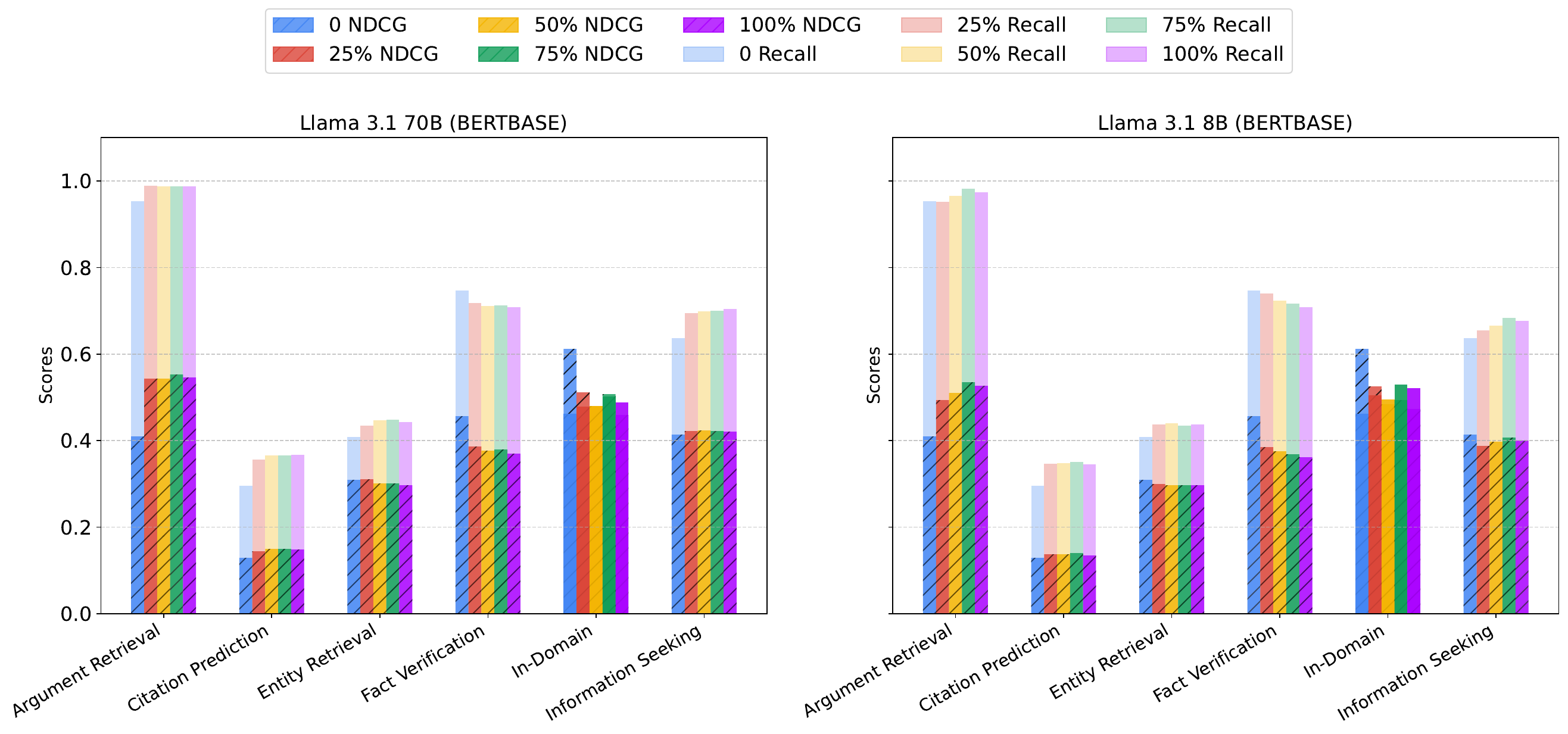}
    \caption{Category-wise Recall @ 100 and NDCG @10 scores on the BEIR benchmark, comparing BERTBASE baselines leveraging different augmentation scales using Llama variants 3.1 70B (left) and 3.1 8B (right).}
    \label{fig:datascale_bert}
\end{figure*}
\section{Main Results}
\label{sec:results_discussions} 
In this section, we describe the prominent findings from our exhaustive experimentation.
We present the task-wise metric scores for our primary experiments in \Cref{sec:primary_tables} and on additional MTEB tasks in \Cref{sec:additional_mteb}.

\subsection{Saturation Effect with increase of Density and Scale of Augmentation}
The over-arching findings from our experiments as depicted in \Cref{fig:datascale_bert,fig:density_bert}, 
indicate that while Doc2Query-style augmentation using LLMs is an effective strategy, its benefits do not scale indefinitely.  From \Cref{fig:datascale_bert,fig:density_bert}, we observe that across nearly all task categories, we observe significant improvements in terms of Recall @ 100, compared to training solely on the provided labeled data, highlighting strong out-of-domain generalization on BEIR. However, for NDCG@10, the results are mixed: some categories experience minor declines, likely due to diluted relevance signals as discussed in \Cref{sec:experiments}, while others see improvements.

\begin{figure*}[t]
    \centering
    \includegraphics[width=0.85\linewidth]{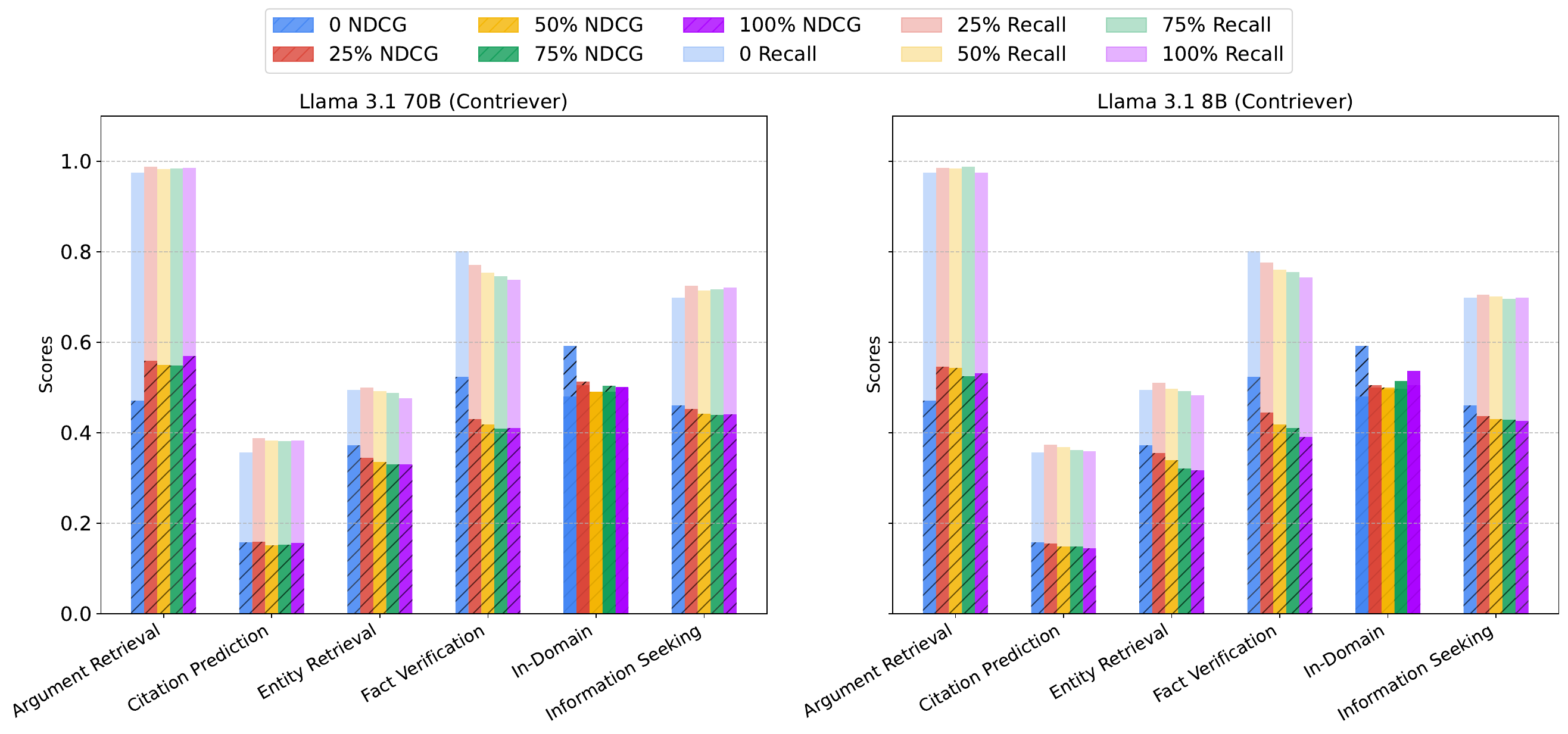}
    \caption{Category-wise Recall @ 100 scores on the BEIR benchmark, comparing Contriever baselines leveraging different augmentation scales (0-100\%) using Llama 3.1 70B and Llama 3.1 8B model respectively.}
    \label{fig:datascale_contriever}
\end{figure*}

\begin{figure*}[!ht]
    \centering
    \includegraphics[width=0.85\linewidth]{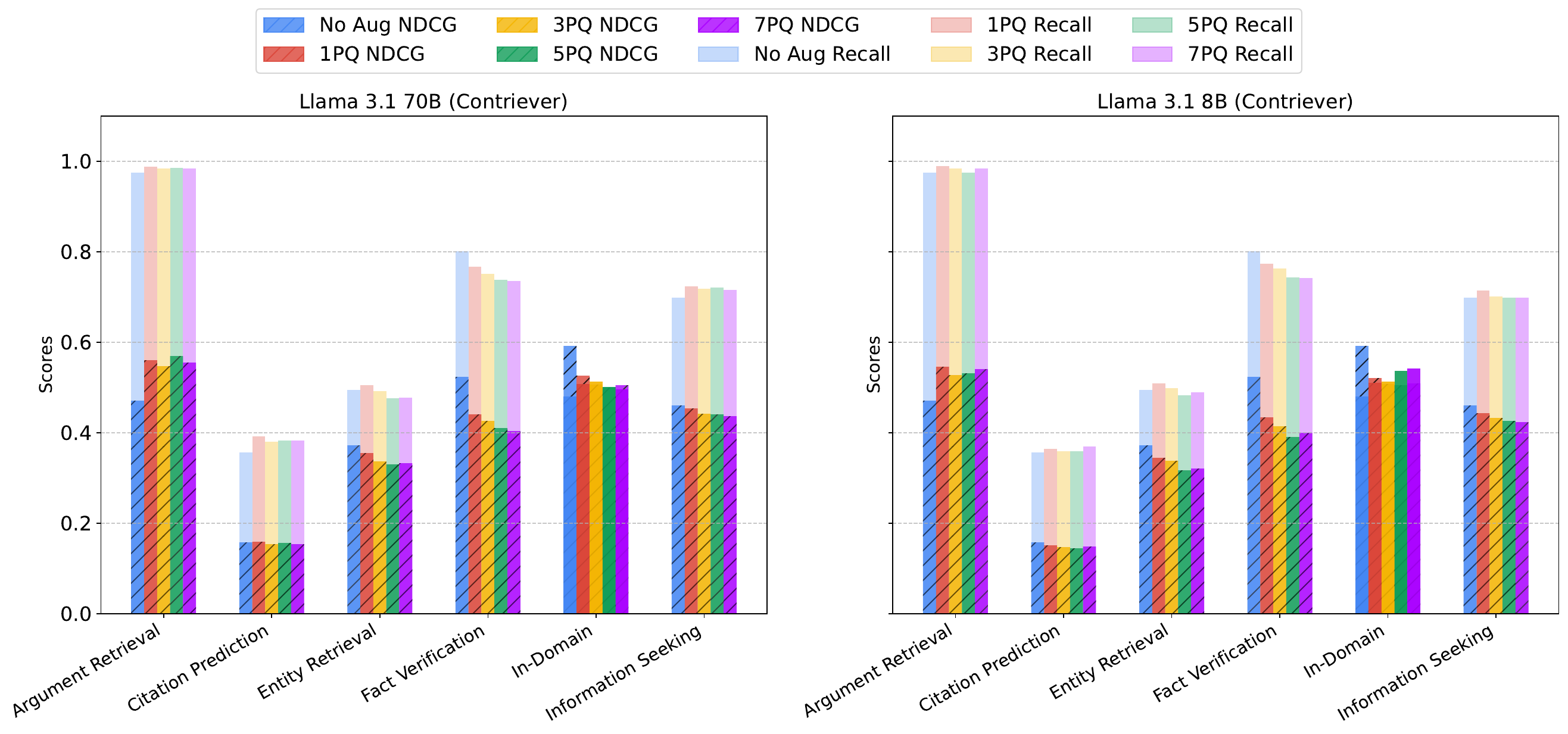}
    \caption{Category-wise Recall @ 100 and NDCG @ 10 scores on the BEIR benchmark, comparing Contriever baselines leveraging different augmentation densities (0-7) using Llama 3.1 70B and Llama 3.1 8B model respectively.}
    \label{fig:mteb_pq_density_contriever}
\end{figure*}

\begin{figure*}[!ht]
    \centering
    \includegraphics[width=0.9\linewidth]{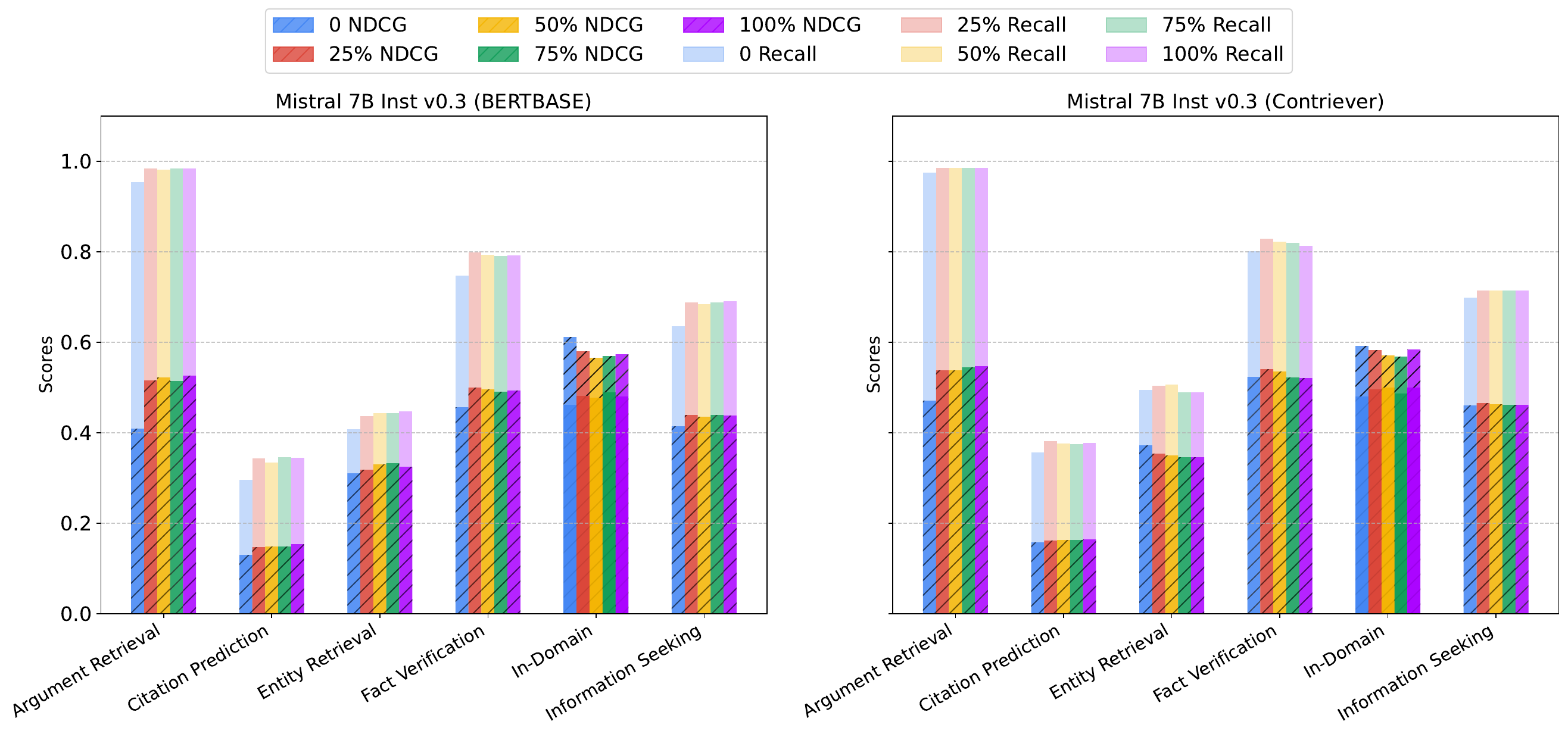}
    \caption{Category-wise Recall @ 100 and NDCG @ 10 scores on the BEIR benchmark, comparing BERTBASE (left) and Contriever (right) baselines leveraging different augmentation scales (0-100\%) using Mistral 7B Instruct v0.3 model.}
    \label{fig:datascale_mistral}
\end{figure*}

\begin{figure*}[!h]
    \centering
    \includegraphics[width=0.9\linewidth]{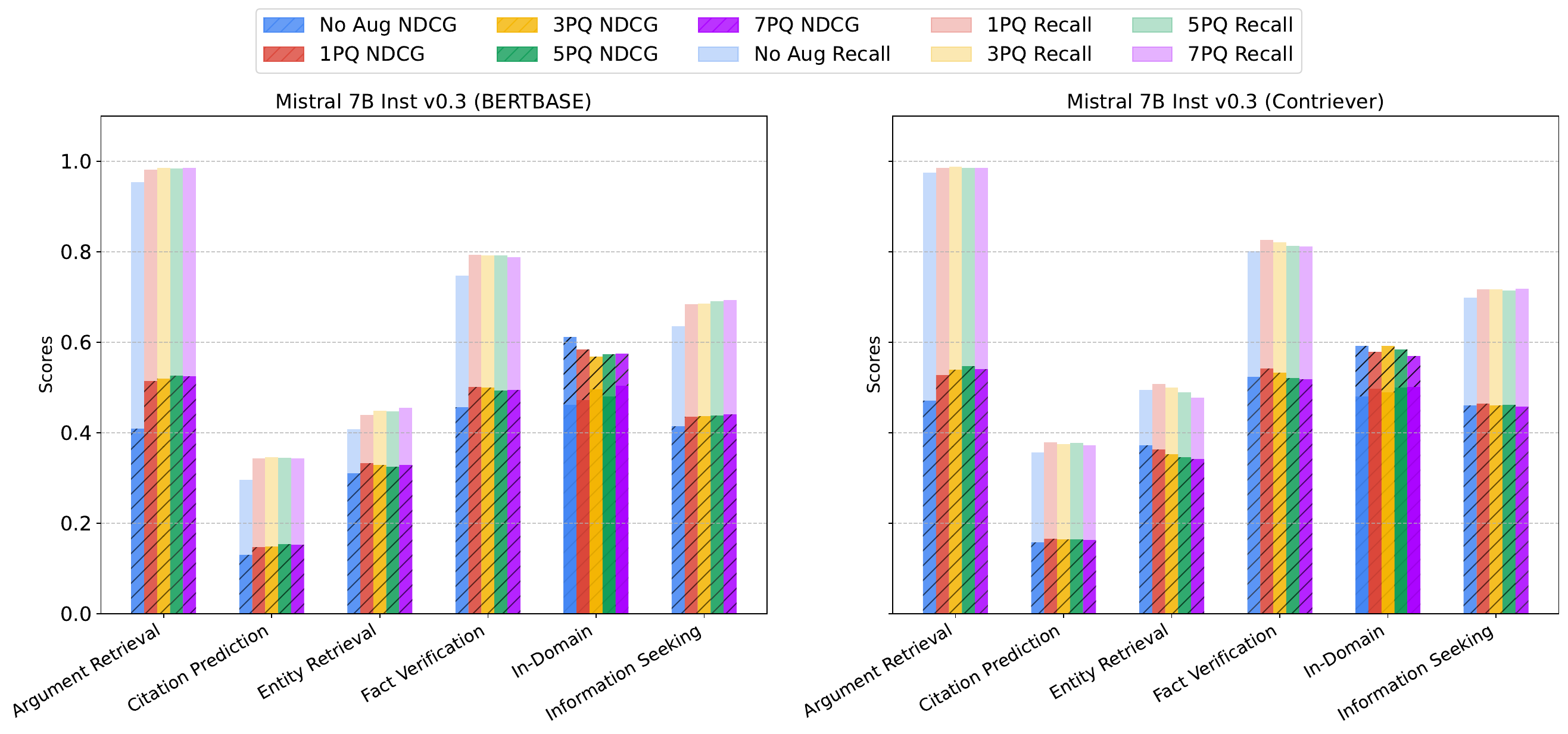}
    \caption{Category-wise Recall @ 100 and NDCG @ 10 scores on the BEIR benchmark, comparing BERTBASE (left) and Contriever (right) baselines leveraging different augmentation densities (0-7) using Mistral 7B Instruct v0.3 model.}
    \label{fig:density_mistral}
\end{figure*}

Crucially, our findings reveal a clear saturation effect across both metrics when scaling both the amount and the density of augmentation. Increasing the scale of augmentation-either by adding more pseudo-queries per passage or by augmenting a larger portion of the dataset-yields diminishing returns beyond a certain threshold. This saturation is \textbf{consistently observed} on BEIR, across multiple models, including DistilBERT, Contriever (\Cref{fig:datascale_contriever,fig:mteb_pq_density_contriever}), and E5 (\Cref{fig:datascale_e5,fig:BEIR_pq_density_e5}), and even when using alternative augmentation models such as Mistral 7B Instruct v0.3 (\Cref{fig:datascale_mistral,fig:density_mistral}) and even validated with supplementary evaluation on 9 additional MTEB tasks (\Cref{sec:additional_mteb}).
These results challenge the common assumption that simply increasing the scale of LLM-based augmentation will lead to continuous performance gains. Instead, our evidence suggests that indiscriminate augmentation of the entire dataset is suboptimal. A more efficient and effective approach may involve selectively augmenting a subset of unlabeled documents, guided by heuristics or a pre-defined budget, to maximize efficiency without significantly compromising retrieval quality.

\subsection{Choice of Augmentation model}
Our empirical analysis reveals critical insights into model selection for Doc2Query-style augmentation. To evaluate the effect of model choice / scale on augmentation, we compare pseudo-query generation using Llama 3.1 70B and Llama 3.1 8B across different fixed augmentation scales while keeping the number of pseudo-queries per document constant. Our results presented in \Cref{fig:datascale_bert,fig:density_bert} indicate that both models effectively generate pseudo-queries and contribute to improved recall. 
The Llama 3.1 70B yields a marginally higher improvement in Recall$@100$ compared to Llama 3.1 8B (~1point) \Cref{tab:PQ_Density_BERTBASE_70B_MTEB,tab:PQ_Density_BERTBASE_8B_MTEB} and \Cref{tab:70B_BERT_density_metrics_BEIR,tab:8B_BERT_scale_metrics_BEIR}. Given the significantly higher computational cost of inference with the 70B model, the marginal gains may not justify its use for large-scale augmentation.  These findings suggest that Llama 3.1 8B serves as a viable alternative for Doc2Query-style augmentation compared to Llama 3.1 70B, particularly for tasks focused on simple retrieval. In \Cref{subsec:distilation}, we also discuss an ablation about distillation from the 70B model to the 8B model to further close the gap. In case of Llama 3.1 models (both 8B and 70B) we observe some drops across NDCG$@10$ across few categories. Also, for task types such as Fact Verification, where the nature of the task differs substantially from standard information-seeking tasks, we notice a sharp drop in performance. In \Cref{subsec:td_augmentation}, we conduct an experiment to investiagate if task-diverse generations can mitigate this problem. In our ablation with with Mistral-7B Instruct v0.3, we observe that using it as the augmentation model outperforms both Llama 3.1 8B as well as 70B models across all metrics, as observed in \Cref{fig:density_mistral,fig:datascale_mistral}. Crucially, Mistral augmentation eliminates drops with respect to NDCG$@10$, or on the other categories like "Fact Verification" and we see a strong improvement in performance across the board. Therefore, amongst the models that were considered for evaluation, Mistral 7B Instruct v0.3, turned out to be the most optimal for our use case maximizing Recall$@100$ as well as NDCG$@10$ across diverse tasks.

\subsection{Trade-off between augmentation scale and density}
Our empirical analysis reveals that Doc2Query-style "passage-conditioned" generation often results in pseudo-queries that exhibit considerable overlap—either lexically or semantically—thereby limiting diversity and reducing the effectiveness of increasing augmentation density. From \Cref{tab:8B_BERT_scale_metrics_BEIR} and \Cref{tab:8B_BERT_density_metrics_BEIR}, we observe that with smaller augmentation models, expanding the scale of augmentation (1PQ) yields better results than significantly increasing the augmentation density with a smaller corpus (25\% - 5PQ). Given a fixed augmentation budget, it is more beneficial to increase the number of augmented documents rather than the number of pseudo-queries per document. This strategy promotes broader corpus coverage and minimizes redundancy among generated queries. Although in theory, boosting query density could enhance performance, current smaller LLMs often produce redundant outputs when multiple queries are generated for the same passage. Consequently, our findings suggest that expanding the set of augmented documents is a more effective way to capture a wider spectrum of potential query intents.

\subsection{Ablation with stronger pre-trained models}
Contriever, trained with contrastive pre-training, serves as a significantly stronger base model that can be used off-the-shelf for unsupervised retrieval. It consistently outperforms the standard BERTBASE model in both zero-shot retrieval and post-MSMARCO fine-tuning.  Given its stronger initialization, we evaluate the utility of data augmentation when using Contriever as the base model compared to BERTBASE. While \Cref{fig:datascale_bert,fig:density_bert} demonstrate that augmentation leads to substantial benefits in Recall @ 100 across both in-domain and out-of-domain settings for BERTBASE, we observe that the corresponding performance gains from augmentation for Contriever depicted in \Cref{fig:datascale_contriever,fig:mteb_pq_density_contriever} are notably lower. To further validate this trend, we replicated our experiments using the E5 Base model \citep{wang2024textembeddingsweaklysupervisedcontrastive}, a more extensively contrastively pre-trained model than Contriever. As shown in \Cref{fig:datascale_e5,fig:BEIR_pq_density_e5}, E5-Base exhibits even smaller improvements with augmentation. These findings suggest that in-distribution Doc2Query-style augmentation offers limited utility for models that have undergone strong contrastive pre-training suggesting that the effectiveness of augmentation varies across different models and depends on the extent of pre-training undergone by the downstream model.

\section{Conclusion}
\label{sec:conclusion}
In this study, we conducted a thorough examination of the scalability of LLM-based data augmentation to enhance the performance of smaller retrieval models. Our analysis concentrated on the impact of augmentation scale and model size. The findings reveal that while augmentation can indeed improve retrieval performance, its advantages do not scale indefinitely. We identified a point of diminishing returns beyond a certain augmentation threshold, indicating that augmenting the entire corpus is not always necessary.
Our results demonstrate that smaller augmentation models, such as those with 8B parameters, can achieve performance comparable to larger models (70B). This finding suggests a more computationally efficient alternative without sacrificing effectiveness. Interestingly, we observe that retrievers with extensive pre-training show less improvement from augmentation, highlighting that augmentation is particularly advantageous for models lacking robust pre-training. These insights emphasize the necessity of adopting strategic augmentation approaches rather than relying on indiscriminate scaling. Future research should aim to develop adaptive augmentation strategies that tailor efforts based on the model's capacity and dataset characteristics. This could enhance efficiency and generalization, optimizing the augmentation process for various retrieval tasks.



\section{Limitations}
\label{sec:limitations}
In our experimental setup, we focus on leveraging LLM-generated anchor-positive pairs to bridge the knowledge gap, without incorporating hard negatives or LLM-based negative augmentation. Our objective is to assess the scalability of this approach and determine to what extent do dual-encoder models benefit from positive augmentation alone. While the integration of hard negatives, LLM-generated negatives, or LLM-based filtering could refine performance, these aspects are beyond the scope of our study. We evaluate all the models using a standard hyperparameter set to gauge the average-case performance improvements resulting from data augmentation, rather than optimizing hyperparameters for each model. Additionally, we employ a standard loss function instead of exploring more complex formulations, with the expectation that the observed augmentation gains would generalize to other training paradigms. Our findings indicate that while augmentation enhances performance, scaling beyond a certain threshold leads to diminishing returns. Identifying an optimal augmentation budget is a promising direction for further research, requiring a more detailed investigation into its trade-offs.

\acknowledgementsvariable

\bibliography{anthology,custom}

\clearpage
\appendix



\clearpage

\section{Hyperparameter Configuration}
\label{app:hparams}
\subsection{Hyperparameter Configuration for Training Dense Retrieval Models}
\Cref{tab:hyperparams} describes the standard hyperparameter configuration used for training all our retrieval models.

\begin{table}[h]
\small
    \centering
    \caption{Hyperparameter Configuration}
    \begin{tabular}{l|c}
        \toprule
        \textbf{Hyperparameter} & \textbf{Value} \\
        \midrule
        Batch Size & 256 \\
        Max. Seq. Length & 300 \\
        No. of Epochs & 15 \\
        Pooling & Mean \\
        Warmup & 1000 \\
        Learning Rate & 2e-5 \\
        Loss  & MultipleNegatives \\
        & RankingLoss \\
        \bottomrule
    \end{tabular}
    \label{tab:hyperparams}
\end{table}

\section{Additional Results and Discussion}

\subsection{Performance trends with E5 as the Downstream model}
\begin{figure*}[h]
    \centering
    \includegraphics[width=\linewidth]{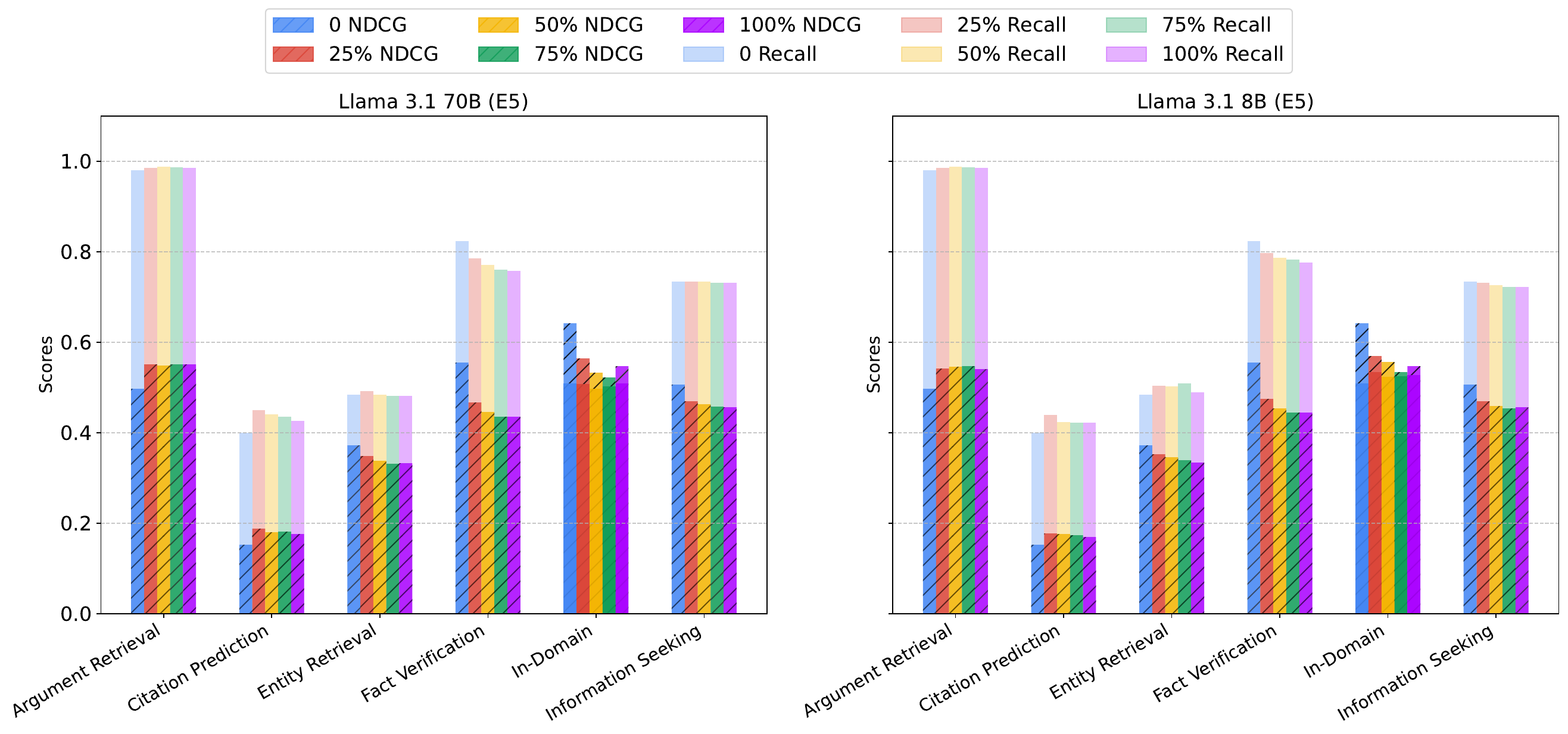}
    \caption{Category-wise Recall @ 100 scores on the BEIR benchmark, comparing E5 baselines leveraging different augmentation scales using Llama 3.1 70B and Llama 3.1 8B model respectively.}
    \label{fig:datascale_e5}
\end{figure*}

\begin{figure*}[h]
    \centering
    \includegraphics[width=\linewidth]{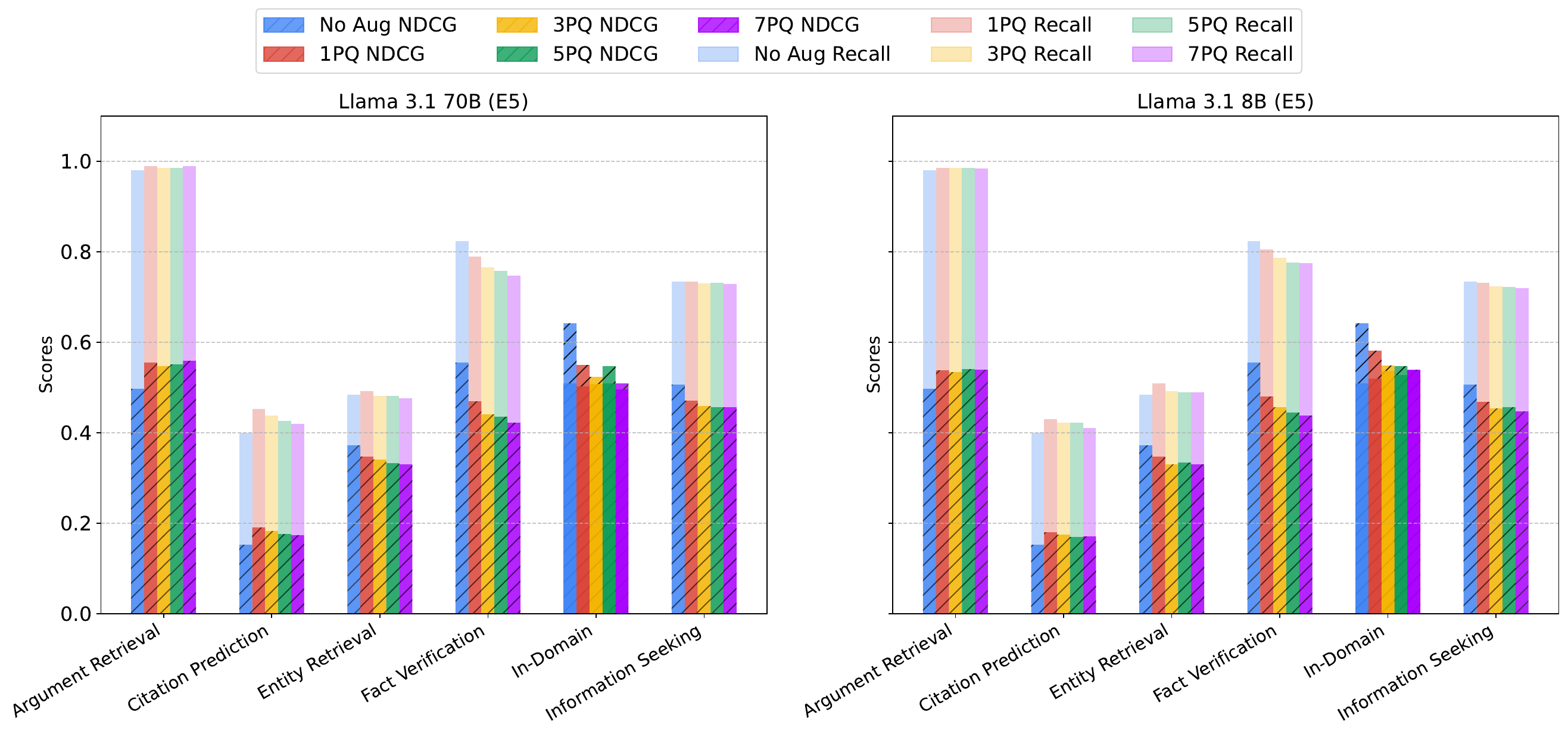}
    \caption{Category-wise Recall @ 100 and NDCG @ 10 scores on the BEIR benchmark, comparing E5 baselines leveraging different augmentation densities using Llama 3.1 70B and Llama 3.1 8B model respectively.}
    \label{fig:BEIR_pq_density_e5}
\end{figure*}

\Cref{fig:datascale_e5} and \Cref{fig:BEIR_pq_density_e5} present the Recall @ 100 and NDCG @10 scores on the BEIR benchmark when using the E5-base-unsupervised\footnote{\url{https://huggingface.co/intfloat/e5-base-unsupervised}}\citep{wang2024textembeddingsweaklysupervisedcontrastive} model as the downstream retriever model and Llama 3.1 8B and Llama 3.1 70B models as the augmentation models.

\clearpage

\section{Analysis: Impact of Model Choice on Query Quality}
\label{sec:model_analysis}
Our experiments reveal performance variations based on the choice of augmentation model. From \Cref{fig:datascale_mistral,fig:density_mistral}, we observe that the Mistral 7B model consistently improves performance across all benchmarks, across both metrics Recall @ 100 and NDCG @10. However, in case of both LLama 3.1 8B as well as 70B models, we observed notable degradation on fact verification tasks (FEVER) and in-domain retrieval (MS MARCO). To further investigate this phenomenon, we conducted a qualitative analysis of pseudo-queries generated by Llama 3.1 8B, Llama 3.1 70B, and Mistral 7B Instruct v0.3.

We randomly sampled passages from our corpus and examined the corresponding pseudo-queries generated by each model. \Cref{tab:query_examples} presents four representative examples that illustrate systematic differences in query generation patterns.

\begin{table*}[t]
\centering
\small
\begin{tabular}{@{}p{0.25\textwidth}p{0.22\textwidth}p{0.22\textwidth}p{0.22\textwidth}@{}}
\toprule
\textbf{Passage} & \textbf{Mistral 7B v0.3} & \textbf{Llama 3.1 8B} & \textbf{Llama 3.1 70B} \\
\midrule
\textit{``The interest to a loan on the street or through a bookie. It is common to not pay back the principle but have to pay the vig or interest weekly to keep your legs intact.''} & 
\textit{Street loans with weekly interest payments and no principle reduction; Underground lending practices involving vigorish} & 
\textit{Weekly interest payments for high-interest loans; Loan shark interest rates vs traditional lenders} & 
\textit{What are the consequences of not paying back a loan from a loan shark; How do loan sharks collect weekly interest payments} \\
\midrule
\textit{``Derek Jeter's first spring training as a retired Yankee has taken him far from George M. Steinbrenner Field in Tampa. First, he joined his former teammate Hideki Matsui for a charity exhibition game in Japan to raise money for survivors of the 2011 earthquake and tsunami.''} & 
\textit{Derek Jeter charity event Japan tsunami survivors; Derek Jeter Hideki Matsui spring training exhibition game} & 
\textit{Yankees charity events; Derek Jeter philanthropic efforts; George M Steinbrenner Field location} & 
\textit{Former MLB players supporting disaster relief efforts; Yankees players participating in international charity games} \\
\midrule
\textit{``An example of a structural mutation in the human serum cholinesterase gene that is a result of a mutation in the protein-coding region is described below. A point mutation in the gene for human serum cholinesterase was identified that changes Asp-70 to Gly in the atypical form of serum cholinesterase.''} & 
\textit{Asp-70 point mutation in human serum cholinesterase gene; Structural consequences of Asp-70 to Gly mutation} & 
\textit{Glycine substitution at position 70 in serum cholinesterase; Impact of genetic mutations on serum cholinesterase function} & 
\textit{Genetic variations in the human serum cholinesterase protein-coding region; Effects of point mutations on enzyme activity} \\
\midrule
\textit{``The SLA and the ABC state that the minimum requirements for liquor license eligibility are: 1) The age of 21 or older. 2) No records of conviction. 3) No police office with arresting powers.''} & 
\textit{Minimum age for liquor license eligibility in SLA and ABC regulations; Requirements for criminal background check for liquor license applicants} & 
\textit{What are the minimum requirements for a liquor license in the state of ABC; Liquor license requirements for individuals 21 and older} & 
\textit{What are the basic requirements to be eligible for a liquor license; How does a criminal record affect liquor license eligibility} \\
\bottomrule
\end{tabular}
\caption{Representative examples of pseudo-queries generated by different models, illustrating systematic differences in specificity and factual grounding.}
\label{tab:query_examples}
\end{table*}

Our analysis reveals three distinct generation patterns across models:

\textbf{Mistral 7B demonstrates superior factual preservation}: Generated queries maintain high specificity by directly incorporating passage terminology. The model comprehensively captures multiple related facts from single passages while preserving unique contextual details that distinguish passages from topically similar content.

\textbf{Llama 3.1 8B exhibits over-generalization tendencies}: Queries tend to fragment complex information into oversimplified components and introduce topically related but irrelevant comparisons. The model over-generalizes specific facts to broader categories, losing the precision necessary for accurate retrieval in knowledge-intensive tasks.

\textbf{Llama 3.1 70B shows hypothetical expansion}: Despite its larger scale, this model creates hypothetical scenarios not grounded in passage content. It frequently expands scope to unrelated concepts within the same domain and dilutes passage specificity through excessive abstraction.

\subsection{Impact on Downstream Performance}

These qualitative differences directly explain the observed performance patterns. Mistral's specific, contextually-aligned queries retrieve passages that are both topically and factually relevant to the original content. In contrast, Llama models' tendency toward generalization results in retrieving passages that are topically similar but factually less precise. This specificity-relevance trade-off particularly impacts ranking performance (NDCG@10) rather than retrieval breadth (Recall@100). The dilution of factual relevance makes it substantially harder for truly optimal passages to achieve high rankings, as the retrieval model must distinguish between multiple topically similar but factually distinct passages. This effect is most pronounced in fact verification tasks like FEVER, where precision of specific factual claims is critical for accurate verification.

\section{Evaluation on additional MTEB tasks}
\label{sec:additional_mteb}

\Cref{tab:PQ_Density_BERTBASE_70B_MTEB,tab:PQ_Density_BERTBASE_8B_MTEB} and \Cref{tab:PQ_Density_Contriever_70B_MTEB,tab:PQ_Density_Contriever_8B_MTEB} demonstrates the performance metrics of the augmentation density experiment using the BERTBASE and Contriever models evaluated on 9 additional tasks from MTEB.

\clearpage

\begin{table*}[ht]
\centering
\tiny
\setlength{\tabcolsep}{3pt}
\begin{tabular}{lrrrrrrrrrr}
\toprule
Task Name & \multicolumn{2}{c}{No Aug} & \multicolumn{2}{c}{1PQ} & \multicolumn{2}{c}{3PQ} & \multicolumn{2}{c}{5PQ} & \multicolumn{2}{c}{7PQ} \\
\cmidrule(lr){2-3} \cmidrule(lr){4-5} \cmidrule(lr){6-7} \cmidrule(lr){8-9} \cmidrule(lr){10-11}
 & NDCG@10 & Recall@100 & NDCG@10 & Recall@100 & NDCG@10 & Recall@100 & NDCG@10 & Recall@100 & NDCG@10 & Recall@100 \\
\midrule
CosQA & 0.0717 & 0.3800 & 0.1414 & 0.5960 & 0.1665 & 0.6500 & 0.1480 & 0.6160 & 0.1689 & 0.6520 \\
HellaSwag & 0.2093 & 0.5321 & 0.2452 & 0.5948 & 0.2542 & 0.6071 & 0.2597 & 0.6145 & 0.2631 & 0.6142 \\
MedicalQARetrieval & 0.5745 & 0.9106 & 0.5596 & 0.9375 & 0.5803 & 0.9409 & 0.5793 & 0.9404 & 0.5841 & 0.9414 \\
NarrativeQARetrieval & 0.2186 & 0.5441 & 0.1943 & 0.5868 & 0.1903 & 0.5839 & 0.1929 & 0.6081 & 0.1963 & 0.6133 \\
PIQA & 0.2071 & 0.5419 & 0.2306 & 0.6072 & 0.2413 & 0.6017 & 0.2428 & 0.6039 & 0.2446 & 0.6072 \\
PublicHealthQA & 0.6365 & 0.9709 & 0.6873 & 0.9826 & 0.7120 & 0.9826 & 0.7542 & 0.9884 & 0.7478 & 0.9884 \\
SIQA & 0.0198 & 0.0696 & 0.0280 & 0.1223 & 0.0311 & 0.1192 & 0.0289 & 0.1244 & 0.0278 & 0.1157 \\
StackOverflowQA & 0.4705 & 0.7653 & 0.5451 & 0.8270 & 0.5473 & 0.8325 & 0.5528 & 0.8425 & 0.5487 & 0.8486 \\
SyntheticText2SQL & 0.3002 & 0.8267 & 0.3378 & 0.8605 & 0.3226 & 0.8243 & 0.3235 & 0.8171 & 0.3146 & 0.8055 \\
\bottomrule
\end{tabular}
\caption{Augmentation Density Experiment: BERTBASE (Augmentation Model = Llama 3.1 70B)}
\label{tab:PQ_Density_BERTBASE_70B_MTEB}
\end{table*}

\begin{table*}[ht]
\centering
\tiny
\setlength{\tabcolsep}{3pt}
\begin{tabular}{lrrrrrrrrrr}
\toprule
Task Name & \multicolumn{2}{c}{No Aug} & \multicolumn{2}{c}{1PQ} & \multicolumn{2}{c}{3PQ} & \multicolumn{2}{c}{5PQ} & \multicolumn{2}{c}{7PQ} \\
\cmidrule(lr){2-3} \cmidrule(lr){4-5} \cmidrule(lr){6-7} \cmidrule(lr){8-9} \cmidrule(lr){10-11}
 & NDCG@10 & Recall@100 & NDCG@10 & Recall@100 & NDCG@10 & Recall@100 & NDCG@10 & Recall@100 & NDCG@10 & Recall@100 \\
\midrule
CosQA & 0.0717 & 0.3800 & 0.1998 & 0.7160 & 0.2014 & 0.7500 & 0.2076 & 0.7280 & 0.2070 & 0.7120 \\
HellaSwag & 0.2093 & 0.5321 & 0.2435 & 0.5868 & 0.2451 & 0.5928 & 0.2474 & 0.5899 & 0.2479 & 0.5904 \\
MedicalQARetrieval & 0.5745 & 0.9106 & 0.5764 & 0.9341 & 0.5821 & 0.9429 & 0.5749 & 0.9458 & 0.5688 & 0.9448 \\
NarrativeQARetrieval & 0.2186 & 0.5441 & 0.2215 & 0.6199 & 0.2146 & 0.6053 & 0.2188 & 0.6248 & 0.2141 & 0.6242 \\
PIQA & 0.2071 & 0.5419 & 0.2038 & 0.5620 & 0.2047 & 0.5669 & 0.2085 & 0.5724 & 0.2068 & 0.5724 \\
PublicHealthQA & 0.6365 & 0.9709 & 0.7359 & 0.9942 & 0.7399 & 0.9942 & 0.7374 & 0.9942 & 0.7417 & 0.9942 \\
SIQA & 0.0198 & 0.0696 & 0.0297 & 0.1244 & 0.0314 & 0.1218 & 0.0306 & 0.1198 & 0.0288 & 0.1274 \\
StackOverflowQA & 0.4705 & 0.7653 & 0.4893 & 0.7889 & 0.4530 & 0.7593 & 0.4519 & 0.7553 & 0.4389 & 0.7558 \\
SyntheticText2SQL & 0.3002 & 0.8267 & 0.3564 & 0.8740 & 0.3260 & 0.8568 & 0.3027 & 0.8341 & 0.3033 & 0.8322 \\
\bottomrule
\end{tabular}
\caption{Augmentation Density Experiment: BERTBASE (Augmentation Model = Llama 3.1 8B)}
\label{tab:PQ_Density_BERTBASE_8B_MTEB}
\end{table*}

\begin{table*}[ht]
\centering
\tiny
\setlength{\tabcolsep}{3pt}
\begin{tabular}{lrrrrrrrrrr}
\toprule
Task Name & \multicolumn{2}{c}{No Aug} & \multicolumn{2}{c}{1PQ} & \multicolumn{2}{c}{3PQ} & \multicolumn{2}{c}{5PQ} & \multicolumn{2}{c}{7PQ} \\
\cmidrule(lr){2-3} \cmidrule(lr){4-5} \cmidrule(lr){6-7} \cmidrule(lr){8-9} \cmidrule(lr){10-11}
 & NDCG@10 & Recall@100 & NDCG@10 & Recall@100 & NDCG@10 & Recall@100 & NDCG@10 & Recall@100 & NDCG@10 & Recall@100 \\
\midrule
CosQA & 0.2000 & 0.5520 & 0.1705 & 0.6240 & 0.1603 & 0.6380 & 0.1772 & 0.6600 & 0.1624 & 0.6420 \\
HellaSwag & 0.2843 & 0.5921 & 0.2732 & 0.6279 & 0.2746 & 0.6289 & 0.2770 & 0.6265 & 0.2781 & 0.6280 \\
MedicalQARetrieval & 0.6227 & 0.9263 & 0.5856 & 0.9429 & 0.5806 & 0.9443 & 0.5747 & 0.9463 & 0.5747 & 0.9463 \\
NarrativeQARetrieval & 0.3193 & 0.6244 & 0.2253 & 0.6247 & 0.2275 & 0.6224 & 0.2137 & 0.6097 & 0.2207 & 0.6271 \\
PIQA & 0.2957 & 0.6001 & 0.2543 & 0.6262 & 0.2529 & 0.6235 & 0.2526 & 0.6181 & 0.2576 & 0.6219 \\
PublicHealthQA & 0.7154 & 0.9767 & 0.6946 & 0.9884 & 0.7225 & 0.9767 & 0.7712 & 0.9942 & 0.7587 & 0.9884 \\
SIQA & 0.0173 & 0.0476 & 0.0301 & 0.1290 & 0.0326 & 0.1377 & 0.0339 & 0.1320 & 0.0275 & 0.1136 \\
StackOverflowQA & 0.6208 & 0.8676 & 0.6199 & 0.8811 & 0.5957 & 0.8706 & 0.6055 & 0.8696 & 0.5988 & 0.8726 \\
SyntheticText2SQL & 0.3997 & 0.8699 & 0.3920 & 0.8918 & 0.3553 & 0.8552 & 0.3599 & 0.8578 & 0.3618 & 0.8636 \\
\bottomrule
\end{tabular}
\caption{Augmentation Density Experiment: Contriever (Augmentation Model = Llama 3.1 70B)}
\label{tab:PQ_Density_Contriever_70B_MTEB}
\end{table*}

\begin{table*}[ht]
\centering
\tiny
\setlength{\tabcolsep}{3pt}
\begin{tabular}{lrrrrrrrrrr}
\toprule
Task Name & \multicolumn{2}{c}{No Aug} & \multicolumn{2}{c}{1PQ} & \multicolumn{2}{c}{3PQ} & \multicolumn{2}{c}{5PQ} & \multicolumn{2}{c}{7PQ} \\
\cmidrule(lr){2-3} \cmidrule(lr){4-5} \cmidrule(lr){6-7} \cmidrule(lr){8-9} \cmidrule(lr){10-11}
 & NDCG@10 & Recall@100 & NDCG@10 & Recall@100 & NDCG@10 & Recall@100 & NDCG@10 & Recall@100 & NDCG@10 & Recall@100 \\
\midrule
CosQA & 0.2000 & 0.5520 & 0.2370 & 0.7720 & 0.2182 & 0.7420 & 0.2314 & 0.7840 & 0.2289 & 0.7380 \\
HellaSwag & 0.2843 & 0.5921 & 0.2709 & 0.6232 & 0.2644 & 0.6181 & 0.2688 & 0.6138 & 0.2633 & 0.6073 \\
MedicalQARetrieval & 0.6227 & 0.9263 & 0.5940 & 0.9458 & 0.5803 & 0.9468 & 0.5750 & 0.9443 & 0.5688 & 0.9458 \\
NarrativeQARetrieval & 0.3193 & 0.6244 & 0.2488 & 0.6263 & 0.2480 & 0.6402 & 0.2476 & 0.6368 & 0.2438 & 0.6417 \\
PIQA & 0.2957 & 0.6001 & 0.2342 & 0.5930 & 0.2315 & 0.5947 & 0.2298 & 0.5941 & 0.2309 & 0.5838 \\
PublicHealthQA & 0.7154 & 0.9767 & 0.7452 & 0.9942 & 0.7610 & 0.9942 & 0.7449 & 0.9942 & 0.7495 & 0.9942 \\
SIQA & 0.0173 & 0.0476 & 0.0343 & 0.1428 & 0.0351 & 0.1366 & 0.0315 & 0.1433 & 0.0308 & 0.1326 \\
StackOverflowQA & 0.6208 & 0.8676 & 0.5744 & 0.8475 & 0.5004 & 0.7969 & 0.4957 & 0.7884 & 0.4892 & 0.7939 \\
SyntheticText2SQL & 0.3997 & 0.8699 & 0.4020 & 0.9125 & 0.3614 & 0.8824 & 0.3769 & 0.8906 & 0.3667 & 0.8792 \\
\bottomrule
\end{tabular}
\caption{Augmentation Density Experiment: Contriever (Augmentation Model = Llama 3.1 8B)}
\label{tab:PQ_Density_Contriever_8B_MTEB}
\end{table*}
\clearpage
\begin{figure*}[t]
    \centering
    \includegraphics[width=0.75\linewidth]{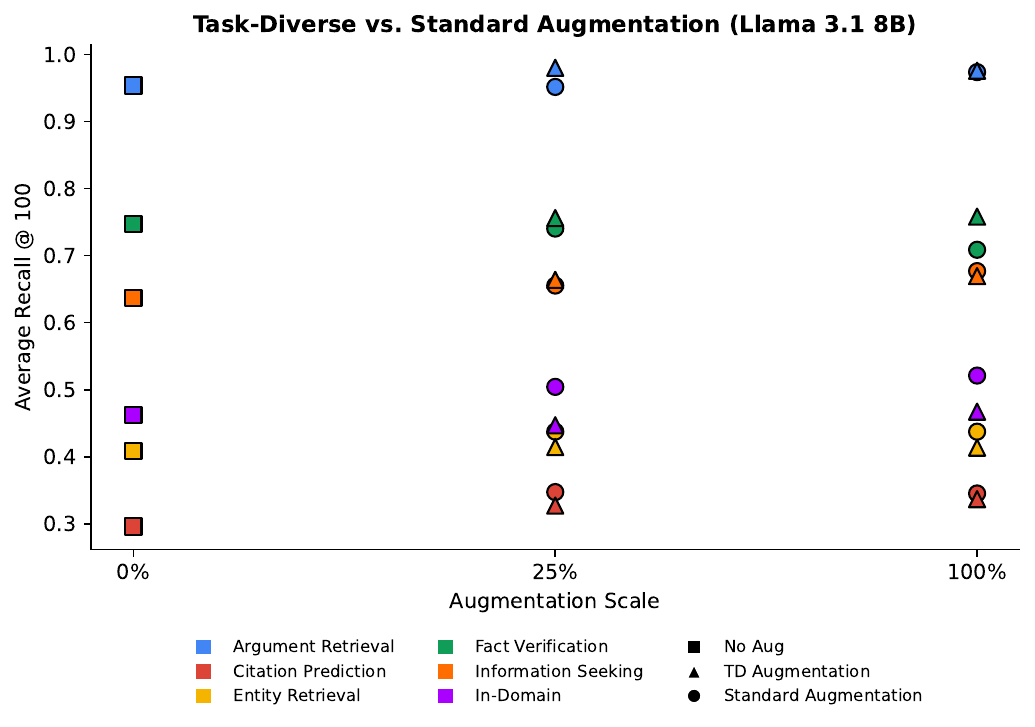}
    \caption{Category-wise Recall @ 100 scores on the BEIR benchmark, comparing BERTBASE baselines leveraging different augmentation scales with ($\triangle$) and without ($\bullet$) task diversity using Llama 3.1 8B model.}
    \label{fig:mteb_td}
\end{figure*}

\section{Ablations on Improving Performance and Efficiency}
We consider two ablations to improve performance and efficiency, one is the use of task-diverse augmentation to mitigate performance drops in case of the Llama models and performing distillation for augmentation efficiency \Cref{subsec:distilation}.

\label{app:qualitative_example}
\begin{table*}[ht]
\centering
\small
\begin{tabular}{p{5cm} p{3cm} p{5cm}}
    \toprule
    \textbf{Passage} & \textbf{Augmentation Type} & \textbf{Sample Query} \\
    \midrule
    Contact Volaris (airline): Find below customer service details of Volaris, including phone and email. Besides contact details, the page also offers information on the airline’s products and services. Reach the Volaris customer service below for queries, complaints or feedback. Volaris Head Office.  
    & Argument Retrieval 
    & Volaris does not provide phone or email contact details on this page. \\
    \midrule
    Many states in the US have shown growth at every census, but Colorado's population is actually growing faster now than it did during the 20s, 30s, and 40s. The current growth is already causing housing shortages and increased traffic as Denver and Colorado Springs try to address the increased immigration.  
    & Citation Prediction 
    & Population surge in Colorado triggers infrastructure strain. \\
    \midrule
    Scroll down for video. Speaking out: Mary Kay Letourneau Fualaau and her husband Vili Fualaau have spoken to Barbara Walters in an interview that will air on 20/20 on Friday - nearly 20 years after Letourneau's arrest for the relationship.  
    & Entity Retrieval 
    & Mary Kay Letourneau Fualaau \\
    \midrule
    In plant cells this is done by synthesising a new cell plate in between the two nuclei, which becomes a part of the new cell wall of each genetically identical daughter cell as the cytoplasm divides and the two cells split. The cell cycle (interphase) and mitosis are vital in both plant and animal cells as they allow growth and repair and asexual reproduction to occur.  
    & Information Seeking 
    & Plant cell division process \\
    \midrule
    Back to TopCauses. Hemorrhagic stroke occurs when a blood vessel bursts inside the brain. The brain is very sensitive to bleeding and damage can occur very rapidly. Bleeding irritates the brain tissue, causing swelling. Bleeding collects into a mass called a hematoma.  
    & Fact Verification 
    & A hematoma forms when bleeding collects in the brain. \\
    \bottomrule
\end{tabular}
\caption{Examples of Passage Augmentation with Sample Queries obtained in the Task Diversity Experiment \Cref{sec:methodology}, the augmentation model is Llama 3.1 8B}
\label{tab:augmentation}
\end{table*}

\section{Task Augmentation Representative Example}
\Cref{tab:augmentation} provides representative examples for each augmentation type considered for the Task-Diversity Experiment.

\subsection{Usage of Task-Diverse Augmentations}
\label{subsec:td_augmentation}
In the standard augmentation setup, pseudo-queries were generated from an "Information-Seeking" perspective for documents. While \Cref{fig:datascale_bert,fig:density_bert} shows an overall improvement on average, a category-wise analysis reveals a considerable drop in performance for "Fact-Checking" tasks compared to the No Augmentation baseline raising the need for more diverse, task-aligned augmentation strategies. Unlike prior works \citep{inpars,inparsv2,dai2023promptagator}, we do not expose the true target distribution but instead simulate an approximation of it. As discussed in \Cref{sec:methodology}, we adopt a fairer and more task-diverse augmentation strategy, that samples augmentations uniformly across task types. Using the Llama 3.1 8B model, we conduct an ablation study on BERTBASE to assess whether diverse augmentations can mitigate or delay saturation effects while maintaining performance across different tasks. From \Cref{fig:mteb_td}, we observe that incorporating task diversity helps counteract the performance drops observed in "Fact-Checking" tasks under the previous setup, leading to more balanced improvements across categories. However, while in-domain performance sees a slight decline, the saturation effects observed in standard augmentation persist in task-diverse augmentations as well.

\clearpage

\begin{table}[h]
\label{tab:distillation}
\centering
\tiny
\begin{tabular}{lcccc}
\toprule
category & No Aug & 70B Aug & 8B Aug & 8B Dist \\
\midrule
Argument Retrieval & 0.9538 & 0.9879 & 0.9737 & 0.9865 \\
Citation Prediction & 0.2959 & 0.3676 & 0.3453 & 0.3576 \\
Entity Retrieval & 0.4083 & 0.4435 & 0.4374 & 0.4484 \\
Fact Verification & 0.7474 & 0.7085 & 0.7087 & 0.7197 \\
In-Domain & 0.4623 & 0.4882 & 0.5211 & 0.4742 \\
Information Seeking & 0.6364 & 0.7036 & 0.6769 & 0.6950 \\
\bottomrule
\end{tabular}
\caption{Category-wise Recall@100 on BEIR for BERTBASE with 100\% Aug (5 PQs) using Llama variants viz. 3.1-70B, 3.1-8B \& 3.1-8B-Dist}
\end{table}

\begin{table}[h]
\small
\centering
\begin{tabular}{ll}
\toprule
Hyperparameter & Value \\
\midrule
model & Llama-3.1-8B-Instruct \\
lora alpha & 64 \\
lora rank & 32 \\
num epochs` & 10 \\
effective batch size & 128 \\
optimizer & adamw \\
learning rate & 5e-4 \\
warmup ratio & 0.05 \\
target modules & "q\_proj","k\_proj","v\_proj", \\
& "o\_proj","fc1","fc2" \\
lora dropout & 0.1 \\
\bottomrule
\end{tabular}
\caption{Hyperparameter used for SFT of the Llama 3.1 8B model}
\label{tab:sft_hyperparameters}
\end{table}

\subsection{Effectiveness of Distillation}
\label{subsec:distilation}
In \Cref{fig:datascale_bert,fig:density_bert}, we previously observed some minor differences in the metric scores obtained using Llama 3.1 8B and Llama 3.1 70B. As a follow-up experiment, we investigate whether small-scale data generation using Llama 3.1 70B, followed by supervised fine-tuning using this data, can bridge the performance gap. We generate 100K data points and do a 90\% data for training and 10\% data for validation split, and perform LORA \citep{hu2022lora} SFT of Llama 3.1 8B model. The hyperparameters used for the fine-tuning are presented in \Cref{tab:sft_hyperparameters}. As shown in \Cref{tab:distillation}, this distillation-based approach demonstrates that leveraging a smaller, fine-tuned model can be a viable and cost-effective alternative to relying solely on large-scale augmentation with a more expensive model. This suggests that fine-tuning on a small-scale curated subset of high-quality synthetic data can provide further performance improvements while significantly reducing computational costs.

\clearpage
\section{Result Tables for Primary Experiments}
\label{sec:primary_tables}

\Cref{tab:70B_BERT_scale_metrics_BEIR,tab:8B_BERT_scale_metrics_BEIR} and \Cref{tab:70B_BERT_density_metrics_BEIR,tab:8B_BERT_density_metrics_BEIR} demonstrate the NDCG @ 10, Recall @ 100, MRR @ 10 scores at different augmentation scales and densities. for BERTBASE experiments for the 70B and 8B models.

\Cref{tab:70B_Contriever_scale_metrics_BEIR,tab:8B_Contriever_scale_metrics_BEIR} and \Cref{tab:70B_BERT_density_metrics_BEIR,tab:8B_BERT_density_metrics_BEIR} demonstrate the NDCG @ 10, Recall @ 100, MRR @ 10 scores at different augmentation scales and densities for Contriever experiments.

\Cref{tab:70B_DB_scale_metrics_BEIR,tab:8B_DB_scale_metrics_BEIR} and \Cref{tab:70B_DB_density_metrics_BEIR,tab:8B_DB_density_metrics_BEIR} demonstrate the NDCG @ 10, Recall @ 100, MRR @ 10 scores at different augmentation scales and densities for DistilBERT experiments for the 70B and 8B models.

\Cref{tab:70B_E5_scale_metrics_BEIR,tab:8B_E5_scale_metrics_BEIR} and \Cref{tab:70B_E5_density_metrics_BEIR,tab:8B_E5_density_metrics_BEIR} demonstrate the NDCG @ 10, Recall @ 100, MRR @ 10 scores at different augmentation scales and densities for E5 experiments for the 70B and 8B models.

\begin{table*}[]
\centering
\tiny
\setlength{\tabcolsep}{3pt}
\begin{tabular}{@{}c|ccc|ccc|ccc|ccc|ccc@{}}
\toprule
                                                                          & \multicolumn{3}{c}{No Aug} & \multicolumn{3}{c}{1PQ}  & \multicolumn{3}{c}{3PQ}  & \multicolumn{3}{c}{5PQ}  & \multicolumn{3}{c}{7PQ}  \\ \midrule
task\_name                                                                & N       & R       & M      & N      & R      & M      & N      & R      & M      & N      & R      & M      & N      & R      & M      \\
\midrule
ArguAna                                                                   & 0.4090  & 0.9538  & 0.3281 & 0.5405 & 0.9900 & 0.4530 & 0.5473 & 0.9908 & 0.4616 & 0.5459 & 0.9879 & 0.4589 & 0.5525 & 0.9893 & 0.4667 \\
\begin{tabular}[c]{@{}c@{}}CQADupstack\\ Android\\ Retrieval\end{tabular} & 0.3577  & 0.6641  & 0.3638 & 0.3886 & 0.7552 & 0.3803 & 0.4075 & 0.7777 & 0.4013 & 0.4123 & 0.7936 & 0.4030 & 0.4127 & 0.7909 & 0.4054 \\
\begin{tabular}[c]{@{}c@{}}Climate\\ FEVER\end{tabular}                   & 0.2116  & 0.4822  & 0.2891 & 0.1914 & 0.5172 & 0.2457 & 0.1781 & 0.5022 & 0.2320 & 0.1763 & 0.5090 & 0.2315 & 0.1777 & 0.5020 & 0.2342 \\
DBPedia                                                                   & 0.3095  & 0.4083  & 0.6369 & 0.3061 & 0.4431 & 0.5812 & 0.2953 & 0.4429 & 0.5634 & 0.2964 & 0.4435 & 0.5733 & 0.2975 & 0.4415 & 0.5649 \\
FEVER                                                                     & 0.6391  & 0.9040  & 0.6131 & 0.3499 & 0.7411 & 0.3093 & 0.3149 & 0.7070 & 0.2735 & 0.3098 & 0.6948 & 0.2711 & 0.3111 & 0.6947 & 0.2718 \\
\begin{tabular}[c]{@{}c@{}}FiQA\\ 2018\end{tabular}                       & 0.2242  & 0.5166  & 0.2782 & 0.2648 & 0.6272 & 0.3166 & 0.2742 & 0.6414 & 0.3288 & 0.2683 & 0.6536 & 0.3179 & 0.2749 & 0.6493 & 0.3291 \\
HotpotQA                                                                  & 0.4557  & 0.5897  & 0.6339 & 0.4055 & 0.6326 & 0.5337 & 0.3835 & 0.6188 & 0.5009 & 0.3752 & 0.6155 & 0.4898 & 0.3649 & 0.6066 & 0.4759 \\
MSMARCO                                                                   & 0.6113  & 0.4623  & 0.8826 & 0.4947 & 0.4727 & 0.7611 & 0.4861 & 0.4952 & 0.7181 & 0.4590 & 0.4882 & 0.6753 & 0.4537 & 0.4971 & 0.6863 \\
NFCorpus                                                                  & 0.2709  & 0.2505  & 0.4654 & 0.3215 & 0.3153 & 0.4994 & 0.3312 & 0.3216 & 0.5103 & 0.3357 & 0.3365 & 0.5197 & 0.3384 & 0.3267 & 0.5225 \\
NQ                                                                        & 0.3467  & 0.8183  & 0.2987 & 0.3032 & 0.8334 & 0.2502 & 0.3081 & 0.8423 & 0.2555 & 0.3088 & 0.8416 & 0.2557 & 0.3095 & 0.8437 & 0.2550 \\
\begin{tabular}[c]{@{}c@{}}Quora\\ Retrieval\end{tabular}                 & 0.8284  & 0.9792  & 0.8221 & 0.8273 & 0.9823 & 0.8193 & 0.8272 & 0.9816 & 0.8186 & 0.8232 & 0.9811 & 0.8136 & 0.8239 & 0.9812 & 0.8148 \\
SCIDOCS                                                                   & 0.1293  & 0.2959  & 0.2432 & 0.1438 & 0.3579 & 0.2633 & 0.1460 & 0.3627 & 0.2595 & 0.1480 & 0.3676 & 0.2708 & 0.1450 & 0.3693 & 0.2624 \\
SciFact                                                                   & 0.5187  & 0.8560  & 0.4867 & 0.6271 & 0.9260 & 0.5886 & 0.6254 & 0.9283 & 0.5907 & 0.6210 & 0.9217 & 0.5843 & 0.6299 & 0.9300 & 0.5922 \\
\midrule
Average                                                                   & 0.4086  & 0.6293  & 0.4878 & 0.3973 & 0.6611 & 0.4617 & 0.3942 & 0.6625 & 0.4549 & 0.3908 & 0.6642 & 0.4511 & 0.3917 & 0.6633 & 0.4524 \\ \bottomrule
\end{tabular}
\caption{Benchmark-wise metric scores on the BEIR benchmark, comparing BERTBASE baselines leveraging different augmentation densities using Llama 3.1 70B model. N indicates NDCG @ 10, R indicates Recall @ 100 and M indicates MRR @ 10}
\label{tab:70B_BERT_density_metrics_BEIR}
\end{table*}

\begin{table*}[]
\centering
\tiny
\setlength{\tabcolsep}{3pt}
\begin{tabular}{@{}c|ccc|ccc|ccc|ccc|ccc@{}}
\toprule
                                                                          & \multicolumn{3}{c}{No Aug} & \multicolumn{3}{c}{1PQ}  & \multicolumn{3}{c}{3PQ}  & \multicolumn{3}{c}{5PQ}           & \multicolumn{3}{c}{7PQ}  \\ \midrule
task\_name                                                                & N       & R       & M      & N      & R      & M      & N      & R      & M      & N      & R               & M      & N      & R      & M      \\
\midrule
ArguAna                                                                   & 0.4090  & 0.9538  & 0.3281 & 0.5147 & 0.9787 & 0.4283 & 0.5215 & 0.9730 & 0.4343 & 0.5261 & 0.9737          & 0.4405 & 0.5180 & 0.9723 & 0.4330 \\
\begin{tabular}[c]{@{}c@{}}CQADupstack\\ Android\\ Retrieval\end{tabular} & 0.3577  & 0.6641  & 0.3638 & 0.3659 & 0.7422 & 0.3553 & 0.3726 & 0.7503 & 0.3631 & 0.3598 & 0.7514          & 0.3512 & 0.3579 & 0.7341 & 0.3479 \\
\begin{tabular}[c]{@{}c@{}}Climate\\ FEVER\end{tabular}                   & 0.2116  & 0.4822  & 0.2891 & 0.1673 & 0.5025 & 0.2191 & 0.1570 & 0.4879 & 0.2050 & 0.1484 & 0.4723          & 0.1945 & 0.1414 & 0.4589 & 0.1835 \\
DBPedia                                                                   & 0.3095  & 0.4083  & 0.6369 & 0.2863 & 0.4461 & 0.5535 & 0.2981 & 0.4421 & 0.5672 & 0.2967 & 0.4374          & 0.5733 & 0.2941 & 0.4345 & 0.5629 \\
FEVER                                                                     & 0.6391  & 0.9040  & 0.6131 & 0.3348 & 0.7714 & 0.2904 & 0.3198 & 0.7393 & 0.2776 & 0.3037 & 0.7256          & 0.2607 & 0.3016 & 0.7237 & 0.2600 \\
\begin{tabular}[c]{@{}c@{}}FiQA\\ 2018\end{tabular}                       & 0.2242  & 0.5166  & 0.2782 & 0.2132 & 0.5592 & 0.2549 & 0.2076 & 0.5573 & 0.2500 & 0.2179 & 0.5717          & 0.2639 & 0.2159 & 0.5727 & 0.2678 \\
HotpotQA                                                                  & 0.4557  & 0.5897  & 0.6339 & 0.3852 & 0.6170 & 0.5090 & 0.3792 & 0.6022 & 0.5003 & 0.3700 & 0.5864          & 0.4884 & 0.3609 & 0.5817 & 0.4778 \\
MSMARCO                                                                   & 0.6113  & 0.4623  & 0.8826 & 0.5089 & 0.4962 & 0.7781 & 0.4809 & 0.5224 & 0.7339 & 0.4719 & \textbf{0.5211} & 0.7091 & 0.4701 & 0.5269 & 0.6913 \\
NFCorpus                                                                  & 0.2709  & 0.2505  & 0.4654 & 0.3184 & 0.3150 & 0.4995 & 0.3304 & 0.3258 & 0.5121 & 0.3292 & 0.3298          & 0.5039 & 0.3289 & 0.3266 & 0.5172 \\
NQ                                                                        & 0.3467  & 0.8183  & 0.2987 & 0.2955 & 0.8319 & 0.2442 & 0.3001 & 0.8391 & 0.2480 & 0.3035 & 0.8442          & 0.2509 & 0.2985 & 0.8345 & 0.2465 \\
\begin{tabular}[c]{@{}c@{}}Quora\\ Retrieval\end{tabular}                 & 0.8284  & 0.9792  & 0.8221 & 0.8256 & 0.9812 & 0.8174 & 0.8223 & 0.9805 & 0.8130 & 0.8176 & 0.9779          & 0.8071 & 0.8136 & 0.9763 & 0.8035 \\
SCIDOCS                                                                   & 0.1293  & 0.2959  & 0.2432 & 0.1338 & 0.3465 & 0.2404 & 0.1360 & 0.3422 & 0.2422 & 0.1347 & 0.3453          & 0.2398 & 0.1334 & 0.3426 & 0.2368 \\
SciFact                                                                   & 0.5187  & 0.8560  & 0.4867 & 0.6280 & 0.9277 & 0.5869 & 0.6210 & 0.9200 & 0.5791 & 0.6316 & 0.9283 & 0.5930 & 0.6198 & 0.9283 & 0.5800 \\
\midrule
Average                                                                   & 0.4086  & 0.6293  & 0.4878 & 0.3829 & 0.6550 & 0.4444 & 0.3805 & 0.6525 & 0.4404 & 0.3778 & 0.6512 & 0.4366 & 0.3734 & 0.6472 & 0.4314 \\ \bottomrule
\end{tabular}
\caption{Benchmark-wise metric scores on the BEIR benchmark, comparing BERTBASE baselines leveraging different augmentation densities using Llama 3.1 8B model. N indicates NDCG @ 10, R indicates Recall @ 100 and M indicates MRR @ 10}
\label{tab:8B_BERT_density_metrics_BEIR}
\end{table*}

\begin{table*}[]
\centering
\tiny
\setlength{\tabcolsep}{3pt}
\begin{tabular}{@{}c|ccc|ccc|ccc|ccc|ccc@{}}
\toprule
                                                                          & \multicolumn{3}{c}{No Aug} & \multicolumn{3}{c}{25\%} & \multicolumn{3}{c}{50\%} & \multicolumn{3}{c}{75\%}          & \multicolumn{3}{c}{100\%} \\ \midrule
task\_name                                                                & N       & R       & M      & N      & R      & M      & N      & R      & M      & N      & R               & M      & N       & R      & M      \\
\midrule
ArguAna                                                                   & 0.4090  & 0.9538  & 0.3281 & 0.5432 & 0.9893 & 0.4596 & 0.5426 & 0.9879 & 0.4541 & 0.5527 & 0.9879          & 0.4640 & 0.5459  & 0.9879 & 0.4589 \\
\begin{tabular}[c]{@{}c@{}}CQADupstack\\ Android\\ Retrieval\end{tabular} & 0.3577  & 0.6641  & 0.3638 & 0.3949 & 0.7609 & 0.3874 & 0.4066 & 0.7815 & 0.3998 & 0.4116 & 0.7860          & 0.4070 & 0.4123  & 0.7936 & 0.4030 \\
\begin{tabular}[c]{@{}c@{}}Climate\\ FEVER\end{tabular}                   & 0.2116  & 0.4822  & 0.2891 & 0.1859 & 0.5122 & 0.2435 & 0.1885 & 0.5077 & 0.2486 & 0.1807 & 0.5044          & 0.2389 & 0.1763  & 0.5090 & 0.2315 \\
DBPedia                                                                   & 0.3095  & 0.4083  & 0.6369 & 0.3100 & 0.4342 & 0.5906 & 0.3004 & 0.4464 & 0.5728 & 0.3012 & 0.4491          & 0.5810 & 0.2964  & 0.4435 & 0.5733 \\
FEVER                                                                     & 0.6391  & 0.9040  & 0.6131 & 0.3454 & 0.7334 & 0.3049 & 0.3170 & 0.7126 & 0.2761 & 0.3267 & 0.7161          & 0.2884 & 0.3098  & 0.6948 & 0.2711 \\
\begin{tabular}[c]{@{}c@{}}FiQA\\ 2018\end{tabular}                       & 0.2242  & 0.5166  & 0.2782 & 0.2682 & 0.6318 & 0.3219 & 0.2768 & 0.6410 & 0.3300 & 0.2679 & 0.6486          & 0.3214 & 0.2683  & 0.6536 & 0.3179 \\
HotpotQA                                                                  & 0.4557  & 0.5897  & 0.6339 & 0.4014 & 0.6292 & 0.5274 & 0.3862 & 0.6232 & 0.5047 & 0.3814 & 0.6157          & 0.4978 & 0.3752  & 0.6155 & 0.4898 \\
MSMARCO                                                                   & 0.6113  & 0.4623  & 0.8826 & 0.5106 & 0.4782 & 0.7443 & 0.4796 & 0.4799 & 0.7088 & 0.5068 & \textbf{0.5018} & 0.7403 & 0.4590  & 0.4882 & 0.6753 \\
NFCorpus                                                                  & 0.2709  & 0.2505  & 0.4654 & 0.3265 & 0.3200 & 0.5035 & 0.3384 & 0.3286 & 0.5246 & 0.3357 & 0.3251          & 0.5169 & 0.3357  & 0.3365 & 0.5197 \\
NQ                                                                        & 0.3467  & 0.8183  & 0.2987 & 0.3077 & 0.8420 & 0.2547 & 0.3055 & 0.8418 & 0.2521 & 0.3120 & 0.8457          & 0.2594 & 0.3088  & 0.8416 & 0.2557 \\
\begin{tabular}[c]{@{}c@{}}Quora\\ Retrieval\end{tabular}                 & 0.8284  & 0.9792  & 0.8221 & 0.8276 & 0.9824 & 0.8196 & 0.8246 & 0.9806 & 0.8158 & 0.8241 & 0.9816          & 0.8148 & 0.8232  & 0.9811 & 0.8136 \\
SCIDOCS                                                                   & 0.1293  & 0.2959  & 0.2432 & 0.1431 & 0.3564 & 0.2580 & 0.1488 & 0.3664 & 0.2688 & 0.1495 & 0.3653          & 0.2690 & 0.1480  & 0.3676 & 0.2708 \\
SciFact                                                                   & 0.5187  & 0.8560  & 0.4867 & 0.6281 & 0.9100 & 0.5948 & 0.6236 & 0.9150 & 0.5885 & 0.6286 & 0.9183 & 0.5890 & 0.6210  & 0.9217 & 0.5843 \\
\midrule
Average                                                                   & 0.4086  & 0.6293  & 0.4878 & 0.3994 & 0.6600 & 0.4623 & 0.3953 & 0.6625 & 0.4573 & 0.3984 & 0.6650 & 0.4606 & 0.3908  & 0.6642 & 0.4511 \\ \bottomrule
\end{tabular}
\caption{Benchmark-wise metric scores on the BEIR benchmark, comparing BERTBASE baselines leveraging different augmentation scales using Llama 3.1 70B model. N indicates NDCG @ 10, R indicates Recall @ 100 and M indicates MRR @ 10}
\label{tab:70B_BERT_scale_metrics_BEIR}
\end{table*}

\begin{table*}[]
\centering
\tiny
\setlength{\tabcolsep}{3pt}
\begin{tabular}{@{}c|ccc|ccc|ccc|lll|ccc@{}}
\toprule
                                                                          & \multicolumn{3}{c}{No Aug} & \multicolumn{3}{c}{25\%} & \multicolumn{3}{c}{50\%} & \multicolumn{3}{c}{75\%}                                                             & \multicolumn{3}{c}{100\%} \\ \midrule
task\_name                                                                & N       & R       & M      & N      & R      & M      & N      & R      & M      & \multicolumn{1}{c}{N}      & \multicolumn{1}{c}{R}      & \multicolumn{1}{c}{M}      & N       & R      & M      \\
\midrule
ArguAna                                                                   & 0.4090  & 0.9538  & 0.3281 & 0.4930 & 0.9516 & 0.4075 & 0.5102 & 0.9652 & 0.4270 & 0.5349                     & 0.9822                     & 0.4451                     & 0.5261  & 0.9737 & 0.4405 \\
\begin{tabular}[c]{@{}c@{}}CQADupstack\\ Android\\ Retrieval\end{tabular} & 0.3577  & 0.6641  & 0.3638 & 0.3401 & 0.7201 & 0.3318 & 0.3553 & 0.7249 & 0.3521 & 0.3537                     & 0.7521                     & 0.3460                     & 0.3598  & 0.7514 & 0.3512 \\
\begin{tabular}[c]{@{}c@{}}Climate\\ FEVER\end{tabular}                   & 0.2116  & 0.4822  & 0.2891 & 0.1800 & 0.4991 & 0.2325 & 0.1650 & 0.4893 & 0.2162 & 0.1526                     & 0.4785                     & 0.1996                     & 0.1484  & 0.4723 & 0.1945 \\
DBPedia                                                                   & 0.3095  & 0.4083  & 0.6369 & 0.2991 & 0.4375 & 0.5715 & 0.2965 & 0.4407 & 0.5684 & 0.2969                     & 0.4346                     & 0.5726                     & 0.2967  & 0.4374 & 0.5733 \\
FEVER                                                                     & 0.6391  & 0.9040  & 0.6131 & 0.3792 & 0.7908 & 0.3349 & 0.3439 & 0.7591 & 0.3021 & 0.3168                     & 0.7376                     & 0.2748                     & 0.3037  & 0.7256 & 0.2607 \\
\begin{tabular}[c]{@{}c@{}}FiQA\\ 2018\end{tabular}                       & 0.2242  & 0.5166  & 0.2782 & 0.1989 & 0.5388 & 0.2403 & 0.2143 & 0.5585 & 0.2628 & 0.2283                     & 0.5699                     & 0.2738                     & 0.2179  & 0.5717 & 0.2639 \\
HotpotQA                                                                  & 0.4557  & 0.5897  & 0.6339 & 0.3643 & 0.5766 & 0.4868 & 0.3789 & 0.5986 & 0.5034 & 0.3887                     & 0.6128                     & 0.5112                     & 0.3700  & 0.5864 & 0.4884 \\
MSMARCO                                                                   & 0.6113  & 0.4623  & 0.8826 & 0.5246 & 0.5042 & 0.7796 & 0.4839 & 0.4958 & 0.7219 & 0.4930                     & 0.5296                     & 0.7419                     & 0.4719  & 0.5211 & 0.7091 \\
NFCorpus                                                                  & 0.2709  & 0.2505  & 0.4654 & 0.3147 & 0.3116 & 0.4945 & 0.3223 & 0.3147 & 0.4995 & 0.3364                     & 0.3285                     & 0.5212                     & 0.3292  & 0.3298 & 0.5039 \\
NQ                                                                        & 0.3467  & 0.8183  & 0.2987 & 0.2899 & 0.8064 & 0.2410 & 0.2921 & 0.8240 & 0.2406 & 0.3163                     & 0.8591                     & 0.2618                     & 0.3035  & 0.8442 & 0.2509 \\
\begin{tabular}[c]{@{}c@{}}Quora\\ Retrieval\end{tabular}                 & 0.8284  & 0.9792  & 0.8221 & 0.8167 & 0.9784 & 0.8076 & 0.8157 & 0.9775 & 0.8075 & 0.8163                     & 0.9781                     & 0.8065                     & 0.8176  & 0.9779 & 0.8071 \\
SCIDOCS                                                                   & 0.1293  & 0.2959  & 0.2432 & 0.1370 & 0.3473 & 0.2502 & 0.1374 & 0.3484 & 0.2516 & 0.1390                     & 0.3506                     & 0.2501                     & 0.1347  & 0.3453 & 0.2398 \\
SciFact                                                                   & 0.5187  & 0.8560  & 0.4867 & 0.5952 & 0.9310 & 0.5444 & 0.6150 & 0.9210 & 0.5727 & 0.6334                     & 0.9350                     & 0.5923                     & 0.6316  & 0.9283 & 0.5930 \\
\midrule
Average                                                                   & 0.4086  & 0.6293  & 0.4878 & 0.3794 & 0.6456 & 0.4402 & 0.3793 & 0.6475 & 0.4404 & \multicolumn{1}{c}{0.3851} & \multicolumn{1}{c}{0.6576} & \multicolumn{1}{c}{0.4459} & 0.3778  & 0.6512 & 0.4366 \\ \bottomrule
\end{tabular}
\caption{Benchmark-wise metric scores on the BEIR benchmark, comparing BERTBASE baselines leveraging different augmentation scales using Llama 3.1 8B model. N indicates NDCG @ 10, R indicates Recall @ 100 and M indicates MRR @ 10}
\label{tab:8B_BERT_scale_metrics_BEIR}
\end{table*}

\begin{table*}[]
\centering
\tiny
\setlength{\tabcolsep}{3pt}
\begin{tabular}{@{}c|ccc|ccc|ccc|ccc|ccc@{}}
\toprule
                                                                          & \multicolumn{3}{c}{No Aug} & \multicolumn{3}{c}{1PQ}  & \multicolumn{3}{c}{3PQ}                                                              & \multicolumn{3}{c}{5PQ}  & \multicolumn{3}{c}{7PQ}  \\ \midrule
task\_name                                                                & N       & R       & M      & N      & R      & M      & \multicolumn{1}{c}{N}      & \multicolumn{1}{c}{R}      & \multicolumn{1}{c}{M}      & N      & R      & M      & N      & R      & M      \\
ArguAna                                                                   & 0.4705  & 0.9751  & 0.3829 & 0.5596 & 0.9879 & 0.4720 & 0.5473                     & 0.9851                     & 0.4600                     & 0.5694 & 0.9865 & 0.4837 & 0.5544 & 0.9844 & 0.4688 \\
\begin{tabular}[c]{@{}c@{}}CQADupstack\\ Android\\ Retrieval\end{tabular} & 0.4181  & 0.7520  & 0.4132 & 0.4246 & 0.7805 & 0.4182 & 0.4181                     & 0.7845                     & 0.4115                     & 0.4247 & 0.8022 & 0.4222 & 0.4223 & 0.7955 & 0.4139 \\
\begin{tabular}[c]{@{}c@{}}Climate\\ FEVER\end{tabular}                   & 0.2370  & 0.5419  & 0.3210 & 0.2039 & 0.5415 & 0.2627 & 0.1984                     & 0.5125                     & 0.2567                     & 0.1768 & 0.5049 & 0.2319 & 0.1691 & 0.5008 & 0.2179 \\
DBPedia                                                                   & 0.3713  & 0.4945  & 0.7022 & 0.3549 & 0.5050 & 0.6498 & 0.3366                     & 0.4925                     & 0.6137                     & 0.3295 & 0.4770 & 0.6144 & 0.3322 & 0.4781 & 0.6129 \\
FEVER                                                                     & 0.7064  & 0.9366  & 0.6869 & 0.4335 & 0.8204 & 0.3910 & 0.4162                     & 0.8046                     & 0.3733                     & 0.3658 & 0.7669 & 0.3233 & 0.3639 & 0.7592 & 0.3222 \\
\begin{tabular}[c]{@{}c@{}}FiQA\\ 2018\end{tabular}                       & 0.2672  & 0.6061  & 0.3283 & 0.2939 & 0.6448 & 0.3548 & 0.2891                     & 0.6489                     & 0.3440                     & 0.2882 & 0.6622 & 0.3458 & 0.2831 & 0.6579 & 0.3404 \\
HotpotQA                                                                  & 0.5409  & 0.6824  & 0.7206 & 0.4939 & 0.7067 & 0.6405 & 0.4505                     & 0.6783                     & 0.5853                     & 0.4272 & 0.6621 & 0.5530 & 0.4244 & 0.6546 & 0.5488 \\
MSMARCO                                                                   & 0.5913  & 0.4800  & 0.8667 & 0.5262 & 0.5078 & 0.7929 & 0.5066                     & 0.5131                     & 0.7644                     & 0.4990 & 0.5018 & 0.7750 & 0.5046 & 0.4963 & 0.7727 \\
NFCorpus                                                                  & 0.3121  & 0.2964  & 0.5020 & 0.3457 & 0.3425 & 0.5345 & 0.3458                     & 0.3500                     & 0.5329                     & 0.3547 & 0.3497 & 0.5436 & 0.3437 & 0.3433 & 0.5349 \\
NQ                                                                        & 0.3762  & 0.8692  & 0.3191 & 0.3313 & 0.8823 & 0.2723 & 0.3187                     & 0.8682                     & 0.2614                     & 0.3202 & 0.8654 & 0.2632 & 0.3162 & 0.8650 & 0.2601 \\
\begin{tabular}[c]{@{}c@{}}Quora\\ Retrieval\end{tabular}                 & 0.8470  & 0.9871  & 0.8385 & 0.8353 & 0.9855 & 0.8267 & 0.8290                     & 0.9830                     & 0.8203                     & 0.8307 & 0.9837 & 0.8212 & 0.8285 & 0.9838 & 0.8195 \\
SCIDOCS                                                                   & 0.1578  & 0.3568  & 0.2898 & 0.1582 & 0.3918 & 0.2803 & 0.1534                     & 0.3803                     & 0.2716                     & 0.1558 & 0.3832 & 0.2713 & 0.1539 & 0.3834 & 0.2700 \\
SciFact                                                                   & 0.6268  & 0.9277  & 0.5843 & 0.6818 & 0.9393 & 0.6454 & 0.6621                     & 0.9360                     & 0.6182                     & 0.6880 & 0.9427 & 0.6551 & 0.6770 & 0.9493 & 0.6372 \\
Average                                                                   & 0.4556  & 0.6851  & 0.5350 & 0.4341 & 0.6951 & 0.5032 & \multicolumn{1}{c}{0.4209} & \multicolumn{1}{c}{0.6875} & \multicolumn{1}{c}{0.4856} & 0.4177 & 0.6837 & 0.4849 & 0.4133 & 0.6809 & 0.4784 \\ \bottomrule
\end{tabular}
\caption{Benchmark-wise metric scores on the BEIR benchmark, comparing Contriever baselines leveraging different augmentation densities using Llama 3.1 70B model. N indicates NDCG @ 10, R indicates Recall @ 100 and M indicates MRR @ 10}
\label{tab:70B_Contriever_density_metrics_BEIR}
\end{table*}

\begin{table*}[]
\centering
\tiny
\setlength{\tabcolsep}{3pt}
\begin{tabular}{@{}c|ccc|ccc|ccc|ccc|ccc@{}}
\toprule
                                                                          & \multicolumn{3}{c}{No Aug} & \multicolumn{3}{c}{1PQ}                                                              & \multicolumn{3}{c}{3PQ}                                                              & \multicolumn{3}{c}{5PQ}                                                              & \multicolumn{3}{c}{7PQ}                                                              \\ \midrule
task\_name                                                                & N       & R       & M      & \multicolumn{1}{c}{N}      & \multicolumn{1}{c}{R}      & \multicolumn{1}{c}{M}      & \multicolumn{1}{c}{N}      & \multicolumn{1}{c}{R}      & \multicolumn{1}{c}{M}      & \multicolumn{1}{c}{N}      & \multicolumn{1}{c}{R}      & \multicolumn{1}{c}{M}      & \multicolumn{1}{c}{N}      & \multicolumn{1}{c}{R}      & \multicolumn{1}{c}{M}      \\
ArguAna                                                                   & 0.4705  & 0.9751  & 0.3829 & 0.5460                     & 0.9900                     & 0.4540                     & 0.5278                     & 0.9844                     & 0.4386                     & 0.5307                     & 0.9751                     & 0.4445                     & 0.5403                     & 0.9851                     & 0.4518                     \\
\begin{tabular}[c]{@{}c@{}}CQADupstack\\ Android\\ Retrieval\end{tabular} & 0.4181  & 0.7520  & 0.4132 & 0.4081                     & 0.7681                     & 0.4004                     & 0.3974                     & 0.7555                     & 0.3965                     & 0.3848                     & 0.7601                     & 0.3741                     & 0.3789                     & 0.7704                     & 0.3714                     \\
\begin{tabular}[c]{@{}c@{}}Climate\\ FEVER\end{tabular}                   & 0.2370  & 0.5419  & 0.3210 & 0.1664                     & 0.5126                     & 0.2156                     & 0.1583                     & 0.4988                     & 0.2021                     & 0.1461                     & 0.4696                     & 0.1904                     & 0.1479                     & 0.4707                     & 0.1893                     \\
DBPedia                                                                   & 0.3713  & 0.4945  & 0.7022 & 0.3438                     & 0.5094                     & 0.6244                     & 0.3373                     & 0.4989                     & 0.6159                     & 0.3170                     & 0.4833                     & 0.5942                     & 0.3204                     & 0.4893                     & 0.6036                     \\
FEVER                                                                     & 0.7064  & 0.9366  & 0.6869 & 0.4480                     & 0.8621                     & 0.3974                     & 0.4106                     & 0.8385                     & 0.3612                     & 0.3743                     & 0.8072                     & 0.3271                     & 0.3718                     & 0.8062                     & 0.3256                     \\
\begin{tabular}[c]{@{}c@{}}FiQA\\ 2018\end{tabular}                       & 0.2672  & 0.6061  & 0.3283 & 0.2375                     & 0.5921                     & 0.2890                     & 0.2357                     & 0.5784                     & 0.2910                     & 0.2376                     & 0.5932                     & 0.2921                     & 0.2324                     & 0.5754                     & 0.2926                     \\
HotpotQA                                                                  & 0.5409  & 0.6824  & 0.7206 & 0.4989                     & 0.7175                     & 0.6451                     & 0.4621                     & 0.6866                     & 0.5998                     & 0.4366                     & 0.6561                     & 0.5699                     & 0.4337                     & 0.6621                     & 0.5661                     \\
MSMARCO                                                                   & 0.5913  & 0.4800  & 0.8667 & 0.5208                     & 0.5107                     & 0.7805                     & 0.5070                     & 0.5135                     & 0.7973                     & 0.5046                     & 0.5369                     & 0.7979                     & 0.5087                     & 0.5422                     & 0.7985                     \\
NFCorpus                                                                  & 0.3121  & 0.2964  & 0.5020 & 0.3475                     & 0.3350                     & 0.5446                     & 0.3467                     & 0.3336                     & 0.5380                     & 0.3536                     & 0.3350                     & 0.5568                     & 0.3532                     & 0.3381                     & 0.5541                     \\
NQ                                                                        & 0.3762  & 0.8692  & 0.3191 & 0.3364                     & 0.8899                     & 0.2785                     & 0.3245                     & 0.8695                     & 0.2650                     & 0.3236                     & 0.8676                     & 0.2644                     & 0.3214                     & 0.8702                     & 0.2643                     \\
\begin{tabular}[c]{@{}c@{}}Quora\\ Retrieval\end{tabular}                 & 0.8470  & 0.9871  & 0.8385 & 0.8323                     & 0.9839                     & 0.8242                     & 0.8262                     & 0.9830                     & 0.8170                     & 0.8192                     & 0.9802                     & 0.8102                     & 0.8181                     & 0.9800                     & 0.8079                     \\
SCIDOCS                                                                   & 0.1578  & 0.3568  & 0.2898 & 0.1510                     & 0.3652                     & 0.2679                     & 0.1466                     & 0.3595                     & 0.2529                     & 0.1444                     & 0.3591                     & 0.2546                     & 0.1477                     & 0.3698                     & 0.2623                     \\
SciFact                                                                   & 0.6268  & 0.9277  & 0.5843 & 0.6861                     & 0.9483                     & 0.6478                     & 0.6727                     & 0.9533                     & 0.6335                     & 0.6512                     & 0.9560                     & 0.6072                     & 0.6772                     & 0.9493                     & 0.6455                     \\
Average                                                                   & 0.4556  & 0.6851  & 0.5350 & \multicolumn{1}{c}{0.4248} & \multicolumn{1}{c}{0.6911} & \multicolumn{1}{c}{0.4900} & \multicolumn{1}{c}{0.4118} & \multicolumn{1}{c}{0.6810} & \multicolumn{1}{c}{0.4776} & \multicolumn{1}{c}{0.4018} & \multicolumn{1}{c}{0.6753} & \multicolumn{1}{c}{0.4679} & \multicolumn{1}{c}{0.4040} & \multicolumn{1}{c}{0.6776} & \multicolumn{1}{c}{0.4718} \\ \bottomrule
\end{tabular}
\caption{Benchmark-wise metric scores on the BEIR benchmark, comparing Contriever baselines leveraging different augmentation densities using Llama 3.1 8B model. N indicates NDCG @ 10, R indicates Recall @ 100 and M indicates MRR @ 10}
\label{tab:8B_Contriever_density_metrics_BEIR}
\end{table*}

\begin{table*}[]
\centering
\tiny
\setlength{\tabcolsep}{3pt}
\begin{tabular}{@{}c|ccc|lll|lll|lll|lll@{}}
\toprule
                                                                          & \multicolumn{3}{c}{No Aug} & \multicolumn{3}{c}{25\%}                                                             & \multicolumn{3}{c}{50\%}                                                             & \multicolumn{3}{c}{75\%}                                                             & \multicolumn{3}{c}{100\%}                                                            \\ \midrule
task\_name                                                                & N       & R       & M      & \multicolumn{1}{c}{N}      & \multicolumn{1}{c}{R}      & \multicolumn{1}{c}{M}      & \multicolumn{1}{c}{N}      & \multicolumn{1}{c}{R}      & \multicolumn{1}{c}{M}      & \multicolumn{1}{c}{N}      & \multicolumn{1}{c}{R}      & \multicolumn{1}{c}{M}      & \multicolumn{1}{c}{N}      & \multicolumn{1}{c}{R}      & \multicolumn{1}{c}{M}      \\
ArguAna                                                                   & 0.4705  & 0.9751  & 0.3829 & 0.5463                     & 0.9858                     & 0.4526                     & 0.5425                     & 0.9851                     & 0.4529                     & 0.5250                     & 0.9879                     & 0.4381                     & 0.5307                     & 0.9751                     & 0.4445                     \\
\begin{tabular}[c]{@{}c@{}}CQADupstack\\ Android\\ Retrieval\end{tabular} & 0.4181  & 0.7520  & 0.4132 & 0.3856                     & 0.7424                     & 0.3827                     & 0.3871                     & 0.7449                     & 0.3843                     & 0.3865                     & 0.7503                     & 0.3808                     & 0.3848                     & 0.7601                     & 0.3741                     \\
\begin{tabular}[c]{@{}c@{}}Climate\\ FEVER\end{tabular}                   & 0.2370  & 0.5419  & 0.3210 & 0.1829                     & 0.5177                     & 0.2364                     & 0.1608                     & 0.5005                     & 0.2038                     & 0.1565                     & 0.4823                     & 0.2021                     & 0.1461                     & 0.4696                     & 0.1904                     \\
DBPedia                                                                   & 0.3713  & 0.4945  & 0.7022 & 0.3546                     & 0.5103                     & 0.6612                     & 0.3384                     & 0.4978                     & 0.6281                     & 0.3205                     & 0.4922                     & 0.5984                     & 0.3170                     & 0.4833                     & 0.5942                     \\
FEVER                                                                     & 0.7064  & 0.9366  & 0.6869 & 0.4741                     & 0.8628                     & 0.4264                     & 0.4151                     & 0.8419                     & 0.3669                     & 0.4009                     & 0.8259                     & 0.3535                     & 0.3743                     & 0.8072                     & 0.3271                     \\
\begin{tabular}[c]{@{}c@{}}FiQA\\ 2018\end{tabular}                       & 0.2672  & 0.6061  & 0.3283 & 0.2325                     & 0.5784                     & 0.2873                     & 0.2310                     & 0.5715                     & 0.2859                     & 0.2417                     & 0.5667                     & 0.2894                     & 0.2376                     & 0.5932                     & 0.2921                     \\
HotpotQA                                                                  & 0.5409  & 0.6824  & 0.7206 & 0.4999                     & 0.7163                     & 0.6489                     & 0.4706                     & 0.6935                     & 0.6105                     & 0.4518                     & 0.6752                     & 0.5888                     & 0.4366                     & 0.6561                     & 0.5699                     \\
MSMARCO                                                                   & 0.5913  & 0.4800  & 0.8667 & 0.5053                     & 0.5000                     & 0.7833                     & 0.4997                     & 0.4966                     & 0.7877                     & 0.4969                     & 0.5147                     & 0.8066                     & 0.5046                     & 0.5369                     & 0.7979                     \\
NFCorpus                                                                  & 0.3121  & 0.2964  & 0.5020 & 0.3424                     & 0.3288                     & 0.5343                     & 0.3491                     & 0.3370                     & 0.5401                     & 0.3498                     & 0.3363                     & 0.5379                     & 0.3536                     & 0.3350                     & 0.5568                     \\
NQ                                                                        & 0.3762  & 0.8692  & 0.3191 & 0.3320                     & 0.8823                     & 0.2741                     & 0.3197                     & 0.8778                     & 0.2625                     & 0.3240                     & 0.8704                     & 0.2673                     & 0.3236                     & 0.8676                     & 0.2644                     \\
\begin{tabular}[c]{@{}c@{}}Quora\\ Retrieval\end{tabular}                 & 0.8470  & 0.9871  & 0.8385 & 0.8270                     & 0.9827                     & 0.8176                     & 0.8247                     & 0.9812                     & 0.8147                     & 0.8205                     & 0.9811                     & 0.8112                     & 0.8192                     & 0.9802                     & 0.8102                     \\
SCIDOCS                                                                   & 0.1578  & 0.3568  & 0.2898 & 0.1546                     & 0.3736                     & 0.2687                     & 0.1475                     & 0.3688                     & 0.2564                     & 0.1485                     & 0.3621                     & 0.2581                     & 0.1444                     & 0.3591                     & 0.2546                     \\
SciFact                                                                   & 0.6268  & 0.9277  & 0.5843 & 0.6757                     & 0.9510                     & 0.6384                     & 0.6770                     & 0.9400                     & 0.6385                     & 0.6718                     & 0.9593                     & 0.6337                     & 0.6512                     & 0.9560                     & 0.6072                     \\
Average                                                                   & 0.4556  & 0.6851  & 0.5350 & \multicolumn{1}{c}{0.4241} & \multicolumn{1}{c}{0.6871} & \multicolumn{1}{c}{0.4932} & \multicolumn{1}{c}{0.4125} & \multicolumn{1}{c}{0.6797} & \multicolumn{1}{c}{0.4794} & \multicolumn{1}{c}{0.4073} & \multicolumn{1}{c}{0.6773} & \multicolumn{1}{c}{0.4743} & \multicolumn{1}{c}{0.4018} & \multicolumn{1}{c}{0.6753} & \multicolumn{1}{c}{0.4679} \\ \bottomrule
\end{tabular}
\caption{Benchmark-wise metric scores on the BEIR benchmark, comparing Contriever baselines leveraging different augmentation scales using Llama 3.1 8B model. N indicates NDCG @ 10, R indicates Recall @ 100 and M indicates MRR @ 10}
\label{tab:8B_Contriever_scale_metrics_BEIR}
\end{table*}

\begin{table*}[]
\centering
\tiny
\setlength{\tabcolsep}{3pt}
\begin{tabular}{@{}c|ccc|ccc|lll|lll|ccc@{}}
\toprule
                                                                          & \multicolumn{3}{c}{No Aug} & \multicolumn{3}{c}{25\%} & \multicolumn{3}{c}{50\%}                                                             & \multicolumn{3}{c}{75\%}                                                             & \multicolumn{3}{c}{100\%} \\ \midrule
task\_name                                                                & N       & R       & M      & N      & R      & M      & \multicolumn{1}{c}{N}      & \multicolumn{1}{c}{R}      & \multicolumn{1}{c}{M}      & \multicolumn{1}{c}{N}      & \multicolumn{1}{c}{R}      & \multicolumn{1}{c}{M}      & N       & R      & M      \\
ArguAna                                                                   & 0.4705  & 0.9751  & 0.3829 & 0.5588 & 0.9879 & 0.4701 & 0.5500                     & 0.9836                     & 0.4614                     & 0.5476                     & 0.9844                     & 0.4615                     & 0.5694  & 0.9865 & 0.4837 \\
\begin{tabular}[c]{@{}c@{}}CQADupstack\\ Android\\ Retrieval\end{tabular} & 0.4181  & 0.7520  & 0.4132 & 0.4256 & 0.7893 & 0.4179 & 0.4176                     & 0.7783                     & 0.4097                     & 0.4252                     & 0.7899                     & 0.4231                     & 0.4247  & 0.8022 & 0.4222 \\
\begin{tabular}[c]{@{}c@{}}Climate\\ FEVER\end{tabular}                   & 0.2370  & 0.5419  & 0.3210 & 0.2001 & 0.5346 & 0.2567 & 0.1893                     & 0.5140                     & 0.2456                     & 0.1834                     & 0.5067                     & 0.2381                     & 0.1768  & 0.5049 & 0.2319 \\
DBPedia                                                                   & 0.3713  & 0.4945  & 0.7022 & 0.3439 & 0.5002 & 0.6297 & 0.3354                     & 0.4922                     & 0.6168                     & 0.3299                     & 0.4878                     & 0.6065                     & 0.3295  & 0.4770 & 0.6144 \\
FEVER                                                                     & 0.7064  & 0.9366  & 0.6869 & 0.4191 & 0.8124 & 0.3760 & 0.4079                     & 0.8022                     & 0.3645                     & 0.3869                     & 0.7797                     & 0.3446                     & 0.3658  & 0.7669 & 0.3233 \\
\begin{tabular}[c]{@{}c@{}}FiQA\\ 2018\end{tabular}                       & 0.2672  & 0.6061  & 0.3283 & 0.2900 & 0.6507 & 0.3561 & 0.2799                     & 0.6438                     & 0.3372                     & 0.2840                     & 0.6510                     & 0.3341                     & 0.2882  & 0.6622 & 0.3458 \\
HotpotQA                                                                  & 0.5409  & 0.6824  & 0.7206 & 0.4840 & 0.7022 & 0.6267 & 0.4585                     & 0.6809                     & 0.5949                     & 0.4371                     & 0.6703                     & 0.5679                     & 0.4272  & 0.6621 & 0.5530 \\
MSMARCO                                                                   & 0.5913  & 0.4800  & 0.8667 & 0.5126 & 0.5059 & 0.7878 & 0.4908                     & 0.4912                     & 0.7785                     & 0.5035                     & 0.4916                     & 0.7876                     & 0.4990  & 0.5018 & 0.7750 \\
NFCorpus                                                                  & 0.3121  & 0.2964  & 0.5020 & 0.3477 & 0.3485 & 0.5369 & 0.3423                     & 0.3431                     & 0.5271                     & 0.3427                     & 0.3452                     & 0.5328                     & 0.3547  & 0.3497 & 0.5436 \\
NQ                                                                        & 0.3762  & 0.8692  & 0.3191 & 0.3282 & 0.8795 & 0.2696 & 0.3188                     & 0.8628                     & 0.2613                     & 0.3173                     & 0.8646                     & 0.2613                     & 0.3202  & 0.8654 & 0.2632 \\
\begin{tabular}[c]{@{}c@{}}Quora\\ Retrieval\end{tabular}                 & 0.8470  & 0.9871  & 0.8385 & 0.8361 & 0.9853 & 0.8283 & 0.8287                     & 0.9831                     & 0.8193                     & 0.8298                     & 0.9835                     & 0.8214                     & 0.8307  & 0.9837 & 0.8212 \\
SCIDOCS                                                                   & 0.1578  & 0.3568  & 0.2898 & 0.1586 & 0.3885 & 0.2769 & 0.1513                     & 0.3833                     & 0.2667                     & 0.1518                     & 0.3812                     & 0.2658                     & 0.1558  & 0.3832 & 0.2713 \\
SciFact                                                                   & 0.6268  & 0.9277  & 0.5843 & 0.6719 & 0.9667 & 0.6328 & 0.6582                     & 0.9467                     & 0.6178                     & 0.6561                     & 0.9527                     & 0.6194                     & 0.6880  & 0.9427 & 0.6551 \\
Average                                                                   & 0.4556  & 0.6851  & 0.5350 & 0.4290 & 0.6963 & 0.4974 & \multicolumn{1}{c}{0.4176} & \multicolumn{1}{c}{0.6850} & \multicolumn{1}{c}{0.4847} & \multicolumn{1}{c}{0.4150} & \multicolumn{1}{c}{0.6837} & \multicolumn{1}{c}{0.4819} & 0.4177  & 0.6837 & 0.4849 \\ \bottomrule
\end{tabular}
\caption{Benchmark-wise metric scores on the BEIR benchmark, comparing Contriever baselines leveraging different augmentation scales using Llama 3.1 70B model. N indicates NDCG @ 10, R indicates Recall @ 100 and M indicates MRR @ 10}
\label{tab:70B_Contriever_scale_metrics_BEIR}
\end{table*}

\begin{table*}[]
\centering
\tiny
\setlength{\tabcolsep}{3pt}
\begin{tabular}{@{}c|lll|lll|lll|ccc|lll@{}}
\toprule
                                                                          & \multicolumn{3}{c}{No Aug}                                                           & \multicolumn{3}{c}{1PQ}                                                              & \multicolumn{3}{c}{3PQ}                                                              & \multicolumn{3}{c}{5PQ}  & \multicolumn{3}{c}{7PQ}                                                              \\ \midrule
task\_name                                                                & \multicolumn{1}{c}{N}      & \multicolumn{1}{c}{R}      & \multicolumn{1}{c}{M}      & \multicolumn{1}{c}{N}      & \multicolumn{1}{c}{R}      & \multicolumn{1}{c}{M}      & \multicolumn{1}{c}{N}      & \multicolumn{1}{c}{R}      & \multicolumn{1}{c}{M}      & N      & R      & M      & \multicolumn{1}{c}{N}      & \multicolumn{1}{c}{R}      & \multicolumn{1}{c}{M}      \\
ArguAna                                                                   & 0.4964                     & 0.9808                     & 0.4114                     & 0.5552                     & 0.9893                     & 0.4716                     & 0.5472                     & 0.9858                     & 0.4592                     & 0.5505 & 0.9865 & 0.4679 & 0.5588                     & 0.9893                     & 0.4712                     \\
\begin{tabular}[c]{@{}c@{}}CQADupstack\\ Android\\ Retrieval\end{tabular} & 0.5036                     & 0.8355                     & 0.5024                     & 0.4617                     & 0.8165                     & 0.4610                     & 0.4511                     & 0.8200                     & 0.4446                     & 0.4486 & 0.8259 & 0.4414 & 0.4438                     & 0.8235                     & 0.4376                     \\
\begin{tabular}[c]{@{}c@{}}Climate\\ FEVER\end{tabular}                   & 0.2541                     & 0.5668                     & 0.3428                     & 0.1976                     & 0.5354                     & 0.2519                     & 0.1805                     & 0.5167                     & 0.2313                     & 0.1925 & 0.5191 & 0.2556 & 0.1696                     & 0.5080                     & 0.2216                     \\
DBPedia                                                                   & 0.3721                     & 0.4842                     & 0.7026                     & 0.3464                     & 0.4928                     & 0.6310                     & 0.3400                     & 0.4815                     & 0.6228                     & 0.3321 & 0.4814 & 0.6167 & 0.3297                     & 0.4769                     & 0.6132                     \\
FEVER                                                                     & 0.7255                     & 0.9372                     & 0.7083                     & 0.5042                     & 0.8687                     & 0.4594                     & 0.4319                     & 0.8253                     & 0.3858                     & 0.4102 & 0.8054 & 0.3638 & 0.3959                     & 0.7900                     & 0.3516                     \\
\begin{tabular}[c]{@{}c@{}}FiQA\\ 2018\end{tabular}                       & 0.3285                     & 0.6784                     & 0.3983                     & 0.2966                     & 0.6618                     & 0.3593                     & 0.2954                     & 0.6680                     & 0.3500                     & 0.2984 & 0.6792 & 0.3664 & 0.2979                     & 0.6748                     & 0.3573                     \\
HotpotQA                                                                  & 0.5460                     & 0.6769                     & 0.7266                     & 0.4950                     & 0.7063                     & 0.6425                     & 0.4561                     & 0.6819                     & 0.5919                     & 0.4411 & 0.6735 & 0.5689 & 0.4389                     & 0.6663                     & 0.5664                     \\
MSMARCO                                                                   & 0.6424                     & 0.5095                     & 0.9399                     & 0.5495                     & 0.5023                     & 0.8209                     & 0.5230                     & 0.5079                     & 0.7547                     & 0.5471 & 0.5094 & 0.8121 & 0.5090                     & 0.4967                     & 0.7516                     \\
NFCorpus                                                                  & 0.3565                     & 0.3392                     & 0.5616                     & 0.3644                     & 0.3504                     & 0.5662                     & 0.3673                     & 0.3501                     & 0.5664                     & 0.3753 & 0.3532 & 0.5638 & 0.3768                     & 0.3548                     & 0.5682                     \\
NQ                                                                        & 0.4299                     & 0.8838                     & 0.3781                     & 0.3536                     & 0.8858                     & 0.2956                     & 0.3390                     & 0.8770                     & 0.2812                     & 0.3346 & 0.8734 & 0.2777 & 0.3361                     & 0.8744                     & 0.2790                     \\
\begin{tabular}[c]{@{}c@{}}Quora\\ Retrieval\end{tabular}                 & 0.8699                     & 0.9927                     & 0.8620                     & 0.8484                     & 0.9872                     & 0.8400                     & 0.8449                     & 0.9863                     & 0.8362                     & 0.8405 & 0.9858 & 0.8311 & 0.8394                     & 0.9851                     & 0.8305                     \\
SCIDOCS                                                                   & 0.1521                     & 0.4005                     & 0.2610                     & 0.1906                     & 0.4523                     & 0.3244                     & 0.1821                     & 0.4377                     & 0.3116                     & 0.1762 & 0.4262 & 0.3019 & 0.1733                     & 0.4196                     & 0.2976                     \\
SciFact                                                                   & 0.6865                     & 0.9700                     & 0.6430                     & 0.7066                     & 0.9633                     & 0.6664                     & 0.7080                     & 0.9560                     & 0.6671                     & 0.7007 & 0.9500 & 0.6598 & 0.6988                     & 0.9433                     & 0.6618                     \\
Average                                                                   & \multicolumn{1}{c}{0.4895} & \multicolumn{1}{c}{0.7119} & \multicolumn{1}{c}{0.5722} & \multicolumn{1}{c}{0.4515} & \multicolumn{1}{c}{0.7086} & \multicolumn{1}{c}{0.5223} & \multicolumn{1}{c}{0.4359} & \multicolumn{1}{c}{0.6996} & \multicolumn{1}{c}{0.5002} & 0.4344 & 0.6976 & 0.5021 & \multicolumn{1}{c}{0.4283} & \multicolumn{1}{c}{0.6925} & \multicolumn{1}{c}{0.4929} \\ \bottomrule
\end{tabular}
\caption{Benchmark-wise metric scores on the BEIR benchmark, comparing E5 baselines leveraging different augmentation densities using Llama 3.1 70B model. N indicates NDCG @ 10, R indicates Recall @ 100 and M indicates MRR @ 10}
\label{tab:70B_E5_density_metrics_BEIR}
\end{table*}

\begin{table*}[]
\centering
\tiny
\setlength{\tabcolsep}{3pt}
\begin{tabular}{@{}c|lll|lll|lll|lll|lll@{}}
\toprule
                                                                          & \multicolumn{3}{c}{No Aug}                                                           & \multicolumn{3}{c}{1PQ}                                                              & \multicolumn{3}{c}{3PQ}                                                              & \multicolumn{3}{c}{5PQ}                                                              & \multicolumn{3}{c}{7PQ}                                                              \\ \midrule
task\_name                                                                & \multicolumn{1}{c}{N}      & \multicolumn{1}{c}{R}      & \multicolumn{1}{c}{M}      & \multicolumn{1}{c}{N}      & \multicolumn{1}{c}{R}      & \multicolumn{1}{c}{M}      & \multicolumn{1}{c}{N}      & \multicolumn{1}{c}{R}      & \multicolumn{1}{c}{M}      & \multicolumn{1}{c}{N}      & \multicolumn{1}{c}{R}      & \multicolumn{1}{c}{M}      & \multicolumn{1}{c}{N}      & \multicolumn{1}{c}{R}      & \multicolumn{1}{c}{M}      \\
ArguAna                                                                   & 0.4964                     & 0.9808                     & 0.4114                     & 0.5379                     & 0.9865                     & 0.4463                     & 0.5337                     & 0.9865                     & 0.4430                     & 0.5402                     & 0.9865                     & 0.4518                     & 0.5393                     & 0.9851                     & 0.4510                     \\
\begin{tabular}[c]{@{}c@{}}CQADupstack\\ Android\\ Retrieval\end{tabular} & 0.5036                     & 0.8355                     & 0.5024                     & 0.4478                     & 0.8121                     & 0.4445                     & 0.4235                     & 0.8041                     & 0.4180                     & 0.4305                     & 0.8056                     & 0.4280                     & 0.4151                     & 0.7959                     & 0.4047                     \\
\begin{tabular}[c]{@{}c@{}}Climate\\ FEVER\end{tabular}                   & 0.2541                     & 0.5668                     & 0.3428                     & 0.2122                     & 0.5624                     & 0.2787                     & 0.1952                     & 0.5417                     & 0.2511                     & 0.1885                     & 0.5340                     & 0.2452                     & 0.1910                     & 0.5323                     & 0.2474                     \\
DBPedia                                                                   & 0.3721                     & 0.4842                     & 0.7026                     & 0.3468                     & 0.5098                     & 0.6333                     & 0.3301                     & 0.4924                     & 0.6097                     & 0.3332                     & 0.4893                     & 0.6108                     & 0.3297                     & 0.4902                     & 0.6136                     \\
FEVER                                                                     & 0.7255                     & 0.9372                     & 0.7083                     & 0.5067                     & 0.8906                     & 0.4575                     & 0.4684                     & 0.8574                     & 0.4181                     & 0.4342                     & 0.8391                     & 0.3854                     & 0.4230                     & 0.8329                     & 0.3752                     \\
\begin{tabular}[c]{@{}c@{}}FiQA\\ 2018\end{tabular}                       & 0.3285                     & 0.6784                     & 0.3983                     & 0.2892                     & 0.6287                     & 0.3437                     & 0.2804                     & 0.6268                     & 0.3403                     & 0.2990                     & 0.6362                     & 0.3623                     & 0.2773                     & 0.6394                     & 0.3327                     \\
HotpotQA                                                                  & 0.5460                     & 0.6769                     & 0.7266                     & 0.4909                     & 0.7063                     & 0.6374                     & 0.4724                     & 0.6991                     & 0.6119                     & 0.4618                     & 0.6883                     & 0.5951                     & 0.4461                     & 0.6764                     & 0.5740                     \\
MSMARCO                                                                   & 0.6424                     & 0.5095                     & 0.9399                     & 0.5815                     & 0.5193                     & 0.8401                     & 0.5488                     & 0.5367                     & 0.8301                     & 0.5467                     & 0.5282                     & 0.7952                     & 0.5395                     & 0.5384                     & 0.8328                     \\
NFCorpus                                                                  & 0.3565                     & 0.3392                     & 0.5616                     & 0.3745                     & 0.3595                     & 0.5749                     & 0.3616                     & 0.3367                     & 0.5483                     & 0.3681                     & 0.3394                     & 0.5616                     & 0.3640                     & 0.3445                     & 0.5559                     \\
NQ                                                                        & 0.4299                     & 0.8838                     & 0.3781                     & 0.3565                     & 0.8941                     & 0.2962                     & 0.3430                     & 0.8883                     & 0.2834                     & 0.3370                     & 0.8838                     & 0.2793                     & 0.3412                     & 0.8826                     & 0.2847                     \\
\begin{tabular}[c]{@{}c@{}}Quora\\ Retrieval\end{tabular}                 & 0.8699                     & 0.9927                     & 0.8620                     & 0.8485                     & 0.9871                     & 0.8394                     & 0.8421                     & 0.9859                     & 0.8339                     & 0.8389                     & 0.9850                     & 0.8289                     & 0.8350                     & 0.9841                     & 0.8255                     \\
SCIDOCS                                                                   & 0.1521                     & 0.4005                     & 0.2610                     & 0.1802                     & 0.4308                     & 0.3061                     & 0.1740                     & 0.4220                     & 0.3003                     & 0.1696                     & 0.4228                     & 0.2911                     & 0.1700                     & 0.4106                     & 0.2932                     \\
SciFact                                                                   & 0.6865                     & 0.9700                     & 0.6430                     & 0.7195                     & 0.9633                     & 0.6794                     & 0.7059                     & 0.9633                     & 0.6682                     & 0.7115                     & 0.9567                     & 0.6725                     & 0.6995                     & 0.9627                     & 0.6575                     \\
Average                                                                   & \multicolumn{1}{c}{0.4895} & \multicolumn{1}{c}{0.7119} & \multicolumn{1}{c}{0.5722} & \multicolumn{1}{c}{0.4532} & \multicolumn{1}{c}{0.7116} & \multicolumn{1}{c}{0.5213} & \multicolumn{1}{c}{0.4368} & \multicolumn{1}{c}{0.7031} & \multicolumn{1}{c}{0.5043} & \multicolumn{1}{c}{0.4353} & \multicolumn{1}{c}{0.6996} & \multicolumn{1}{c}{0.5006} & \multicolumn{1}{c}{0.4285} & \multicolumn{1}{c}{0.6981} & \multicolumn{1}{c}{0.4960} \\ \bottomrule
\end{tabular}
\caption{Benchmark-wise metric scores on the BEIR benchmark, comparing E5 baselines leveraging different augmentation densities using Llama 3.1 8B model. N indicates NDCG @ 10, R indicates Recall @ 100 and M indicates MRR @ 10}
\label{tab:8B_E5_density_metrics_BEIR}
\end{table*}

\begin{table*}[]
\centering
\tiny
\setlength{\tabcolsep}{3pt}
\begin{tabular}{@{}c|lll|lll|lll|lll|ccc@{}}
\toprule
                                                                          & \multicolumn{3}{c}{No Aug}                                                           & \multicolumn{3}{c}{25\%}                                                             & \multicolumn{3}{c}{50\%}                                                             & \multicolumn{3}{c}{75\%}                                                             & \multicolumn{3}{c}{100\%} \\ \midrule
task\_name                                                                & \multicolumn{1}{c}{N}      & \multicolumn{1}{c}{R}      & \multicolumn{1}{c}{M}      & \multicolumn{1}{c}{N}      & \multicolumn{1}{c}{R}      & \multicolumn{1}{c}{M}      & \multicolumn{1}{c}{N}      & \multicolumn{1}{c}{R}      & \multicolumn{1}{c}{M}      & \multicolumn{1}{c}{N}      & \multicolumn{1}{c}{R}      & \multicolumn{1}{c}{M}      & N       & R      & M      \\
ArguAna                                                                   & 0.4964                     & 0.9808                     & 0.4114                     & 0.5514                     & 0.9858                     & 0.4683                     & 0.5485                     & 0.9886                     & 0.4627                     & 0.5512                     & 0.9872                     & 0.4649                     & 0.5505  & 0.9865 & 0.4679 \\
\begin{tabular}[c]{@{}c@{}}CQADupstack\\ Android\\ Retrieval\end{tabular} & 0.5036                     & 0.8355                     & 0.5024                     & 0.4584                     & 0.8233                     & 0.4559                     & 0.4513                     & 0.8235                     & 0.4412                     & 0.4529                     & 0.8290                     & 0.4491                     & 0.4486  & 0.8259 & 0.4414 \\
\begin{tabular}[c]{@{}c@{}}Climate\\ FEVER\end{tabular}                   & 0.2541                     & 0.5668                     & 0.3428                     & 0.1971                     & 0.5356                     & 0.2507                     & 0.1891                     & 0.5303                     & 0.2478                     & 0.1804                     & 0.5148                     & 0.2309                     & 0.1925  & 0.5191 & 0.2556 \\
DBPedia                                                                   & 0.3721                     & 0.4842                     & 0.7026                     & 0.3481                     & 0.4922                     & 0.6238                     & 0.3376                     & 0.4843                     & 0.6191                     & 0.3312                     & 0.4817                     & 0.6099                     & 0.3321  & 0.4814 & 0.6167 \\
FEVER                                                                     & 0.7255                     & 0.9372                     & 0.7083                     & 0.4981                     & 0.8613                     & 0.4538                     & 0.4494                     & 0.8307                     & 0.4052                     & 0.4241                     & 0.8205                     & 0.3792                     & 0.4102  & 0.8054 & 0.3638 \\
\begin{tabular}[c]{@{}c@{}}FiQA\\ 2018\end{tabular}                       & 0.3285                     & 0.6784                     & 0.3983                     & 0.3002                     & 0.6676                     & 0.3607                     & 0.3059                     & 0.6771                     & 0.3724                     & 0.3003                     & 0.6745                     & 0.3625                     & 0.2984  & 0.6792 & 0.3664 \\
HotpotQA                                                                  & 0.5460                     & 0.6769                     & 0.7266                     & 0.4929                     & 0.7041                     & 0.6391                     & 0.4680                     & 0.6877                     & 0.6051                     & 0.4519                     & 0.6813                     & 0.5835                     & 0.4411  & 0.6735 & 0.5689 \\
MSMARCO                                                                   & 0.6424                     & 0.5095                     & 0.9399                     & 0.5644                     & 0.5075                     & 0.8367                     & 0.5323                     & 0.4976                     & 0.7864                     & 0.5224                     & 0.5035                     & 0.7752                     & 0.5471  & 0.5094 & 0.8121 \\
NFCorpus                                                                  & 0.3565                     & 0.3392                     & 0.5616                     & 0.3701                     & 0.3498                     & 0.5674                     & 0.3689                     & 0.3555                     & 0.5615                     & 0.3644                     & 0.3469                     & 0.5571                     & 0.3753  & 0.3532 & 0.5638 \\
NQ                                                                        & 0.4299                     & 0.8838                     & 0.3781                     & 0.3496                     & 0.8792                     & 0.2927                     & 0.3394                     & 0.8750                     & 0.2823                     & 0.3345                     & 0.8767                     & 0.2786                     & 0.3346  & 0.8734 & 0.2777 \\
\begin{tabular}[c]{@{}c@{}}Quora\\ Retrieval\end{tabular}                 & 0.8699                     & 0.9927                     & 0.8620                     & 0.8467                     & 0.9862                     & 0.8385                     & 0.8437                     & 0.9862                     & 0.8346                     & 0.8410                     & 0.9855                     & 0.8315                     & 0.8405  & 0.9858 & 0.8311 \\
SCIDOCS                                                                   & 0.1521                     & 0.4005                     & 0.2610                     & 0.1870                     & 0.4506                     & 0.3192                     & 0.1800                     & 0.4407                     & 0.3112                     & 0.1810                     & 0.4351                     & 0.3124                     & 0.1762  & 0.4262 & 0.3019 \\
SciFact                                                                   & 0.6865                     & 0.9700                     & 0.6430                     & 0.7066                     & 0.9600                     & 0.6644                     & 0.6999                     & 0.9533                     & 0.6599                     & 0.7013                     & 0.9467                     & 0.6626                     & 0.7007  & 0.9500 & 0.6598 \\
Average                                                                   & \multicolumn{1}{c}{0.4895} & \multicolumn{1}{c}{0.7119} & \multicolumn{1}{c}{0.5722} & \multicolumn{1}{c}{0.4516} & \multicolumn{1}{c}{0.7079} & \multicolumn{1}{c}{0.5209} & \multicolumn{1}{c}{0.4395} & \multicolumn{1}{c}{0.7024} & \multicolumn{1}{c}{0.5069} & \multicolumn{1}{c}{0.4336} & \multicolumn{1}{c}{0.6987} & \multicolumn{1}{c}{0.4998} & 0.4344  & 0.6976 & 0.5021 \\ \bottomrule
\end{tabular}
\caption{Benchmark-wise metric scores on the BEIR benchmark, comparing E5 baselines leveraging different augmentation scales using Llama 3.1 70B model. N indicates NDCG @ 10, R indicates Recall @ 100 and M indicates MRR @ 10}
\label{tab:70B_E5_scale_metrics_BEIR}
\end{table*}

\begin{table*}[]
\centering
\tiny
\setlength{\tabcolsep}{3pt}
\begin{tabular}{@{}c|lll|lll|lll|lll|lll@{}}
\toprule
                                                                          & \multicolumn{3}{c}{No Aug}                                                           & \multicolumn{3}{c}{25\%}                                                             & \multicolumn{3}{c}{50\%}                                                             & \multicolumn{3}{c}{75\%}                                                             & \multicolumn{3}{c}{100\%}                                                            \\ \midrule
task\_name                                                                & \multicolumn{1}{c}{N}      & \multicolumn{1}{c}{R}      & \multicolumn{1}{c}{M}      & \multicolumn{1}{c}{N}      & \multicolumn{1}{c}{R}      & \multicolumn{1}{c}{M}      & \multicolumn{1}{c}{N}      & \multicolumn{1}{c}{R}      & \multicolumn{1}{c}{M}      & \multicolumn{1}{c}{N}      & \multicolumn{1}{c}{R}      & \multicolumn{1}{c}{M}      & \multicolumn{1}{c}{N}      & \multicolumn{1}{c}{R}      & \multicolumn{1}{c}{M}      \\
ArguAna                                                                   & 0.4964                     & 0.9808                     & 0.4114                     & 0.5417                     & 0.9865                     & 0.4526                     & 0.5456                     & 0.9879                     & 0.4578                     & 0.5469                     & 0.9872                     & 0.4572                     & 0.5402                     & 0.9865                     & 0.4518                     \\
\begin{tabular}[c]{@{}c@{}}CQADupstack\\ Android\\ Retrieval\end{tabular} & 0.5036                     & 0.8355                     & 0.5024                     & 0.4490                     & 0.8084                     & 0.4442                     & 0.4287                     & 0.7963                     & 0.4213                     & 0.4188                     & 0.7928                     & 0.4132                     & 0.4305                     & 0.8056                     & 0.4280                     \\
\begin{tabular}[c]{@{}c@{}}Climate\\ FEVER\end{tabular}                   & 0.2541                     & 0.5668                     & 0.3428                     & 0.2092                     & 0.5546                     & 0.2703                     & 0.1850                     & 0.5402                     & 0.2410                     & 0.1857                     & 0.5364                     & 0.2397                     & 0.1885                     & 0.5340                     & 0.2452                     \\
DBPedia                                                                   & 0.3721                     & 0.4842                     & 0.7026                     & 0.3523                     & 0.5043                     & 0.6379                     & 0.3460                     & 0.5027                     & 0.6380                     & 0.3393                     & 0.5088                     & 0.6219                     & 0.3332                     & 0.4893                     & 0.6108                     \\
FEVER                                                                     & 0.7255                     & 0.9372                     & 0.7083                     & 0.4981                     & 0.8770                     & 0.4486                     & 0.4644                     & 0.8614                     & 0.4141                     & 0.4481                     & 0.8516                     & 0.3970                     & 0.4342                     & 0.8391                     & 0.3854                     \\
\begin{tabular}[c]{@{}c@{}}FiQA\\ 2018\end{tabular}                       & 0.3285                     & 0.6784                     & 0.3983                     & 0.2924                     & 0.6372                     & 0.3511                     & 0.2867                     & 0.6326                     & 0.3444                     & 0.2821                     & 0.6296                     & 0.3354                     & 0.2990                     & 0.6362                     & 0.3623                     \\
HotpotQA                                                                  & 0.5460                     & 0.6769                     & 0.7266                     & 0.5062                     & 0.7159                     & 0.6557                     & 0.4850                     & 0.7024                     & 0.6292                     & 0.4713                     & 0.6982                     & 0.6084                     & 0.4618                     & 0.6883                     & 0.5951                     \\
MSMARCO                                                                   & 0.6424                     & 0.5095                     & 0.9399                     & 0.5689                     & 0.5344                     & 0.8207                     & 0.5558                     & 0.5236                     & 0.8043                     & 0.5341                     & 0.5254                     & 0.8357                     & 0.5467                     & 0.5282                     & 0.7952                     \\
NFCorpus                                                                  & 0.3565                     & 0.3392                     & 0.5616                     & 0.3715                     & 0.3516                     & 0.5634                     & 0.3678                     & 0.3519                     & 0.5509                     & 0.3671                     & 0.3428                     & 0.5551                     & 0.3681                     & 0.3394                     & 0.5616                     \\
NQ                                                                        & 0.4299                     & 0.8838                     & 0.3781                     & 0.3507                     & 0.8944                     & 0.2903                     & 0.3417                     & 0.8878                     & 0.2826                     & 0.3395                     & 0.8846                     & 0.2789                     & 0.3370                     & 0.8838                     & 0.2793                     \\
\begin{tabular}[c]{@{}c@{}}Quora\\ Retrieval\end{tabular}                 & 0.8699                     & 0.9927                     & 0.8620                     & 0.8479                     & 0.9877                     & 0.8398                     & 0.8427                     & 0.9853                     & 0.8336                     & 0.8402                     & 0.9855                     & 0.8309                     & 0.8389                     & 0.9850                     & 0.8289                     \\
SCIDOCS                                                                   & 0.1521                     & 0.4005                     & 0.2610                     & 0.1775                     & 0.4397                     & 0.3059                     & 0.1751                     & 0.4242                     & 0.3030                     & 0.1732                     & 0.4228                     & 0.3022                     & 0.1696                     & 0.4228                     & 0.2911                     \\
SciFact                                                                   & 0.6865                     & 0.9700                     & 0.6430                     & 0.7176                     & 0.9627                     & 0.6783                     & 0.7093                     & 0.9600                     & 0.6724                     & 0.6983                     & 0.9627                     & 0.6594                     & 0.7115                     & 0.9567                     & 0.6725                     \\
Average                                                                   & \multicolumn{1}{c}{0.4895} & \multicolumn{1}{c}{0.7119} & \multicolumn{1}{c}{0.5722} & \multicolumn{1}{c}{0.4525} & \multicolumn{1}{c}{0.7119} & \multicolumn{1}{c}{0.5199} & \multicolumn{1}{c}{0.4410} & \multicolumn{1}{c}{0.7043} & \multicolumn{1}{c}{0.5071} & \multicolumn{1}{c}{0.4342} & \multicolumn{1}{c}{0.7022} & \multicolumn{1}{c}{0.5027} & \multicolumn{1}{c}{0.4353} & \multicolumn{1}{c}{0.6996} & \multicolumn{1}{c}{0.5006} \\ \bottomrule
\end{tabular}
\caption{Benchmark-wise metric scores on the BEIR benchmark, comparing E5 baselines leveraging different augmentation scales using Llama 3.1 8B model. N indicates NDCG @ 10, R indicates Recall @ 100 and M indicates MRR @ 10}
\label{tab:8B_E5_scale_metrics_BEIR}
\end{table*}

\begin{table*}[]
\centering
\tiny
\setlength{\tabcolsep}{3pt}
\begin{tabular}{@{}c|ccc|ccc|ccc|ccc|ccc@{}}
\toprule
                                                                          & \multicolumn{3}{c}{No Aug} & \multicolumn{3}{c}{25\%} & \multicolumn{3}{c}{50\%} & \multicolumn{3}{c}{75\%} & \multicolumn{3}{c}{100\%} \\ \midrule
task\_name                                                                & N       & R       & M      & N      & R      & M      & N      & R      & M      & N      & R      & M      & N       & R      & M      \\
ArguAna                                                                   & 0.4173  & 0.9566  & 0.3345 & 0.5185 & 0.9808 & 0.4316 & 0.5223 & 0.9822 & 0.4386 & 0.5191 & 0.9765 & 0.4342 & 0.5020  & 0.9673 & 0.4155 \\
\begin{tabular}[c]{@{}c@{}}CQADupstack\\ Android\\ Retrieval\end{tabular} & 0.3569  & 0.6680  & 0.3642 & 0.3312 & 0.7105 & 0.3247 & 0.3307 & 0.7289 & 0.3164 & 0.3268 & 0.7161 & 0.3222 & 0.3343  & 0.7206 & 0.3272 \\
\begin{tabular}[c]{@{}c@{}}Climate\\ FEVER\end{tabular}                   & 0.2160  & 0.4921  & 0.2966 & 0.1692 & 0.5147 & 0.2202 & 0.1532 & 0.5030 & 0.1971 & 0.1527 & 0.4902 & 0.1992 & 0.1397  & 0.4739 & 0.1813 \\
DBPedia                                                                   & 0.3119  & 0.4128  & 0.6412 & 0.2981 & 0.4466 & 0.5745 & 0.2922 & 0.4432 & 0.5602 & 0.2922 & 0.4448 & 0.5676 & 0.2914  & 0.4324 & 0.5779 \\
FEVER                                                                     & 0.6261  & 0.9088  & 0.5973 & 0.3854 & 0.7985 & 0.3391 & 0.3662 & 0.7870 & 0.3220 & 0.3308 & 0.7541 & 0.2869 & 0.3160  & 0.7418 & 0.2728 \\
\begin{tabular}[c]{@{}c@{}}FiQA\\ 2018\end{tabular}                       & 0.2169  & 0.5248  & 0.2651 & 0.2166 & 0.5652 & 0.2627 & 0.2133 & 0.5633 & 0.2636 & 0.2067 & 0.5588 & 0.2523 & 0.2159  & 0.5667 & 0.2574 \\
HotpotQA                                                                  & 0.4677  & 0.6063  & 0.6477 & 0.4055 & 0.6213 & 0.5361 & 0.4031 & 0.6153 & 0.5326 & 0.3900 & 0.6030 & 0.5171 & 0.3795  & 0.5976 & 0.5034 \\
MSMARCO                                                                   & 0.5996  & 0.4473  & 0.8886 & 0.4824 & 0.4598 & 0.7474 & 0.4665 & 0.4858 & 0.7890 & 0.4570 & 0.4790 & 0.7118 & 0.4636  & 0.4890 & 0.7297 \\
NFCorpus                                                                  & 0.2720  & 0.2490  & 0.4704 & 0.3060 & 0.2963 & 0.4909 & 0.3198 & 0.3120 & 0.5090 & 0.3173 & 0.3111 & 0.4990 & 0.3153  & 0.3122 & 0.4951 \\
NQ                                                                        & 0.3380  & 0.8066  & 0.2913 & 0.2829 & 0.8169 & 0.2337 & 0.2904 & 0.8199 & 0.2396 & 0.2950 & 0.8266 & 0.2407 & 0.2837  & 0.8150 & 0.2335 \\
\begin{tabular}[c]{@{}c@{}}Quora\\ Retrieval\end{tabular}                 & 0.8299  & 0.9803  & 0.8240 & 0.8233 & 0.9804 & 0.8167 & 0.8194 & 0.9783 & 0.8127 & 0.8178 & 0.9787 & 0.8094 & 0.8205  & 0.9789 & 0.8121 \\
SCIDOCS                                                                   & 0.1297  & 0.2990  & 0.2412 & 0.1395 & 0.3455 & 0.2530 & 0.1388 & 0.3444 & 0.2519 & 0.1399 & 0.3413 & 0.2561 & 0.1391  & 0.3423 & 0.2481 \\
SciFact                                                                   & 0.5322  & 0.8610  & 0.4932 & 0.6138 & 0.9193 & 0.5770 & 0.6250 & 0.9127 & 0.5896 & 0.6203 & 0.9193 & 0.5811 & 0.6253  & 0.9283 & 0.5927 \\
Average                                                                   & 0.4088  & 0.6317  & 0.4889 & 0.3825 & 0.6504 & 0.4467 & 0.3801 & 0.6520 & 0.4479 & 0.3743 & 0.6461 & 0.4367 & 0.3713  & 0.6435 & 0.4344 \\ \bottomrule
\end{tabular}
\caption{Benchmark-wise metric scores on the BEIR benchmark, comparing DistilBERT baselines leveraging different augmentation scales using Llama 3.1 8B model. N indicates NDCG @ 10, R indicates Recall @ 100 and M indicates MRR @ 10}
\label{tab:8B_DB_scale_metrics_BEIR}
\end{table*}

\begin{table*}[]
\centering
\tiny
\setlength{\tabcolsep}{3pt}
\begin{tabular}{@{}c|ccc|ccc|ccc|ccc|ccc@{}}
\toprule
                                                                          & \multicolumn{3}{c}{No Aug} & \multicolumn{3}{c}{25\%} & \multicolumn{3}{c}{50\%} & \multicolumn{3}{c}{75\%} & \multicolumn{3}{c}{100\%} \\ \midrule
task\_name                                                                & N       & R       & M      & N      & R      & M      & N      & R      & M      & N      & R      & M      & N       & R      & M      \\
ArguAna                                                                   & 0.4173  & 0.9566  & 0.3345 & 0.5314 & 0.9829 & 0.4450 & 0.5327 & 0.9836 & 0.4466 & 0.5298 & 0.9829 & 0.4431 & 0.5437  & 0.9865 & 0.4588 \\
\begin{tabular}[c]{@{}c@{}}CQADupstack\\ Android\\ Retrieval\end{tabular} & 0.3569  & 0.6680  & 0.3642 & 0.3633 & 0.7261 & 0.3602 & 0.3681 & 0.7393 & 0.3610 & 0.3754 & 0.7564 & 0.3687 & 0.3875  & 0.7667 & 0.3801 \\
\begin{tabular}[c]{@{}c@{}}Climate\\ FEVER\end{tabular}                   & 0.2160  & 0.4921  & 0.2966 & 0.1916 & 0.5215 & 0.2479 & 0.1882 & 0.5086 & 0.2446 & 0.1721 & 0.5007 & 0.2240 & 0.1675  & 0.5002 & 0.2179 \\
DBPedia                                                                   & 0.3119  & 0.4128  & 0.6412 & 0.3030 & 0.4403 & 0.5895 & 0.2979 & 0.4348 & 0.5673 & 0.2980 & 0.4395 & 0.5740 & 0.3055  & 0.4425 & 0.5767 \\
FEVER                                                                     & 0.6261  & 0.9088  & 0.5973 & 0.3893 & 0.7762 & 0.3478 & 0.3397 & 0.7370 & 0.2984 & 0.3183 & 0.7151 & 0.2807 & 0.3016  & 0.6949 & 0.2623 \\
\begin{tabular}[c]{@{}c@{}}FiQA\\ 2018\end{tabular}                       & 0.2169  & 0.5248  & 0.2651 & 0.2498 & 0.5993 & 0.3023 & 0.2598 & 0.6123 & 0.3115 & 0.2545 & 0.6212 & 0.3066 & 0.2698  & 0.6240 & 0.3227 \\
HotpotQA                                                                  & 0.4677  & 0.6063  & 0.6477 & 0.4094 & 0.6280 & 0.5404 & 0.3942 & 0.6167 & 0.5208 & 0.3901 & 0.6168 & 0.5128 & 0.3855  & 0.6127 & 0.5052 \\
MSMARCO                                                                   & 0.5996  & 0.4473  & 0.8886 & 0.4980 & 0.4590 & 0.7497 & 0.4835 & 0.4644 & 0.7336 & 0.4890 & 0.4765 & 0.7448 & 0.4806  & 0.4892 & 0.7291 \\
NFCorpus                                                                  & 0.2720  & 0.2490  & 0.4704 & 0.3179 & 0.3021 & 0.5100 & 0.3205 & 0.3093 & 0.5035 & 0.3226 & 0.3184 & 0.5021 & 0.3312  & 0.3262 & 0.5154 \\
NQ                                                                        & 0.3380  & 0.8066  & 0.2913 & 0.2893 & 0.8193 & 0.2394 & 0.2885 & 0.8172 & 0.2369 & 0.2881 & 0.8226 & 0.2361 & 0.2958  & 0.8282 & 0.2433 \\
\begin{tabular}[c]{@{}c@{}}Quora\\ Retrieval\end{tabular}                 & 0.8299  & 0.9803  & 0.8240 & 0.8250 & 0.9803 & 0.8183 & 0.8263 & 0.9799 & 0.8193 & 0.8246 & 0.9808 & 0.8173 & 0.8233  & 0.9812 & 0.8149 \\
SCIDOCS                                                                   & 0.1297  & 0.2990  & 0.2412 & 0.1399 & 0.3475 & 0.2554 & 0.1417 & 0.3481 & 0.2529 & 0.1422 & 0.3539 & 0.2545 & 0.1438  & 0.3548 & 0.2579 \\
SciFact                                                                   & 0.5322  & 0.8610  & 0.4932 & 0.6013 & 0.9150 & 0.5670 & 0.6200 & 0.9217 & 0.5845 & 0.6167 & 0.9317 & 0.5820 & 0.6326  & 0.9210 & 0.5978 \\
Average                                                                   & 0.4088  & 0.6317  & 0.4889 & 0.3930 & 0.6537 & 0.4594 & 0.3893 & 0.6518 & 0.4524 & 0.3863 & 0.6551 & 0.4498 & 0.3899  & 0.6560 & 0.4525 \\ \bottomrule
\end{tabular}
\caption{Benchmark-wise metric scores on the BEIR benchmark, comparing DistilBERT baselines leveraging different augmentation scales using Llama 3.1 70B model. N indicates NDCG @ 10, R indicates Recall @ 100 and M indicates MRR @ 10}
\label{tab:70B_DB_scale_metrics_BEIR}
\end{table*}

\begin{table*}[]
\centering
\tiny
\setlength{\tabcolsep}{3pt}
\begin{tabular}{@{}c|lll|lll|lll|lll|lll@{}}
\toprule
                                                                          & \multicolumn{3}{c}{No Aug}                                                           & \multicolumn{3}{c}{1PQ}                                                             & \multicolumn{3}{c}{3PQ}                                                             & \multicolumn{3}{c}{5PQ}                                                             & \multicolumn{3}{c}{7PQ}                                                            \\ \midrule
task\_name                                                                & \multicolumn{1}{c}{N}      & \multicolumn{1}{c}{R}      & \multicolumn{1}{c}{M}      & \multicolumn{1}{c}{N}      & \multicolumn{1}{c}{R}      & \multicolumn{1}{c}{M}      & \multicolumn{1}{c}{N}      & \multicolumn{1}{c}{R}      & \multicolumn{1}{c}{M}      & \multicolumn{1}{c}{N}      & \multicolumn{1}{c}{R}      & \multicolumn{1}{c}{M}      & \multicolumn{1}{c}{N}      & \multicolumn{1}{c}{R}      & \multicolumn{1}{c}{M}      \\
ArguAna                                                                   & 0.4173                     & 0.9566                     & 0.3345                     & 0.5201                     & 0.9815                     & 0.4324                     & 0.5070                     & 0.9644                     & 0.4200                     & 0.5020                     & 0.9673                     & 0.4155                     & 0.5021                     & 0.9666                     & 0.4153                     \\
\begin{tabular}[c]{@{}c@{}}CQADupstack\\ Android\\ Retrieval\end{tabular} & 0.3569                     & 0.6680                     & 0.3642                     & 0.3432                     & 0.7231                     & 0.3343                     & 0.3322                     & 0.7219                     & 0.3206                     & 0.3343                     & 0.7206                     & 0.3272                     & 0.3401                     & 0.7045                     & 0.3381                     \\
\begin{tabular}[c]{@{}c@{}}Climate\\ FEVER\end{tabular}                   & 0.2160                     & 0.4921                     & 0.2966                     & 0.1663                     & 0.5052                     & 0.2178                     & 0.1517                     & 0.4808                     & 0.1985                     & 0.1397                     & 0.4739                     & 0.1813                     & 0.1355                     & 0.4650                     & 0.1718                     \\
DBPedia                                                                   & 0.3119                     & 0.4128                     & 0.6412                     & 0.2969                     & 0.4473                     & 0.5610                     & 0.2939                     & 0.4316                     & 0.5633                     & 0.2914                     & 0.4324                     & 0.5779                     & 0.2909                     & 0.4375                     & 0.5565                     \\
FEVER                                                                     & 0.6261                     & 0.9088                     & 0.5973                     & 0.3760                     & 0.7991                     & 0.3282                     & 0.3418                     & 0.7674                     & 0.2975                     & 0.3160                     & 0.7418                     & 0.2728                     & 0.3088                     & 0.7360                     & 0.2649                     \\
\begin{tabular}[c]{@{}c@{}}FiQA\\ 2018\end{tabular}                       & 0.2169                     & 0.5248                     & 0.2651                     & 0.2201                     & 0.5614                     & 0.2662                     & 0.2091                     & 0.5612                     & 0.2451                     & 0.2159                     & 0.5667                     & 0.2574                     & 0.2049                     & 0.5499                     & 0.2492                     \\
HotpotQA                                                                  & 0.4677                     & 0.6063                     & 0.6477                     & 0.4212                     & 0.6372                     & 0.5575                     & 0.3923                     & 0.6049                     & 0.5215                     & 0.3795                     & 0.5976                     & 0.5034                     & 0.3687                     & 0.5917                     & 0.4898                     \\
MSMARCO                                                                   & 0.5996                     & 0.4473                     & 0.8886                     & 0.4955                     & 0.4580                     & 0.7126                     & 0.4754                     & 0.4830                     & 0.7186                     & 0.4636                     & 0.4890                     & 0.7297                     & 0.4809                     & 0.5106                     & 0.7223                     \\
NFCorpus                                                                  & 0.2720                     & 0.2490                     & 0.4704                     & 0.3179                     & 0.3023                     & 0.5205                     & 0.3064                     & 0.3005                     & 0.4981                     & 0.3153                     & 0.3122                     & 0.4951                     & 0.3194                     & 0.3161                     & 0.4900                     \\
NQ                                                                        & 0.3380                     & 0.8066                     & 0.2913                     & 0.2978                     & 0.8240                     & 0.2472                     & 0.2750                     & 0.8078                     & 0.2240                     & 0.2837                     & 0.8150                     & 0.2335                     & 0.2865                     & 0.8194                     & 0.2360                     \\
\begin{tabular}[c]{@{}c@{}}Quora\\ Retrieval\end{tabular}                 & 0.8299                     & 0.9803                     & 0.8240                     & 0.8253                     & 0.9804                     & 0.8181                     & 0.8187                     & 0.9802                     & 0.8114                     & 0.8205                     & 0.9789                     & 0.8121                     & 0.8180                     & 0.9791                     & 0.8107                     \\
SCIDOCS                                                                   & 0.1297                     & 0.2990                     & 0.2412                     & 0.1425                     & 0.3495                     & 0.2573                     & 0.1373                     & 0.3404                     & 0.2447                     & 0.1391                     & 0.3423                     & 0.2481                     & 0.1385                     & 0.3428                     & 0.2470                     \\
SciFact                                                                   & 0.5322                     & 0.8610                     & 0.4932                     & 0.6119                     & 0.9150                    & 0.5710                     & 0.6090                     & 0.9210                     & 0.5692                     & 0.6253                     & 0.9283                     & 0.5927                     & 0.6201                     & 0.9377                     & 0.5853                     \\
Average                                                                   & \multicolumn{1}{c}{0.4088} & \multicolumn{1}{c}{0.6317} & \multicolumn{1}{c}{0.4889} & \multicolumn{1}{c}{0.3873} & \multicolumn{1}{c}{0.6526} & \multicolumn{1}{c}{0.4480} & \multicolumn{1}{c}{0.3731} & \multicolumn{1}{c}{0.6435} & \multicolumn{1}{c}{0.4333} & \multicolumn{1}{c}{0.3713} & \multicolumn{1}{c}{0.6435} & \multicolumn{1}{c}{0.4344} & \multicolumn{1}{c}{0.3703} & \multicolumn{1}{c}{0.6428} & \multicolumn{1}{c}{0.4290} \\ \bottomrule
\end{tabular}
\caption{Benchmark-wise metric scores on the BEIR benchmark, comparing DistilBERT baselines leveraging different augmentation densities using Llama 3.1 8B model. N indicates NDCG @ 10, R indicates Recall @ 100 and M indicates MRR @ 10}
\label{tab:8B_DB_density_metrics_BEIR}
\end{table*}

\begin{table*}[]
\centering
\tiny
\setlength{\tabcolsep}{3pt}
\begin{tabular}{@{}c|lll|lll|lll|lll|lll@{}}
\toprule
                                                                          & \multicolumn{3}{c}{No Aug}                                                           & \multicolumn{3}{c}{1PQ}                                                             & \multicolumn{3}{c}{3PQ}                                                             & \multicolumn{3}{c}{5PQ}                                                             & \multicolumn{3}{c}{7PQ}                                                            \\ \midrule
task\_name                                                                & \multicolumn{1}{c}{N}      & \multicolumn{1}{c}{R}      & \multicolumn{1}{c}{M}      & \multicolumn{1}{c}{N}      & \multicolumn{1}{c}{R}      & \multicolumn{1}{c}{M}      & \multicolumn{1}{c}{N}      & \multicolumn{1}{c}{R}      & \multicolumn{1}{c}{M}      & \multicolumn{1}{c}{N}      & \multicolumn{1}{c}{R}      & \multicolumn{1}{c}{M}      & \multicolumn{1}{c}{N}      & \multicolumn{1}{c}{R}      & \multicolumn{1}{c}{M}      \\
ArguAna                                                                   & 0.4173                     & 0.9566                     & 0.3345                     & 0.5348                     & 0.9836                     & 0.4489                     & 0.5337                     & 0.9865                     & 0.4449                     & 0.5437                     & 0.9865                     & 0.4588                     & 0.5350                     & 0.9836                     & 0.4488                     \\
\begin{tabular}[c]{@{}c@{}}CQADupstack\\ Android\\ Retrieval\end{tabular} & 0.3569                     & 0.6680                     & 0.3642                     & 0.3547                     & 0.7284                     & 0.3457                     & 0.3746                     & 0.7485                     & 0.3670                     & 0.3875                     & 0.7667                     & 0.3801                     & 0.3902                     & 0.7565                     & 0.3863                     \\
\begin{tabular}[c]{@{}c@{}}Climate\\ FEVER\end{tabular}                   & 0.2160                     & 0.4921                     & 0.2966                     & 0.1833                     & 0.5136                     & 0.2418                     & 0.1739                     & 0.5011                     & 0.2276                     & 0.1675                     & 0.5002                     & 0.2179                     & 0.1692                     & 0.5001                     & 0.2204                     \\
DBPedia                                                                   & 0.3119                     & 0.4128                     & 0.6412                     & 0.3054                     & 0.4385                     & 0.5919                     & 0.3007                     & 0.4458                     & 0.5837                     & 0.3055                     & 0.4425                     & 0.5767                     & 0.3004                     & 0.4487                     & 0.5770                     \\
FEVER                                                                     & 0.6261                     & 0.9088                     & 0.5973                     & 0.3906                     & 0.7800                     & 0.3496                     & 0.3241                     & 0.7299                     & 0.2848                     & 0.3016                     & 0.6949                     & 0.2623                     & 0.3048                     & 0.7016                     & 0.2648                     \\
\begin{tabular}[c]{@{}c@{}}FiQA\\ 2018\end{tabular}                       & 0.2169                     & 0.5248                     & 0.2651                     & 0.2463                     & 0.5965                     & 0.3003                     & 0.2576                     & 0.6101                     & 0.3093                     & 0.2698                     & 0.6240                     & 0.3227                     & 0.2679                     & 0.6159                     & 0.3229                     \\
HotpotQA                                                                  & 0.4677                     & 0.6063                     & 0.6477                     & 0.4111                     & 0.6285                     & 0.5451                     & 0.3855                     & 0.6140                     & 0.5082                     & 0.3855                     & 0.6127                     & 0.5052                     & 0.3776                     & 0.6134                     & 0.4941                     \\
MSMARCO                                                                   & 0.5996                     & 0.4473                     & 0.8886                     & 0.4981                     & 0.4539                     & 0.7498                     & 0.4811                     & 0.4831                     & 0.7129                     & 0.4806                     & 0.4892                     & 0.7291                     & 0.4487                     & 0.4827                     & 0.7122                     \\
NFCorpus                                                                  & 0.2720                     & 0.2490                     & 0.4704                     & 0.3115                     & 0.2933                     & 0.5076                     & 0.3310                     & 0.3164                     & 0.5306                     & 0.3312                     & 0.3262                     & 0.5154                     & 0.3330                     & 0.3286                     & 0.5068                     \\
NQ                                                                        & 0.3380                     & 0.8066                     & 0.2913                     & 0.2884                     & 0.8140                     & 0.2382                     & 0.2845                     & 0.8176                     & 0.2333                     & \multicolumn{1}{c}{0.2958} & \multicolumn{1}{c}{0.8282} & \multicolumn{1}{c}{0.2433} & 0.2939                     & 0.8237                     & 0.2430                     \\
\begin{tabular}[c]{@{}c@{}}Quora\\ Retrieval\end{tabular}                 & 0.8299                     & 0.9803                     & 0.8240                     & 0.8246                     & 0.9799                     & 0.8175                     & 0.8256                     & 0.9799                     & 0.8183                     & 0.8233                     & 0.9812                     & 0.8149                     & 0.8233                     & 0.9807                     & 0.8161                     \\
SCIDOCS                                                                   & 0.1297                     & 0.2990                     & 0.2412                     & 0.1393                     & 0.3477                     & 0.2516                     & 0.1434                     & 0.3537                     & 0.2583                     & 0.1438                     & 0.3548                     & 0.2579                     & 0.1412                     & 0.3566                     & 0.2552                     \\
SciFact                                                                   & 0.5322                     & 0.8610                     & 0.4932                     & 0.6163                     & 0.9033                     & 0.5786                     & 0.6193                     & 0.9183                     & 0.5849                     & 0.6326                     & 0.9210                     & 0.5978                     & 0.6265                     & 0.9177                     & 0.5935                     \\
Average                                                                   & \multicolumn{1}{c}{0.4088} & \multicolumn{1}{c}{0.6317} & \multicolumn{1}{c}{0.4889} & \multicolumn{1}{c}{0.3926} & \multicolumn{1}{c}{0.6509} & \multicolumn{1}{c}{0.4590} & \multicolumn{1}{c}{0.3873} & \multicolumn{1}{c}{0.6542} & \multicolumn{1}{c}{0.4511} & \multicolumn{1}{c}{0.3899} & \multicolumn{1}{c}{0.6560} & \multicolumn{1}{c}{0.4525} & \multicolumn{1}{c}{0.3855} & \multicolumn{1}{c}{0.6546} & \multicolumn{1}{c}{0.4493} \\ \bottomrule
\end{tabular}
\caption{Benchmark-wise metric scores on the BEIR benchmark, comparing DistilBERT baselines leveraging different augmentation densities using Llama 3.1 70B model. N indicates NDCG @ 10, R indicates Recall @ 100 and M indicates MRR @ 10}
\label{tab:70B_DB_density_metrics_BEIR}
\end{table*}

\clearpage

\begin{table*}[]
\centering
\tiny
\setlength{\tabcolsep}{3pt}
\begin{tabular}{@{}ccccllllllllllll@{}}
\toprule
                                                                          & \multicolumn{3}{c}{No Aug}                          & \multicolumn{3}{c}{1PQ}                                                              & \multicolumn{3}{c}{3PQ}                                                              & \multicolumn{3}{c}{5PQ}                                                              & \multicolumn{3}{c}{7PQ}                                                              \\ \midrule
task\_name                                                                & N               & R               & M               & \multicolumn{1}{c}{N}      & \multicolumn{1}{c}{R}      & \multicolumn{1}{c}{M}      & \multicolumn{1}{c}{N}      & \multicolumn{1}{c}{R}      & \multicolumn{1}{c}{M}      & \multicolumn{1}{c}{N}      & \multicolumn{1}{c}{R}      & \multicolumn{1}{c}{M}      & \multicolumn{1}{c}{N}      & \multicolumn{1}{c}{R}      & \multicolumn{1}{c}{M}      \\
ArguAna                                                                   & 0.4090          & 0.9538          & 0.3281          & 0.5144                     & 0.9815                     & 0.4272                     & 0.5194                     & 0.9858                     & 0.4291                     & 0.5260                     & 0.9851                     & 0.4382                     & 0.5242                     & 0.9865                     & 0.4374                     \\
\begin{tabular}[c]{@{}c@{}}CQADupstack\\ Android\\ Retrieval\end{tabular} & 0.3577          & 0.6641          & 0.3638          & 0.3620                     & 0.7239                     & 0.3549                     & 0.3606                     & 0.7135                     & 0.3539                     & 0.3675                     & 0.7319                     & 0.3624                     & 0.3755                     & 0.7369                     & 0.3678                     \\
\begin{tabular}[c]{@{}c@{}}Climate\\ FEVER\end{tabular}                   & 0.2116          & 0.4822          & 0.2891          & 0.2408                     & 0.5573                     & 0.3126                     & 0.2334                     & 0.5470                     & 0.3042                     & 0.2306                     & 0.5436                     & 0.3046                     & 0.2328                     & 0.5397                     & 0.3067                     \\
DBPedia                                                                   & 0.3095          & 0.4083          & 0.6369          & 0.3326                     & 0.4394                     & 0.6766                     & 0.3289                     & 0.4485                     & 0.6670                     & 0.3245                     & 0.4481                     & 0.6539                     & 0.3279                     & 0.4553                     & 0.6500                     \\
FEVER                                                                     & 0.6391          & 0.9040          & 0.6131          & 0.6453                     & 0.9243                     & 0.6126                     & 0.6331                     & 0.9215                     & 0.5974                     & 0.6273                     & 0.9191                     & 0.5906                     & 0.6239                     & 0.9152                     & 0.5863                     \\
\begin{tabular}[c]{@{}c@{}}FiQA\\ 2018\end{tabular}                       & 0.2242          & 0.5166          & 0.2782          & 0.2332                     & 0.5879                     & 0.2853                     & 0.2412                     & 0.5971                     & 0.2950                     & 0.2404                     & 0.5984                     & 0.2949                     & 0.2432                     & 0.6040                     & 0.2956                     \\
HotpotQA                                                                  & 0.4557          & 0.5897          & 0.6339          & 0.5001                     & 0.6592                     & 0.6739                     & 0.4983                     & 0.6675                     & 0.6712                     & 0.4995                     & 0.6644                     & 0.6711                     & 0.4961                     & 0.6685                     & 0.6637                     \\
MSMARCO                                                                   & 0.6113          & 0.4623          & 0.8826          & 0.5844                     & 0.4724                     & 0.9078                     & 0.5684                     & 0.4965                     & 0.8004                     & 0.5736                     & 0.4800                     & 0.8760                     & 0.5744                     & 0.5040                     & 0.8624                     \\
NFCorpus                                                                  & 0.2709          & 0.2505          & 0.4654          & 0.3010                     & 0.2892                     & 0.5016                     & 0.2998                     & 0.2849                     & 0.4998                     & 0.3064                     & 0.3030                     & 0.4972                     & 0.3089                     & 0.2999                     & 0.5095                     \\
NQ                                                                        & 0.3467          & 0.8183          & 0.2987          & 0.3948                     & 0.8632                     & 0.3439                     & 0.3958                     & 0.8717                     & 0.3422                     & 0.3946                     & 0.8686                     & 0.3405                     & 0.3999                     & 0.8765                     & 0.3450                     \\
\begin{tabular}[c]{@{}c@{}}Quora\\ Retrieval\end{tabular}                 & 0.8284          & 0.9792          & 0.8221          & 0.8228                     & 0.9809                     & 0.8150                     & 0.8201                     & 0.9789                     & 0.8120                     & 0.8186                     & 0.9785                     & 0.8108                     & 0.8210                     & 0.9801                     & 0.8125                     \\
SCIDOCS                                                                   & 0.1293          & 0.2959          & 0.2432          & 0.1471                     & 0.3432                     & 0.2639                     & 0.1479                     & 0.3467                     & 0.2639                     & 0.1528                     & 0.3452                     & 0.2795                     & 0.1520                     & 0.3436                     & 0.2771                     \\
SciFact                                                                   & 0.5187          & 0.8560          & 0.4867          & 0.6146                     & 0.9010                     & 0.5753                     & 0.6314                     & 0.9077                     & 0.5869                     & 0.6226                     & 0.9150                     & 0.5797                     & 0.6259                     & 0.9100                     & 0.5811                     \\
Average                                                                   & 0.4086 & 0.6293 & 0.4878 & \multicolumn{1}{c}{0.4379} & \multicolumn{1}{c}{0.6710} & \multicolumn{1}{c}{0.5193} & \multicolumn{1}{c}{0.4368} & \multicolumn{1}{c}{0.6744} & \multicolumn{1}{c}{0.5095} & \multicolumn{1}{c}{0.4373} & \multicolumn{1}{c}{0.6754} & \multicolumn{1}{c}{0.5153} & \multicolumn{1}{c}{0.4389} & \multicolumn{1}{c}{0.6785} & \multicolumn{1}{c}{0.5150} \\ \bottomrule
\end{tabular}
\caption{Benchmark-wise metric scores on the BEIR benchmark, comparing BERTBASE baselines leveraging different augmentation densities using Mistral-7B Instruct v0.3. N indicates NDCG @ 10, R indicates Recall @ 100 and M indicates MRR @ 10}
\label{tab:70B_BERT_mistral_density_metrics_BEIR}
\end{table*}

\begin{table*}[]
\centering
\tiny
\setlength{\tabcolsep}{3pt}
\begin{tabular}{@{}ccccllllllllllll@{}}
\toprule
                                                                          & \multicolumn{3}{c}{No Aug} & \multicolumn{3}{c}{1PQ}                                                              & \multicolumn{3}{c}{3PQ}                                                              & \multicolumn{3}{c}{5PQ}                                                              & \multicolumn{3}{c}{7PQ}                                                              \\ \midrule
task\_name                                                                & N       & R       & M      & \multicolumn{1}{c}{N}      & \multicolumn{1}{c}{R}      & \multicolumn{1}{c}{M}      & \multicolumn{1}{c}{N}      & \multicolumn{1}{c}{R}      & \multicolumn{1}{c}{M}      & \multicolumn{1}{c}{N}      & \multicolumn{1}{c}{R}      & \multicolumn{1}{c}{M}      & \multicolumn{1}{c}{N}      & \multicolumn{1}{c}{R}      & \multicolumn{1}{c}{M}      \\
ArguAna                                                                   & 0.4090  & 0.9538  & 0.3281 & 0.5147                     & 0.9844                     & 0.4244                     & 0.5214                     & 0.9822                     & 0.4310                     & 0.5145                     & 0.9851                     & 0.4273                     & 0.5260                     & 0.9851                     & 0.4382                     \\
\begin{tabular}[c]{@{}c@{}}CQADupstack\\ Android\\ Retrieval\end{tabular} & 0.3577  & 0.6641  & 0.3638 & 0.3746                     & 0.7364                     & 0.3688                     & 0.3600                     & 0.7177                     & 0.3560                     & 0.3752                     & 0.7338                     & 0.3705                     & 0.3675                     & 0.7319                     & 0.3624                     \\
\begin{tabular}[c]{@{}c@{}}Climate\\ FEVER\end{tabular}                   & 0.2116  & 0.4822  & 0.2891 & 0.2478                     & 0.5682                     & 0.3276                     & 0.2367                     & 0.5489                     & 0.3151                     & 0.2357                     & 0.5478                     & 0.3129                     & 0.2306                     & 0.5436                     & 0.3046                     \\
DBPedia                                                                   & 0.3095  & 0.4083  & 0.6369 & 0.3173                     & 0.4368                     & 0.6463                     & 0.3304                     & 0.4436                     & 0.6697                     & 0.3323                     & 0.4439                     & 0.6702                     & 0.3245                     & 0.4481                     & 0.6539                     \\
FEVER                                                                     & 0.6391  & 0.9040  & 0.6131 & 0.6314                     & 0.9224                     & 0.5954                     & 0.6389                     & 0.9181                     & 0.6040                     & 0.6254                     & 0.9129                     & 0.5897                     & 0.6273                     & 0.9191                     & 0.5906                     \\
\begin{tabular}[c]{@{}c@{}}FiQA\\ 2018\end{tabular}                       & 0.2242  & 0.5166  & 0.2782 & 0.2409                     & 0.5904                     & 0.2862                     & 0.2395                     & 0.5929                     & 0.2976                     & 0.2415                     & 0.5956                     & 0.2935                     & 0.2404                     & 0.5984                     & 0.2949                     \\
HotpotQA                                                                  & 0.4557  & 0.5897  & 0.6339 & \multicolumn{1}{c}{0.4976} & 0.6634                     & 0.6674                     & 0.4951                     & 0.6608                     & 0.6680                     & 0.4949                     & 0.6633                     & 0.6652                     & 0.4995                     & 0.6644                     & 0.6711                     \\
MSMARCO                                                                   & 0.6113  & 0.4623  & 0.8826 & 0.5792                     & 0.4822                     & 0.8864                     & 0.5650                     & 0.4783                     & 0.8333                     & 0.5687                     & 0.4891                     & 0.8636                     & 0.5736                     & 0.4800                     & 0.8760                     \\
NFCorpus                                                                  & 0.2709  & 0.2505  & 0.4654 & 0.2975                     & 0.2874                     & 0.4887                     & 0.2978                     & 0.2913                     & 0.4841                     & 0.3048                     & 0.2925                     & 0.5025                     & 0.3064                     & 0.3030                     & 0.4972                     \\
NQ                                                                        & 0.3467  & 0.8183  & 0.2987 & 0.3998                     & 0.8701                     & 0.3475                     & 0.3942                     & 0.8613                     & 0.3417                     & 0.3964                     & 0.8700                     & 0.3423                     & 0.3946                     & 0.8686                     & 0.3405                     \\
\begin{tabular}[c]{@{}c@{}}Quora\\ Retrieval\end{tabular}                 & 0.8284  & 0.9792  & 0.8221 & 0.8223                     & 0.9803                     & 0.8143                     & 0.8225                     & 0.9805                     & 0.8152                     & 0.8197                     & 0.9780                     & 0.8112                     & 0.8186                     & 0.9785                     & 0.8108                     \\
SCIDOCS                                                                   & 0.1293  & 0.2959  & 0.2432 & 0.1464                     & 0.3442                     & 0.2644                     & 0.1475                     & 0.3350                     & 0.2665                     & 0.1485                     & 0.3462                     & 0.2720                     & 0.1528                     & 0.3452                     & 0.2795                     \\
SciFact                                                                   & 0.5187  & 0.8560  & 0.4867 & 0.6204                     & 0.9077                     & 0.5813                     & 0.6116                     & 0.9133                     & 0.5660                     & 0.6102                     & 0.9133                     & 0.5705                     & 0.6226                     & 0.9150                     & 0.5797                     \\
Average                                                                   & 0.4086  & 0.6293  & 0.4878 & \multicolumn{1}{c}{0.4377} & \multicolumn{1}{c}{0.6749} & \multicolumn{1}{c}{0.5153} & \multicolumn{1}{c}{0.4354} & \multicolumn{1}{c}{0.6711} & \multicolumn{1}{c}{0.5114} & \multicolumn{1}{c}{0.4360} & \multicolumn{1}{c}{0.6747} & \multicolumn{1}{c}{0.5147} & \multicolumn{1}{c}{0.4373} & \multicolumn{1}{c}{0.6754} & \multicolumn{1}{c}{0.5153} \\ \bottomrule
\end{tabular}
\caption{Benchmark-wise metric scores on the BEIR benchmark, comparing BERTBASE baselines leveraging different augmentation scales using Mistral-7B Instruct v0.3. N indicates NDCG @ 10, R indicates Recall @ 100 and M indicates MRR @ 10}
\label{tab:70B_BERT_mistral_datascale_metrics_BEIR}
\end{table*}

\begin{table*}[]
\centering
\tiny
\setlength{\tabcolsep}{3pt}
\begin{tabular}{@{}ccccllllllllllll@{}}
\toprule
                                                                          & \multicolumn{3}{c}{No Aug}                                                           & \multicolumn{3}{c}{1PQ}                                               & \multicolumn{3}{c}{3PQ}                                               & \multicolumn{3}{c}{5PQ}                                               & \multicolumn{3}{c}{7PQ}                                               \\ \midrule
task\_name                                                                & N                          & R                          & M                          & \multicolumn{1}{c}{N} & \multicolumn{1}{c}{R} & \multicolumn{1}{c}{M} & \multicolumn{1}{c}{N} & \multicolumn{1}{c}{R} & \multicolumn{1}{c}{M} & \multicolumn{1}{c}{N} & \multicolumn{1}{c}{R} & \multicolumn{1}{c}{M} & \multicolumn{1}{c}{N} & \multicolumn{1}{c}{R} & \multicolumn{1}{c}{M} \\
ArguAna                                                                   & 0.4705                     & 0.9751                     & 0.3829                     & 0.5270                & 0.9858                & 0.4347                & 0.5385                & 0.9879                & 0.4483                & 0.5468                & 0.9858                & 0.4554                & 0.5404                & 0.9858                & 0.4500                \\
\begin{tabular}[c]{@{}c@{}}CQADupstack\\ Android\\ Retrieval\end{tabular} & 0.4181                     & 0.7520                     & 0.4132                     & 0.3962                & 0.7647                & 0.3934                & 0.3909                & 0.7538                & 0.3881                & 0.3958                & 0.7543                & 0.3947                & 0.3868                & 0.7650                & 0.3823                \\
\begin{tabular}[c]{@{}c@{}}Climate\\ FEVER\end{tabular}                   & 0.2370                     & 0.5419                     & 0.3210                     & 0.2624                & 0.5897                & 0.3458                & 0.2514                & 0.5772                & 0.3292                & 0.2419                & 0.5689                & 0.3183                & 0.2358                & 0.5548                & 0.3139                \\
DBPedia                                                                   & 0.3713                     & 0.4945                     & 0.7022                     & 0.3624                & 0.5075                & 0.6972                & 0.3526                & 0.5005                & 0.6948                & 0.3454                & 0.4893                & 0.6794                & 0.3415                & 0.4773                & 0.6648                \\
FEVER                                                                     & 0.7064                     & 0.9366                     & 0.6869                     & 0.6883                & 0.9414                & 0.6554                & 0.6753                & 0.9378                & 0.6419                & 0.6529                & 0.9329                & 0.6166                & 0.6480                & 0.9312                & 0.6110                \\
\begin{tabular}[c]{@{}c@{}}FiQA\\ 2018\end{tabular}                       & 0.2672                     & 0.6061                     & 0.3283                     & 0.2560                & 0.6206                & 0.3136                & 0.2564                & 0.6279                & 0.3165                & 0.2611                & 0.6192                & 0.3174                & 0.2558                & 0.6274                & 0.3104                \\
HotpotQA                                                                  & 0.5409                     & 0.6824                     & 0.7206                     & 0.5690                & 0.7255                & 0.7450                & 0.5603                & 0.7259                & 0.7322                & 0.5560                & 0.7211                & 0.7278                & 0.5530                & 0.7204                & 0.7241                \\
MSMARCO                                                                   & 0.5913                     & 0.4800                     & 0.8667                     & 0.5782                & 0.4969                & 0.8492                & 0.5919                & 0.4899                & 0.8760                & 0.5838                & 0.5006                & 0.8820                & 0.5697                & 0.5017                & 0.8806                \\
NFCorpus                                                                  & 0.3121                     & 0.2964                     & 0.5020                     & 0.3204                & 0.3123                & 0.5199                & 0.3204                & 0.3225                & 0.5168                & 0.3269                & 0.3176                & 0.5313                & 0.3216                & 0.3205                & 0.5190                \\
NQ                                                                        & 0.3762                     & 0.8692                     & 0.3191                     & 0.4088                & 0.8960                & 0.3527                & 0.4058                & 0.8951                & 0.3493                & 0.4060                & 0.8930                & 0.3485                & 0.4051                & 0.8938                & 0.3486                \\
\begin{tabular}[c]{@{}c@{}}Quora\\ Retrieval\end{tabular}                 & 0.8470                     & 0.9871                     & 0.8385                     & 0.8332                & 0.9855                & 0.8250                & 0.8300                & 0.9829                & 0.8215                & 0.8255                & 0.9821                & 0.8186                & 0.8242                & 0.9818                & 0.8153                \\
SCIDOCS                                                                   & 0.1578                     & 0.3568                     & 0.2898                     & 0.1647                & 0.3792                & 0.2942                & 0.1636                & 0.3746                & 0.2906                & 0.1641                & 0.3776                & 0.2872                & 0.1627                & 0.3729                & 0.2869                \\
SciFact                                                                   & 0.6268                     & 0.9277                     & 0.5843                     & 0.6730                & 0.9500                & 0.6349                & 0.6707                & 0.9500                & 0.6312                & 0.6674                & 0.9383                & 0.6276                & 0.6691                & 0.9500                & 0.6313                \\
Average                                                                   & \multicolumn{1}{l}{0.4556} & \multicolumn{1}{l}{0.6851} & \multicolumn{1}{l}{0.5350} & 0.4646                & 0.7042                & 0.5432                & 0.4621                & 0.7020                & 0.5413                & 0.4595                & 0.6985                & 0.5388                & 0.4549                & 0.6987                & 0.5337                \\ \bottomrule
\end{tabular}
\caption{Benchmark-wise metric scores on the BEIR benchmark, comparing Contriever baselines leveraging different augmentation scales using Mistral-7B Instruct v0.3. N indicates NDCG @ 10, R indicates Recall @ 100 and M indicates MRR @ 10}
\label{tab:70B_Contriever_mistral_datascale_metrics_BEIR}
\end{table*}

\begin{table*}[]
\centering
\tiny
\setlength{\tabcolsep}{3pt}
\begin{tabular}{@{}ccccllllllllllll@{}}
\toprule
                                                                          & \multicolumn{3}{c}{No Aug}                                                           & \multicolumn{3}{c}{1PQ}                                               & \multicolumn{3}{c}{3PQ}                                               & \multicolumn{3}{c}{5PQ}                                               & \multicolumn{3}{c}{7PQ}                                               \\ \midrule
task\_name                                                                & N                          & R                          & M                          & \multicolumn{1}{c}{N} & \multicolumn{1}{c}{R} & \multicolumn{1}{c}{M} & \multicolumn{1}{c}{N} & \multicolumn{1}{c}{R} & \multicolumn{1}{c}{M} & \multicolumn{1}{c}{N} & \multicolumn{1}{c}{R} & \multicolumn{1}{c}{M} & \multicolumn{1}{c}{N} & \multicolumn{1}{c}{R} & \multicolumn{1}{c}{M} \\
ArguAna                                                                   & 0.4705                     & 0.9751                     & 0.3829                     & 0.5270                & 0.9858                & 0.4347                & 0.5385                & 0.9879                & 0.4483                & 0.5468                & 0.9858                & 0.4554                & 0.5404                & 0.9858                & 0.4500                \\
\begin{tabular}[c]{@{}c@{}}CQADupstack\\ Android\\ Retrieval\end{tabular} & 0.4181                     & 0.7520                     & 0.4132                     & 0.3962                & 0.7647                & 0.3934                & 0.3909                & 0.7538                & 0.3881                & 0.3958                & 0.7543                & 0.3947                & 0.3868                & 0.7650                & 0.3823                \\
\begin{tabular}[c]{@{}c@{}}Climate\\ FEVER\end{tabular}                   & 0.2370                     & 0.5419                     & 0.3210                     & 0.2624                & 0.5897                & 0.3458                & 0.2514                & 0.5772                & 0.3292                & 0.2419                & 0.5689                & 0.3183                & 0.2358                & 0.5548                & 0.3139                \\
DBPedia                                                                   & 0.3713                     & 0.4945                     & 0.7022                     & 0.3624                & 0.5075                & 0.6972                & 0.3526                & 0.5005                & 0.6948                & 0.3454                & 0.4893                & 0.6794                & 0.3415                & 0.4773                & 0.6648                \\
FEVER                                                                     & 0.7064                     & 0.9366                     & 0.6869                     & 0.6883                & 0.9414                & 0.6554                & 0.6753                & 0.9378                & 0.6419                & 0.6529                & 0.9329                & 0.6166                & 0.6480                & 0.9312                & 0.6110                \\
\begin{tabular}[c]{@{}c@{}}FiQA\\ 2018\end{tabular}                       & 0.2672                     & 0.6061                     & 0.3283                     & 0.2560                & 0.6206                & 0.3136                & 0.2564                & 0.6279                & 0.3165                & 0.2611                & 0.6192                & 0.3174                & 0.2558                & 0.6274                & 0.3104                \\
HotpotQA                                                                  & 0.5409                     & 0.6824                     & 0.7206                     & 0.5690                & 0.7255                & 0.7450                & 0.5603                & 0.7259                & 0.7322                & 0.5560                & 0.7211                & 0.7278                & 0.5530                & 0.7204                & 0.7241                \\
MSMARCO                                                                   & 0.5913                     & 0.4800                     & 0.8667                     & 0.5782                & 0.4969                & 0.8492                & 0.5919                & 0.4899                & 0.8760                & 0.5838                & 0.5006                & 0.8820                & 0.5697                & 0.5017                & 0.8806                \\
NFCorpus                                                                  & 0.3121                     & 0.2964                     & 0.5020                     & 0.3204                & 0.3123                & 0.5199                & 0.3204                & 0.3225                & 0.5168                & 0.3269                & 0.3176                & 0.5313                & 0.3216                & 0.3205                & 0.5190                \\
NQ                                                                        & 0.3762                     & 0.8692                     & 0.3191                     & 0.4088                & 0.8960                & 0.3527                & 0.4058                & 0.8951                & 0.3493                & 0.4060                & 0.8930                & 0.3485                & 0.4051                & 0.8938                & 0.3486                \\
\begin{tabular}[c]{@{}c@{}}Quora\\ Retrieval\end{tabular}                 & 0.8470                     & 0.9871                     & 0.8385                     & 0.8332                & 0.9855                & 0.8250                & 0.8300                & 0.9829                & 0.8215                & 0.8255                & 0.9821                & 0.8186                & 0.8242                & 0.9818                & 0.8153                \\
SCIDOCS                                                                   & 0.1578                     & 0.3568                     & 0.2898                     & 0.1647                & 0.3792                & 0.2942                & 0.1636                & 0.3746                & 0.2906                & 0.1641                & 0.3776                & 0.2872                & 0.1627                & 0.3729                & 0.2869                \\
SciFact                                                                   & 0.6268                     & 0.9277                     & 0.5843                     & 0.6730                & 0.9500                & 0.6349                & 0.6707                & 0.9500                & 0.6312                & 0.6674                & 0.9383                & 0.6276                & 0.6691                & 0.9500                & 0.6313                \\
Average                                                                   & \multicolumn{1}{l}{0.4556} & \multicolumn{1}{l}{0.6851} & \multicolumn{1}{l}{0.5350} & 0.4646                & 0.7042                & 0.5432                & 0.4621                & 0.7020                & 0.5413                & 0.4595                & 0.6985                & 0.5388                & 0.4549                & 0.6987                & 0.5337                \\ \bottomrule
\end{tabular}
\caption{Benchmark-wise metric scores on the BEIR benchmark, comparing Contriever baselines leveraging different augmentation scales using Mistral-7B Instruct v0.3. N indicates NDCG @ 10, R indicates Recall @ 100 and M indicates MRR @ 10}
\label{tab:70B_Contriever_mistral_density_metrics_BEIR}
\end{table*}

\clearpage
\section{Prompts used for Data Augmentation}
\label{app:prompt}

\subsection{Information Seeking}
\begin{tcolorbox}[
    colback=promptbg,
    colframe=promptborder,
    sharp corners,
    boxrule=1pt,
    width=\textwidth,
    title={\textbf{DOC2QUERY SYSTEM PROMPT}},
    fonttitle=\bfseries\large,
    coltitle=black
]
\textbf{INSTRUCTION:} You are provided with a passage and tasked with generating targeted, unique search queries for a retrieval and search ads platform.

Your goal is to create up to \texttt{\{max\_pseudo\_queries\}} diverse search queries to maximize the visibility and engagement potential of the passage.

Each generated query should focus on unique facets of the passage content without duplicating the same search intent or generating trivial, surface-level queries based solely on text overlap.

\textbf{Guidelines:}
\begin{itemize}
    \item \textbf{Goal:} Enable efficient content retrieval and improve ad relevance by creating queries that capture distinct, meaningful aspects of the passage.
    \item \textbf{Diversity:} Each query should reflect a different perspective or information focus within the passage, ensuring that no two queries serve the exact same user intent.
    \item \textbf{Avoid Redundancy:} Do not include trivial or repetitive queries that a basic model could generate using mere text overlap; aim for sophisticated, contextually rich queries that reveal nuanced insights within the passage.
\end{itemize}

\textbf{Output:} The output should consist strictly of search queries without any bullet points or numbers, separated by newlines. Begin with the queries immediately.

\end{tcolorbox}

\subsection{Fact Checking}
\begin{tcolorbox}[
    colback=promptbg,
    colframe=promptborder,
    sharp corners,
    boxrule=1pt,
    width=\textwidth,
    title={\textbf{FACTCHECK SYSTEM PROMPT}},
    fonttitle=\bfseries\large,
    coltitle=black
]
\textbf{INSTRUCTION:} 

You are tasked with generating up to \texttt{\{max\_pseudo\_queries\}} diverse pseudo queries for a given passage that capture all unique facts, specifically focusing on fact-checking intents rather than solely information-seeking or question-answering ones. 

Your goal is to develop queries that act as claims, each of which can be verified or refuted with evidence directly found in the passage. 

These queries should challenge a retrieval system's ability to discern truth from falsehood based on the passage content.

To achieve this, ensure that each pseudo query presents a statement or claim that requires verification and the passage supports the statement. 

Leverage your world knowledge to craft queries that may necessitate more than traditional Dense retrievers for accurate resolution, particularly due to the need for nuanced contextual understanding or external knowledge about the passage.

Remember, your pseudo queries will serve as gold standard Query-Positive pairs to enhance dense retrieval models, aiming to improve their fact-checking capabilities. 

The final pseudo queries should be diverse, insightful, and directly verifiable with evidence in the passage, ensuring they serve as valuable training data for refining retrieval systems.

You should strictly avoid redundancy, ensuring that each pseudo query is distinct in its claim.

Do not generate any information-seeking or question-answering queries; focus solely on fact-checking intents that require evidence grounded in the passage.

Output should consist strictly of pseudo queries formatted as factual claims without any bullet points or numbers, separated by newlines. Return only the final set of pseudo queries, with each pseudo query separated by \texttt{"\textbackslash n"}.
\end{tcolorbox}

\clearpage

\subsection{Argument Retrieval}
\begin{tcolorbox}[
    colback=promptbg,
    colframe=promptborder,
    sharp corners,
    boxrule=1pt,
    width=\textwidth,
    title={\textbf{ARG SYSTEM PROMPT}},
    fonttitle=\bfseries\large,
    coltitle=black
]
\textbf{INSTRUCTION:}

You are tasked with generating up to \texttt{\{max\_pseudo\_queries\}} diverse pseudo queries for a given passage that capture fact-checking intents related to the passage. These queries should aim to verify the authenticity, accuracy, or credibility of claims within the passage by seeking evidence directly grounded in it. 

Develop queries that can challenge assertions in the passage, prompting verification and fact-checking rather than simple information retrieval or question-answering. 

Leverage your understanding of the passage and broader world knowledge to create queries that require nuanced interpretation and critical analysis, ensuring they go beyond the capabilities of traditional Dense retrievers due to their need for contextual understanding or external validation.

Your final pseudo queries should be diverse, insightful, and focused on fact-checking, ensuring they serve as valuable training data for improving retrieval systems. They should act as claims that require fact-checking, with evidence found in the passage that refutes the claim, to enhance dense retrieval models focused on verification tasks.

Strictly avoid redundancy, ensuring that each pseudo query is distinct in its claim and the evidence to refute it is strictly available in the passage.

Start directly with the counter-argument claim and do not introduce the counter-argument with any phrases. 

Output should consist strictly of search queries without any bullet points or numbers, separated by newlines. Return only the final set of pseudo queries, with each pseudo query separated by \texttt{"\textbackslash n"}. Do not provide any other prefix or suffix text and do not introduce the motivation behind the pseudo query.
\end{tcolorbox}

\clearpage

\subsection{Citation Prediction}
\begin{tcolorbox}[
    colback=promptbg,
    colframe=promptborder,
    sharp corners,
    boxrule=1pt,
    width=\textwidth,
    title={\textbf{TITLE SYSTEM PROMPT}},
    fonttitle=\bfseries\large,
    coltitle=black
]
\textbf{INSTRUCTION:}

You are tasked with generating up to \texttt{\{max\_pseudo\_queries\}} diverse pseudo queries for a given passage that strictly capture citation-prediction intents. These queries should serve as references or citations to the passage, challenging assertions and prompting nuanced interpretation. 

Your goal is to develop queries that require critical analysis and contextual understanding, surpassing simple information retrieval or question-answering.

Identify key elements that could be cited in academic or professional contexts and accurately reflect the title of the work.

The pseudo query should encapsulate the essence of the passage, acting as a reference or citation that could be used to locate the passage in a broader context.

It should not reflect content that is not present in the passage but should be a concise and accurate representation of the passage's main idea or argument.

Each pseudo query should be distinct, insightful, and diverse in focus. Avoid redundancy by ensuring that no two queries address the same aspect of the passage.

Ensure that the pseudo queries are directly related to the passage content and can be used as titles or references in academic or professional contexts.

Present the final set of pseudo queries which act as titles in plain text, each on a new line without bullet points or numbers. Directly start with the final pseudo queries, ensuring they are suitable for training citation prediction systems. Do not provide any other prefix or suffix text and do not introduce the motivation behind the pseudo query.
\end{tcolorbox}

\clearpage

\subsection{Entity Retrieval}
\begin{tcolorbox}[
    colback=promptbg,
    colframe=promptborder,
    sharp corners,
    boxrule=1pt,
    width=\textwidth,
    title={\textbf{ENTITY SYSTEM PROMPT}},
    fonttitle=\bfseries\large,
    coltitle=black
]

\textbf{INSTRUCTION:}

You are tasked with generating up to 10 diverse pseudo queries for a given passage that focus specifically on entity retrieval. These pseudo queries should capture the intent of retrieving principal entities referenced in the content of the passage and serve as a reference or citation to the passage.

Your goal is to develop queries that prioritize the retrieval of entities mentioned in the passage, rather than seeking factual information or direct answers. 

The queries should require a contextual understanding of the passage, emphasizing the retrieval and understanding of entities within the passage.

\clearpage

\textbf{To ensure the pseudo queries are effective:}
\begin{itemize}
    \item \textbf{Focus on Entities:} Each pseudo query must strictly aim to retrieve entities explicitly mentioned in the passage. Avoid any queries that have an information-seeking or question-answering nature.
    \item \textbf{Contextual Relevance:} Ensure that each query is contextually grounded in the passage, capturing unique intents related to retrieving entities. Avoid generic queries that could apply broadly and are not tied to the specific entities within the passage.
    \item \textbf{Diverse Intents:} Each pseudo query should represent a distinct retrieval intent. Avoid redundancy by ensuring no two queries reflect the same retrieval intent or focus on the same aspect of an entity.
    \item \textbf{Avoid Traditional Question Formats:} The queries should not be well-formed questions but should act as pointers to entities, reflecting a need for entity retrieval rather than factual verification or seeking direct answers.
\end{itemize}

Output should consist strictly of search queries without any bullet points or numbers, separated by newlines. Return only the final set of pseudo queries, with each pseudo query separated by a newline. 

Do not provide any other prefix or suffix text, and do not use the word "entity" in pseudo queries.
\end{tcolorbox}
\clearpage

\clearpage

\end{document}